\documentclass[longauth]{aa}

\usepackage{newtxtext,newtxmath,enumitem,multirow}
\usepackage[utf8]{inputenc}
\usepackage[T1]{fontenc}
\usepackage[breaklinks,colorlinks,allcolors=blue,citecolor=blue,urlcolor=blue]{hyperref}
\usepackage{orcidlink}
\usepackage{ctable}   
\pdfoutput=1 
\usepackage{xcolor}
\usepackage{graphicx}

\usepackage{subcaption}
\usepackage[normalem]{ulem}
\renewcommand{\linenumbers}{}

\usepackage{chngcntr}

\usepackage{titlesec}
\titleformat{\paragraph}[runin]{\bfseries}{\theparagraph}{1em}{}
\titlespacing*{\paragraph}{0pt}{1ex}{1em}

\titlespacing*{\section}{0pt}{2ex}{1ex}
\titlespacing*{\subsection}{0pt}{1.5ex}{0.8ex}
\titlespacing*{\subsubsection}{0pt}{1ex}{0.5ex}

\newcommand{\tess}{\textit{TESS}}

\newcommand{\allesfitter}{\texttt{allesfitter}}
\newcommand{\triceratops}{\texttt{TRICERATOPS}}
\newcommand{\lightkurve}{\texttt{lightkurve}}

\newcommand{\edit}[1]{\textbf{\textcolor{red}{#1}}}

\newcommand{\brrA}{$0.07170_{-0.00061}^{+0.00065}$}
\newcommand{\brsumaA}{$0.01191_{-0.00021}^{+0.00037}$} 
\newcommand{\bcosiA}{$0.0024\pm0.0013$}
\newcommand{\bepochA}{$2459669.24457\pm0.00054$} 
\newcommand{\bperiodA}{$104.616063\pm0.000081$} 
\newcommand{\bfcA}{--} 
\newcommand{\bfsA}{--} 
 
\newcommand{\hostldcqoneTESSA}{$0.300_{-0.081}^{+0.088}$}  
\newcommand{\hostldcqtwoTESSA}{$0.086_{-0.056}^{+0.091}$}

\newcommand{\lnerrfluxTESSA}{$-6.61892\pm0.00098$} 
\newcommand{\lnerrfluxASTEPtwozerotwothreeA}{$-5.591\pm0.018$} 
\newcommand{\lnerrfluxASTEPtwozerotwothreeBA}{$-6.122\pm0.045$} 
\newcommand{\lnerrfluxASTEPtwozerotwofiveA}{$-4.6622\pm0.0093$}
\newcommand{\lnjitterrvFEROSA}{$-3.49\pm0.31$} 
\newcommand{\lnjitterrvHARPSoneA}{$-4.87\pm0.28$} 
\newcommand{\lnjitterrvHARPStwoA}{$-4.44_{-0.27}^{+0.29}$} 
\newcommand{\baselinegpmaternthreetwolnsigmafluxTESSA}{$-7.463_{-0.019}^{+0.018}$} 
\newcommand{\baselinegpmaternthreetwolnrhofluxTESSA}{$-0.811\pm0.025$} 
\newcommand{\baselineoffsetrvFEROSA}{$26.051_{-0.011}^{+0.010}$} 
\newcommand{\baselineoffsetrvHARPSoneA}{$26.0584\pm0.0027$} 
\newcommand{\baselineoffsetrvHARPStwoA}{$26.0593_{-0.0070}^{+0.0080}$} 
\newcommand{\stellarvarslopervA}{$-0.027_{-0.015}^{+0.013}$}

\newcommand{\bRstaraA}{$0.01111_{-0.00019}^{+0.00034}$} 
\newcommand{\baRstarA}{$90.0_{-2.7}^{+1.6}$} 
\newcommand{\bRcompanionaA}{$0.000796_{-0.000017}^{+0.000030}$} 
\newcommand{\bRcompanionRearthA}{$8.13\pm0.29$} 
\newcommand{\bRcompanionRjupA}{$0.725\pm0.026$} 
\newcommand{\baRsunA}{$93.1_{-4.0}^{+3.8}$} 
\newcommand{\baAUA}{$0.433_{-0.019}^{+0.018}$} 
\newcommand{\biA}{$89.865\pm0.077$}

\newcommand{\bbtraA}{$0.21\pm0.12$} 
\newcommand{\bTtratotA}{$9.329_{-0.043}^{+0.048}$} 
\newcommand{\bTtrafullA}{$8.018_{-0.070}^{+0.052}$}

\newcommand{\bTeqA}{$425.9_{-8.1}^{+8.9}$}

\newcommand{\combinedhostdensityA}{$1.260_{-0.11}^{+0.067}$}

\newcommand{\BrrB}{$0.117_{-0.028}^{+0.18}$} 
\newcommand{\BrsumaB}{$0.0157_{-0.0015}^{+0.0040}$} 
\newcommand{\BcosiB}{$0.0139_{-0.0016}^{+0.0044}$} 
\newcommand{\BepochB}{$2459842.83508_{-0.00050}^{+0.00053}$} 
\newcommand{\BperiodB}{$39.623945_{-0.000041}^{+0.000043}$}

\newcommand{\BsbratioTESSB}{$0.0095_{-0.0064}^{+0.011}$}

\newcommand{\hostldcqoneTESSB}{$0.77_{-0.13}^{+0.12}$}  
\newcommand{\hostldcqtwoTESSB}{$0.25_{-0.16}^{+0.22}$}

\newcommand{\lnerrfluxTESSB}{$-6.4523_{-0.0020}^{+0.0022}$} 
\newcommand{\lnerrfluxASTEPtwozerotwooneB}{$-6.830\pm0.058$}
\newcommand{\lnerrfluxASTEPtwozerotwothreeB}{$-5.919\pm0.026$} 
 
\newcommand{\lnerrfluxASTEPtwozerotwofourB}{$-5.470_{-0.014}^{+0.015}$} 

\newcommand{\lnerrfluxCTIOGPB}{$-6.177_{-0.060}^{+0.063}$}

\newcommand{\baselinegpmaternthreetwolnsigmafluxTESSB}{$-8.180\pm0.023$} 
\newcommand{\baselinegpmaternthreetwolnrhofluxTESSB}{$-2.179_{-0.057}^{+0.054}$}

\newcommand{\BRstaraB}{$0.01399_{-0.0010}^{+0.00093}$} 
\newcommand{\BaRstarB}{$71.5_{-4.4}^{+5.5}$} 
\newcommand{\BRcompanionaB}{$0.00164_{-0.00049}^{+0.0029}$} 
\newcommand{\BRcompanionRearthB}{$20.2_{-4.9}^{+32}$} 
\newcommand{\BRcompanionRjupB}{$1.80_{-0.44}^{+2.8}$} 
\newcommand{\BaRsunB}{$113.0_{-7.5}^{+9.1}$} 
\newcommand{\BaAUB}{$0.525_{-0.035}^{+0.042}$} 
\newcommand{\BiB}{$89.204_{-0.25}^{+0.093}$} 
\newcommand{\BbtraB}{$0.990_{-0.047}^{+0.22}$} 
\newcommand{\BTtratotB}{$2.184\pm0.061$} 
\newcommand{\BTtrafullB}{$0.32_{-0.16}^{+0.15}$}

\newcommand{\BrsumaC}{$0.0196_{-0.0021}^{+0.0027}$} 
\newcommand{\BcosiC}{$0.0186_{-0.0024}^{+0.0030}$} 
\newcommand{\BepochC}{$2459678.0782_{-0.0020}^{+0.0021}$} 
\newcommand{\BperiodC}{$74.60681\pm0.00021$} 
\newcommand{\BfcC}{$-0.4457\pm0.0013$} 
\newcommand{\BfsC}{$0.076_{-0.017}^{+0.014}$}

\newcommand{\dilTESSC}{$< 0.8$}

\newcommand{\hostldcqoneTESSC}{$0.46\pm0.25$} 
\newcommand{\hostldcqtwoTESSC}{$0.43_{-0.27}^{+0.32}$}

\newcommand{\lnerrfluxTESSC}{$-6.3261_{-0.0011}^{+0.0011}$} 
\newcommand{\lnerrfluxASTEPtwozerotwotwoC}{$-4.960\pm0.030$} 
\newcommand{\lnerrfluxASTEPtwozerotwothreeaC}{$-4.569_{-0.037}^{+0.039}$} 
\newcommand{\lnerrfluxASTEPtwozerotwothreecC}{$-6.025_{-0.039}^{+0.043}$} 
\newcommand{\lnerrfluxASTEPtwozerotwothreeaBC}{$-5.384_{-0.037}^{+0.040}$} 
\newcommand{\lnerrfluxASTEPtwozerotwothreecBC}{$-6.308_{-0.039}^{+0.041}$} 
\newcommand{\lnerrfluxCTIOIPC}{$-6.301\pm0.037$} 
\newcommand{\lnerrfluxMoanatwozerotwothreeaRC}{$-5.723\pm0.028$}

\newcommand{\baselinegpmaternthreetwolnsigmafluxTESSC}{$-8.485\pm0.025$} 
\newcommand{\baselinegpmaternthreetwolnrhofluxTESSC}{$-1.416\pm0.075$}

\newcommand{\BiC}{$88.93_{-0.18}^{+0.14}$}
\newcommand{\BeC}{$0.2044\pm0.0012$} 
\newcommand{\BwC}{$170.3_{-1.8}^{+2.2}$} 
 
\newcommand{\BTtratotC}{$5.01\pm0.27$} 
\newcommand{\BTtrafullC}{$1.175\pm0.025$}

\newcommand{\brrD}{$0.06617\pm0.00079$} 
\newcommand{\brsumaD}{$0.03650_{-0.00085}^{+0.00092}$} 
\newcommand{\bcosiD}{$0.0290_{-0.0080}^{+0.0075}$} 
\newcommand{\bepochD}{$2459531.16898\pm0.00048$} 
\newcommand{\bperiodD}{$32.600268_{-0.000030}^{+0.000028}$} 
\newcommand{\bfcD}{$-0.509_{-0.099}^{+0.11}$}  
\newcommand{\bfsD}{$0.649_{-0.073}^{+0.062}$}

\newcommand{\hostldcqoneTESSthreeD}{$0.3683$} 
\newcommand{\hostldcqtwoTESSthreeD}{$0.1153$}

\newcommand{\lnerrfluxTESSthreeD}{$-6.2147_{-0.0026}^{+0.0022}$} 
\newcommand{\lnerrfluxASTEPtwozerotwothreeD}{$-5.902_{-0.024}^{+0.025}$} 
\newcommand{\lnerrfluxASTEPtwozerotwothreeBD}{$-6.247\pm0.027$} 
\newcommand{\lnerrfluxASTEPtwozerotwofourD}{$-6.144\pm0.028$} 
\newcommand{\lnerrfluxCTIOIPD}{$-6.341\pm0.032$} 
\newcommand{\lnerrfluxMoanaRD}{$-5.758\pm0.028$} 
\newcommand{\lnjitterrvFEROSD}{$-3.69_{-0.56}^{+0.41}$}

\newcommand{\baselinegpmaternthreetwolnsigmafluxTESSthreeD}{$-8.7149$} 
\newcommand{\baselinegpmaternthreetwolnrhofluxTESSthreeD}{$-1.3420$} 
\newcommand{\baselineoffsetrvFEROSD}{$10.049_{-0.011}^{+0.011}$}

\newcommand{\bRstaraD}{$0.03424_{-0.00080}^{+0.00087}$} 
\newcommand{\baRstarD}{$29.21\pm0.72$} 
\newcommand{\bRcompanionaD}{$0.002266_{-0.000056}^{+0.000062}$} 
\newcommand{\bRcompanionRearthD}{$10.64\pm0.25$} 
\newcommand{\bRcompanionRjupD}{$0.949\pm0.022$} 
\newcommand{\baRsunD}{$43.0\pm1.4$} 
\newcommand{\baAUD}{$0.2001\pm0.0064$} 
\newcommand{\biD}{$88.34_{-0.43}^{+0.46}$}
 
\newcommand{\bwD}{$128.1_{-9.1}^{+8.5}$}

\newcommand{\bbtraD}{$0.291\pm0.086$} 
\newcommand{\bTtratotD}{$4.149_{-0.032}^{+0.035}$} 
\newcommand{\bTtrafullD}{$3.585_{-0.035}^{+0.031}$}

\newcommand{\bTeqD}{$754\pm17$}

\newcommand{\combinedhostdensityD}{$0.444\pm0.033$}

\newcommand{\brrE}{$0.02266\pm0.00079$} 
\newcommand{\brsumaE}{$0.01691_{-0.00050}^{+0.00055}$} 
\newcommand{\bcosiE}{$0.0070_{-0.0014}^{+0.0013}$} 
\newcommand{\bepochE}{$2459295.9113_{-0.0017}^{+0.0016}$} 
\newcommand{\bperiodE}{$52.79918_{-0.00016}^{+0.00021}$}

\newcommand{\hostldcqoneTESSE}{$0.46_{-0.19}^{+0.27}$} 
\newcommand{\hostldcqtwoTESSE}{$0.55\pm0.28$}

\newcommand{\lnerrfluxTESSE}{$-6.86056\pm0.00090$} 
\newcommand{\lnerrfluxASTEPE}{$-6.159\pm0.016$} 
\newcommand{\lnerrfluxASTEPBE}{$-6.639\pm0.028$} 
\newcommand{\baselinegpmaternthreetwolnsigmafluxTESSE}{$-9.161\pm0.021$} 
\newcommand{\baselinegpmaternthreetwolnrhofluxTESSE}{$-1.489\pm0.058$}

\newcommand{\bRstaraE}{$0.01654_{-0.00049}^{+0.00053}$} 
\newcommand{\baRstarE}{$60.5\pm1.9$} 
\newcommand{\bRcompanionaE}{$0.000375\pm0.000020$} 
\newcommand{\bRcompanionRearthE}{$2.416\pm0.098$} 
\newcommand{\bRcompanionRjupE}{$0.2155\pm0.0087$} 
\newcommand{\baRsunE}{$59.1\pm2.2$} 
\newcommand{\baAUE}{$0.275\pm0.010$} 
\newcommand{\biE}{$89.602_{-0.072}^{+0.081}$}
\newcommand{\bbtraE}{$0.421_{-0.076}^{+0.062}$} 
\newcommand{\bTtratotE}{$6.215_{-0.079}^{+0.092}$} 
\newcommand{\bTtrafullE}{$5.880_{-0.086}^{+0.10}$} 
 
\newcommand{\bTeqE}{$456\pm47$}

\newcommand{\combinedhostdensityE}{$1.50\pm0.14$}

\newcommand{\TeqA}{$464.8_{-4.1}^{+7.1}$}

\newcommand{\TeqD}{$824.2_{-9.9}^{+10.3}$}
\newcommand{\TeqE}{$491.7_{-7.4}^{+7.9}$}

\renewcommand{\edit}[1]{#1}

\date{1 October 2025 / Accepted 15 May 2026}

\begin{document}

\title{High five from ASTEP}
\subtitle{Three validated planets and two eclipsing binaries in a diverse set of long-period candidates}

\author{
Erika~Rea\,\orcidlink{0009-0009-4849-9764}\inst{\ref{estec}}
\and 
Maximilian~N.~G{\"u}nther\,\orcidlink{0000-0002-3164-9086}\inst{\ref{estec}}
\and
George~Dransfield\,\orcidlink{0000-0002-3937-630X}\inst{\ref{birmingham},\ref{oxford}}
\and
Tristan~Guillot\,\orcidlink{0000-0002-7188-8428}\inst{\ref{nice}}
\and
Amaury~H.M.J.~Triaud\,\orcidlink{0000-0002-5510-8751}\inst{\ref{birmingham}}
\and
Keivan~G.~Stassun\,\orcidlink{0000-0002-3481-9052}\inst{\ref{vanderbilt}}
\and
Juan~I.~Espinoza-Retamal\,\orcidlink{0000-0001-9480-8526}\inst{\ref{princeton},\ref{catolica},\ref{millennium}}
\and
Rafael~Brahm\,\orcidlink{0000-0002-9158-7315}\inst{\ref{ibanez},\ref{millennium},\ref{dof}}
\and
Solène~Ulmer-Moll\,\orcidlink{0000-0003-2417-7006}\inst{\ref{leiden},\ref{geneva}}
\and
Matteo~Beltrame\inst{\ref{torino},\ref{ipev}}
\and
Vincent~Deloupy\,\orcidlink{0009-0007-5876-546X}\inst{\ref{nice},\ref{paris}}
\and
Mathilde~Timmermans\,\orcidlink{0009-0008-2214-5039}\inst{\ref{birmingham}}
\and
Lyu~Abe\,\orcidlink{0000-0002-0856-4527}\inst{\ref{nice}}
\and
Karim~Agabi\,\orcidlink{0000-0001-7948-6493}\inst{\ref{nice}}
\and
Philippe~Bendjoya\,\orcidlink{0000-0002-4278-1437}\inst{\ref{nice}}
\and
Djamel~Mekarnia\,\orcidlink{0000-0001-5000-7292}\inst{\ref{nice}}
\and
Francois-Xavier~Schmider\,\orcidlink{0000-0003-3914-3546}\inst{\ref{nice}}
\and
Olga~Suarez\,\orcidlink{0000-0002-3503-3617}\inst{\ref{nice}}
\and
Ana~M.~Heras\,\orcidlink{0000-0002-6342-9600}\inst{\ref{estec}}
\and
Theresa~Lüftinger\,\orcidlink{0009-0000-4946-6942}\inst{\ref{estec}}
\and
Bruno~Merín\,\orcidlink{0000-0002-8555-3012}\inst{\ref{esac}}
\and
François~Bouchy\,\orcidlink{0000-0002-7613-393X}\inst{\ref{geneva}}
\and
Thomas~Henning\,\orcidlink{0000-0002-1493-300X}\inst{\ref{mpia}}
\and
Andrés~Jordán\,\orcidlink{0000-0002-5389-3944}\inst{\ref{ibanez},\ref{millennium},\ref{dof}}
\and
Monika~Lendl\,\orcidlink{0000-0001-9699-1459}\inst{\ref{geneva}}
\and
Marcelo~Tala-Pinto\,\orcidlink{0009-0004-8891-4057}\inst{\ref{ohio}}
\and
Trifon~Trifonov\,\orcidlink{0000-0002-0236-775X}\inst{\ref{lsw},\ref{sofia}}
\and
Khalid~Barkaoui\,\orcidlink{0000-0003-1464-9276}\inst{\ref{iac},\ref{liege},\ref{miteaps}}
\and
Luke~G.~Bouma\,\orcidlink{0000-0002-0514-5538}\inst{\ref{carnegie}}
\and
Gavin~Boyle\,\orcidlink{0009-0009-2966-7507}\inst{\ref{elsauce},\ref{cambridge}}
\and
C\'{e}sar~Brice\~{n}o\,\orcidlink{0000-0001-7124-4094}\inst{\ref{tololo}}
\and
Amadeo~Castro-González\,\orcidlink{0000-0001-7439-3618}\inst{\ref{geneva}}
\and
Alastair~Claringbold\,\orcidlink{0000-0003-1309-5558}\inst{\ref{warwick-exo},\ref{warwick-physics}}
\and
Karen~A.~Collins\,\orcidlink{0000-0001-6588-9574}\inst{\ref{harvard}}
\and
Keith~Horne\,\orcidlink{0000-0003-1728-0304}\inst{\ref{standrews}}
\and
Steve~B.~Howell\,\orcidlink{0000-0002-2532-2853}\inst{\ref{ames}}
\and
Andrew~W.~Mann\,\orcidlink{0000-0003-3654-1602}\inst{\ref{chapelhill}}
\and
Felipe~Murgas\,\orcidlink{0000-0001-9087-1245}\inst{\ref{iac},\ref{ull}}
\and
Enric~Palle\,\orcidlink{0000-0003-0987-1593}\inst{\ref{iac},\ref{ull}}
\and
Samuel~Quinn\,\orcidlink{0000-0002-8964-8377}\inst{\ref{harvard}}
\and
Joseph~E.~Rodriguez\,\orcidlink{0000-0001-8812-0565}\inst{\ref{michigan}}
\and
Richard~P.~Schwarz\,\orcidlink{0000-0001-8227-1020}\inst{\ref{harvard}}
\and
T.G.~Tan\,\orcidlink{0000-0001-5603-6895}\inst{\ref{perth}}
\and
George~Zhou\,\orcidlink{0000-0001-5603-6895}\inst{\ref{uso}}
\and
Carl~Ziegler\,\orcidlink{0000-0002-0619-7639}\inst{\ref{nacog}}
\and
Sara~Seager\,\orcidlink{0000-0002-6892-6948}\inst{\ref{mitkavli},\ref{miteaps},\ref{mitaero}}
\and
Joshua~N.~Winn\,\orcidlink{0000-0002-4265-047X}\inst{\ref{princeton}}
}

\institute{
European Space Agency (ESA), European Space Research and Technology Centre (ESTEC), Keplerlaan 1, 2201 AZ Noordwijk, The Netherlands\label{estec}
\and
School of Physics \& Astronomy, University of Birmingham, Edgbaston, Birmingham B15 2TT, UK\label{birmingham}
\and
Department of Astrophysics, University of Oxford, Denys Wilkinson Building, Keble Road, Oxford OX1 3RH, UK\label{oxford}
\and
Laboratoire Lagrange, Université Côte d’Azur, CNRS, Nice, France\label{nice}
\and
Department of Physics and Astronomy, Vanderbilt University, VU Station 1807, Nashville, TN 37235, USA\label{vanderbilt}
\and
Department of Astrophysical Sciences, Princeton University, 4 Ivy Lane, Princeton, NJ 08540, USA\label{princeton}
\and
Instituto de Astrofísica, Pontificia Universidad Católica de Chile, Av. Vicuña Mackenna 4860, 782-0436 Macul, Santiago, Chile\label{catolica}
\and
Millennium Institute for Astrophysics, Chile\label{millennium}
\and
Facultad de Ingeniera y Ciencias, Universidad Adolfo Ibáñez, Av. Diagonal las Torres 2640, Peñalolén, Santiago, Chile\label{ibanez}
\and
Data Observatory Foundation, Chile\label{dof}
\and
Leiden Observatory, University of Leiden, Einsteinweg 55, 2333 CA Leiden, The Netherlands\label{leiden}
\and
Observatoire Astronomique de l´Université de Genève, 51 Chemin Pegasi, 1290 Versoix, Switzerland\label{geneva}
\and
Istituto di Scienze Polari del CNR (ISP-CNR), Università Ca' Foscari, Via Torino n. 155, 30172 Venezia Mestre (VE), Italy\label{torino}
\and
Programma Nazionale di Ricerche in Antartide (PNRA), Institut polaire français Paul-Émile Victor (IPEV)\label{ipev}
\and
École Normale Supérieure, Département de Physique, Rue d’Ulm, 75005 Paris Cedex 5, France\label{paris}
\and
European Space Agency (ESA), European Space Astronomy Centre (ESAC), Camino Bajo del Castillo s/n, 28692 Villanueva de la Cañada, Madrid, Spain\label{esac}
\and
Max-Planck-Institut für Astronomie, Königstuhl 17, D-69117 Heidelberg, Germany\label{mpia}
\and
Department of Astronomy, The Ohio State University, 140 West 18th Avenue, 43210, Columbus, OH, USA\label{ohio}
\and
Landessternwarte, Zentrum f\"ur Astronomie der Universt\"at Heidelberg, K\"onigstuhl 12, 69117 Heidelberg, Germany\label{lsw}
\and
Department of Astronomy, Sofia University “St Kliment Ohridski,” 5 James Bourchier Boulevard, BG-1164 Sofia, Bulgaria\label{sofia}
\and
Instituto de Astrof\'isica de Canarias (IAC), Calle V\'ia L\'actea s/n, 38200, La Laguna, Tenerife, Spain\label{iac}
\and
Astrobiology Research Unit, Universit\'e de Li\`ege, All\'ee du 6 Ao\^ut 19C, B-4000 Li\`ege, Belgium\label{liege}
\and
Department of Earth, Atmospheric and Planetary Science, Massachusetts Institute of Technology, 77 Massachusetts Avenue, Cambridge, MA 02139, USA\label{miteaps}
\and
Carnegie Science Observatories, 813 Santa Barbara Street, Pasadena, CA 91101, USA\label{carnegie}
\and
El Sauce Observatory, Coquimbo Province, Chile\label{elsauce}
\and
Cavendish Laboratory, J J Thomson Avenue, Cambridge, CB3 0HE, UK\label{cambridge}
\and
Cerro Tololo Inter-American Observatory, Casilla 603, La Serena, Chile\label{tololo}
\and
Centre for Exoplanets and Habitability, University of Warwick, Gibbet Hill Road, Coventry CV4 7AL, UK\label{warwick-exo}
\and
Department of Physics, University of Warwick, Gibbet Hill Road, Coventry CV4 7AL, UK\label{warwick-physics}
\and
Center for Astrophysics | Harvard \& Smithsonian, 60 Garden Street, Cambridge, MA 02138, USA\label{harvard}
\and
SUPA Physics and Astronomy, University of St. Andrews, Fife, KY16 9SS Scotland, UK\label{standrews}
\and
NASA Ames Research Center, Moffett Field, CA 94035, USA\label{ames}
\and
Department of Physics and Astronomy, The University of North Carolina at Chapel Hill, Chapel Hill, NC 27599-3255, USA\label{chapelhill}
\and
Departamento de Astrofísica, Universidad de La Laguna (ULL), E-38206 La Laguna, Tenerife, Spain\label{ull}
\and
Center for Data Intensive and Time Domain Astronomy, Department of Physics and Astronomy, Michigan State University, East Lansing, MI 48824, USA\label{michigan}
\and
Perth Exoplanet Survey Telescope, Perth, Western Australia, Australia\label{perth}
\and
University of Southern Queensland, Centre for Astrophysics, West Street, Toowoomba, QLD 4350, Australia\label{uso}
\and
Department of Physics, Engineering and Astronomy, Stephen F. Austin State University, 1936 North St, Nacogdoches, TX 75962, USA\label{nacog}
\and
Department of Physics and Kavli Institute for Astrophysics and Space Research, Massachusetts Institute of Technology, Cambridge, MA 02139, USA\label{mitkavli}
\and
Department of Aeronautics and Astronautics, MIT, 77 Massachusetts Avenue, Cambridge, MA 02139, USA\label{mitaero}
}

\abstract{We present the analysis of five long-period TESS Objects of Interest (TOIs), \edit{all orbiting Sun-like stars}, with orbital periods exceeding one month. Initially identified by the Transiting Exoplanet Survey Satellite (TESS), we extensively monitored these targets with the Antarctic Search for Transiting Exoplanets (ASTEP), supported by other facilities in the TESS Follow-up Observing Program (TFOP) network.}{These targets occupy a relatively underexplored region of the period-radius parameter space, offering valuable primordial probes for planetary formation and migration as warm planets better maintain their evolutionary fingerprints.}{To characterise these systems, we leveraged high-resolution speckle imaging to search for nearby stellar companions, and refine stellar parameters using both reconnaissance spectroscopy and spectral energy distribution (SED) fitting. We combined TESS photometry with high-precision ground-based observations from ASTEP, and when available, included additional photometry and radial velocity data. We applied statistical validation to assess the planetary nature of each candidate and used \allesfitter{} to jointly model the photometric and spectroscopic datasets. 
}{We validate the planetary nature of three TOIs, including the two warm Saturns TOI-4507\,b (8.2\,$\mathrm{R_\oplus}$, 104\,d) and TOI-3457\,b (10.0\,$\mathrm{R_\oplus}$, 32.6\,d), as well as the warm sub-Neptune TOI-707\,b (2.4\,$\mathrm{R_\oplus}$, 52.8\,d). The remaining two candidates \edit{are most consistent with} eclipsing binaries, namely TOI-2404 and TOI-4404.}
{These results help populate the sparse regime of warm planets, which serve as key tracers of planetary evolution, and demonstrate ASTEP’s effectiveness as a ground-based follow-up instrument for long-period systems.}

\keywords{Exoplanets, TESS, ASTEP, TOI-4507, TOI-2404, TOI-707, TOI-4404, TOI-3457.}

\titlerunning{High five from ASTEP}
\authorrunning{Rea et al.}
\maketitle

\section{Introduction}
\label{sec:intro}

The study of exoplanets has evolved rapidly over the past three decades, driven by the growing number of discoveries and improved observational techniques. While thousands of transiting exoplanets have been identified, most orbit their stars with short periods, typically less than a month. In contrast, long-period planets remain underexplored due to observational challenges such as rare transit alignments and the need for extended monitoring.
Transiting planets with orbital periods longer than one month are particularly valuable, as they provide both photometric radius measurements and, when combined with radial velocities (RVs), mass estimates. This enables the determination of bulk densities and compositions, while their weaker atmospheric erosion and tidal effects allow them to retain key signatures of their formation and long-term evolution \citep[see reviews by][]{Madhusudhan2014, Dawson2018}.
In this work, we investigate five long-period transiting planet candidates, spanning a range of sizes and orbital configurations, initially identified as TESS Objects of Interest (TOIs): TOI-4507.01, TOI-4404.01, TOI-2404.02, TOI-3457.01, and TOI-707.01\footnote{see ExoFOP: \url{https://exofop.ipac.caltech.edu/tess/}}. Where validated and confirmed, they offer a rare opportunity to probe the physical and dynamical properties of planets in a regime where observational insights are sparse and evolutionary processes may leave clearer imprints than in short-period counterparts.
Warm giants - planets the size of Saturn or Jupiter with orbital periods between 10 and 200 days - can play a key role in understanding their hot and cold siblings.
In particular, the nature of giant planets inside the snowline is still debated, with multiple evolutionary processes likely contributing \citep[see reviews by][]{Dawson2018, Fortney2021}. 
These planets are hypothesised to form either through core accretion \citep{Pollack1996} or gravitational collapse \citep{Cameron1978, Boss1997}; and either via in situ accretion close to their host stars \citep[e.g.][]{Batygin2016, Boley2016, Huang2016} or beyond the snowline followed by inward migration \citep[e.g.][]{Rafikov2005}.
The latter may be driven by smooth disk migration \citep[e.g.][]{Lin1996, Ward1997, Walsh2011, Nelson2018} or high-eccentricity mechanisms such as planet-planet scattering followed by tidal interactions \cite[e.g.][]{Rasio1996, Juric2008, Ford2008, Jackson2008, Wu2011, Petrovich2015}. 
An increasing number of discoveries of hot giants on eccentric orbits indicate that high-eccentricity mechanisms may indeed play a prominent role \citep[e.g.][]{Kossakowski2019, Jordan2020, Schulte2024}.
Yet, hot giants' radii are often inflated due to high irradiation and their orbits are often circularised by tidal forces, obscuring evolutionary features \citep[e.g.][]{Fortney2010, Albrecht2012}.
Warm giants, in contrast, experience less extreme environments, with less irradiation, weaker tidal interactions, and a broader range of orbital eccentricities \citep[][]{Schlecker2020}. They are more likely to preserve initial conditions, allowing direct comparisons with their counterparts \citep[e.g.][]{Huang2016, Espinoza-Retamal2025}.
As a result, they retain a clearer fingerprint of their primordial properties and can be a more direct probe of evolutionary pathways for all types of giant planets. 

Equally puzzling are warm (sub-)Neptunes - planets with radii between 1.6- 4\,$R_\oplus$ - which have no analogue in our Solar System yet are among the most common exoplanets discovered. Despite their prevalence, their bulk composition and formation pathways remain uncertain, particularly at longer orbital periods \citep[e.g.,][]{Raymond2022}. 
These planets may host substantial H/He envelopes or be rich in heavier volatiles such as water. However, the mass-radius relationship is degenerate: drastically different internal compositions can yield similar bulk properties, especially once the H/He fraction exceeds $\sim$1\% of the total mass \citep[e.g.][]{2014_Lopez_Fortney}. This ambiguity is particularly relevant in the 2-4\,$R_\oplus$ range, where planets may be 'gas dwarfs' or 'water worlds' (e.g., \citealt{Zeng2019}). 
The radius valley, an observed deficit of planets between 1.6-1.8\,$R_\oplus$, offers key insights into planetary origins. 
This transition region between rocky super-Earths and volatile-rich sub-Neptunes has been identified in multiple independent studies \citep[e.g.][]{Youdin2011,Lopez2012,Owen2012}, and confirmed through precise radius measurements and occurrence statistics \citep[e.g.][]{2015_Rogers,Fulton2017,Van_Eylen_2018}. 
Models attribute this bimodality to differences in core composition relative to the water ice line, combined with limited gas accretion or atmospheric loss \citep[e.g.][]{Venturini2020,2021_Bean}. 
However, the location and shape of the gap vary with stellar type and metallicity, suggesting distinct formation and migration pathways. For instance, metal-rich stars tend to host larger planets at short periods \citep[e.g.][]{Petigura2018}, while around M dwarfs, the gap may instead separate rocky super-Earths from water-rich sub-Neptunes \citep[][]{Luque2022}.
Formation frameworks incorporate a range of physical processes, such as disk evolution, planetesimal growth, migration, and gas accretion \citep[see review by][]{Raymond2022}. Some models propose late-stage, in situ accretion from inward-drifting solids in gas-poor environments that limits envelope growth and migration; while others favour formation beyond the ice line followed by disk-driven migration that enables volatile-rich planets \citep[e.g.][]{Lee2016,2021_Bean}. Moreover, dynamical instabilities after disk dispersal may disrupt resonant chains and trigger collisions, potentially explaining both the radius valley and intra-system size uniformity \citep{Izidoro2022}. 
Sub-Neptunes on short orbits undergo intense stellar irradiation, driving atmospheric escape and shaping present-day radii \citep{Owen2017,Ginzburg2018,Kubyshkina_2019}. At longer periods, reduced flux allows warm sub-Neptunes to retain more of their primordial atmospheres and sustain interior-atmosphere coupling over Gyr timescales \citep[e.g.][]{2020_Kite}, preserving signatures of their initial composition, thermal evolution, and volatile inventory. 

However, detecting and characterising long-period planets is observationally challenging, requiring extensive follow-up with high-precision photometry and radial velocity instruments to confirm signals and refine planet properties. 
Space-based missions such as the Transiting Exoplanet Survey Satellite (TESS; \citealt{Ricker2015}) significantly expand the number of detected exoplanets, but the relatively short observation windows in each sector ($\sim$27\,days) limit the ability to capture multiple transits of long-period TOIs. The extended missions partly mitigate the effect by revisiting sectors about a year later, although many transits still fall into observational gaps, underlining the need for complementary instruments. As such, ground-based facilities capable of long-term monitoring play a crucial role in validating and refining these detections. Among these, the Antarctic Search for Transiting ExoPlanets (ASTEP; \citealt{Guillot2015, Crouzet2020}) telescope, located at Dome C in Antarctica, offers a unique advantage due to its stable atmospheric conditions and the ability to provide continuous high-precision photometry during the winter season. 
Combining TESS with ASTEP observations enhances our ability to characterise planetary systems, especially those with long orbital periods \citep{2022SPIE_Dransfield}. 
We structure this study as follows. 
In Section~\ref{sec:Targets_and_observations}, we describe the target selection and observations.
Next, Section~\ref{sec:stellar_characterisation} describes the stellar characterisation, followed by Section~\ref{sec:validation} that details the statistical validation of the TOIs based on all this input.
In Section~\ref{sec:Analysis}, we detail the methods used to derive planetary parameters, including the joint modelling of photometric and spectroscopic data.
Section~\ref{sec:results_and_discussion} presents the results for each target, along with a discussion of their physical and orbital properties in the context of planetary formation and evolution.
Finally, Section~\ref{sec:conclusions} summarises the main findings and their broader implications.

\section{Targets and observations}
\label{sec:Targets_and_observations}

\subsection{Target selection}
\label{subsec:Target_selection}

We selected the most suitable targets from TOIs observed by ASTEP between 2020 and 2023, considering only those classified as Planet Candidates (PC) in the NASA Exoplanet Archive \citep{christiansen2025nasaexoplanetarchiveexoplanet}. Targets under active investigation or associated with ongoing publications were excluded based on literature reviews and collaborative discussions. We further narrowed the sample to planets with orbital periods longer than one month and verified that ASTEP data included at least partial transit coverage. Where possible, we obtained additional ASTEP observations from 2024 and 2025 to improve coverage and refine the analysis. 
Our final sample includes five targets:
TOI-4507.01 (orbital period of 104.6\,d), 
TOI-2404.02/03 (74.6\,d), 
TOI-707.01 (52.8\,d),
TOI-4404.01 (39.6\,d), and
TOI-3457.01 (32.6\,d).
All signals initially appeared consistent with Neptune- to Jupiter-sized planet candidates. Notably, at the start of our observing campaign, TOI-2404.03 was believed to represent a single or double transit event with a much longer orbital period. It was only later revealed that TOI-2404.02 and TOI-2404.03 share a common physical origin (see further discussion in this paper).

Two of these targets also exhibit shorter-period signals: TOI-2404.01 (20.4\,d) and TOI-707.02 (17.5\,d). Although these were not part of our observing campaign, we briefly address them in the relevant sections of this paper.

\subsection{Photometric observations}
\label{subsec:Photometric_observations}

\subsubsection{TESS photometry}
\label{subsubsec:TESS_photometry}

\begin{figure*}
\centering
\includegraphics[width=0.85\textwidth]{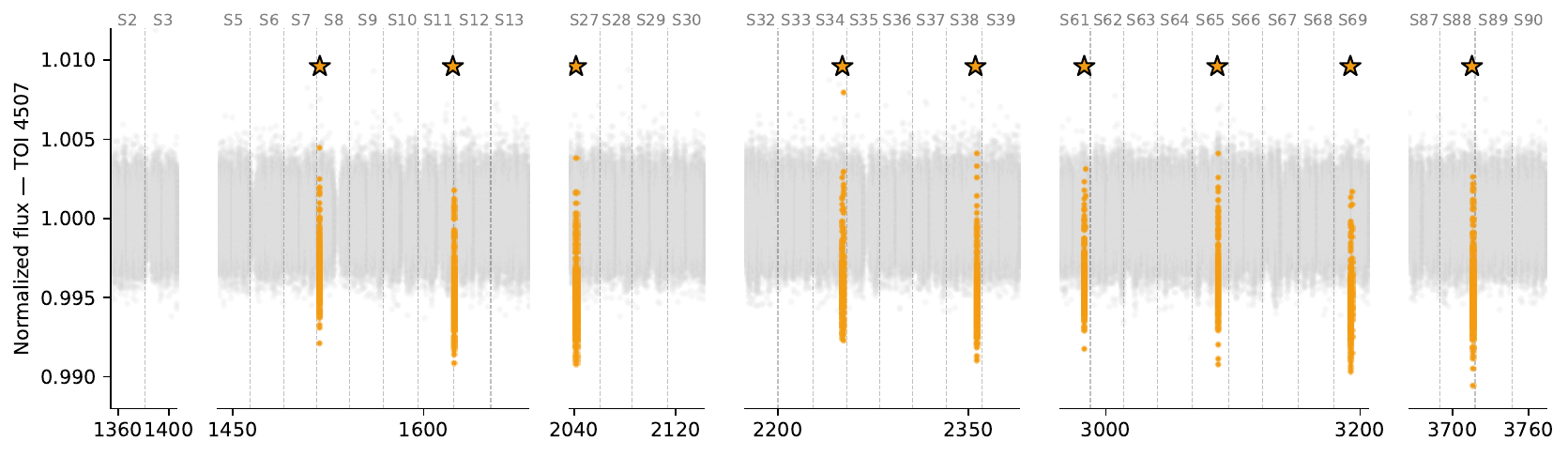}
\includegraphics[width=0.85\textwidth]{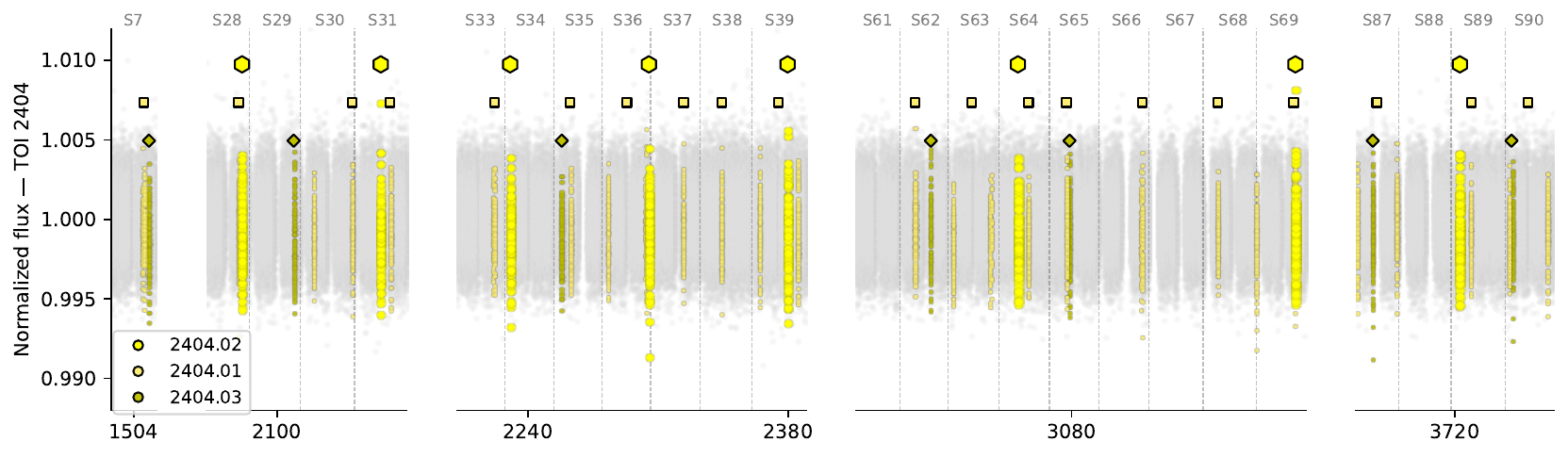}
\includegraphics[width=0.85\textwidth]{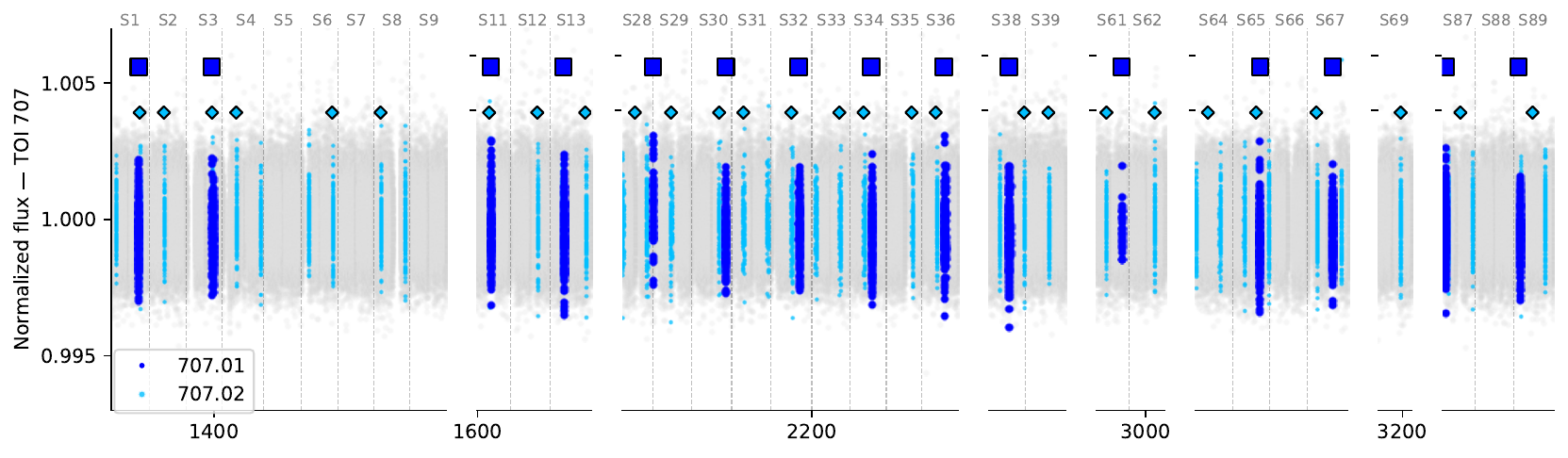}
\includegraphics[width=0.85\textwidth]{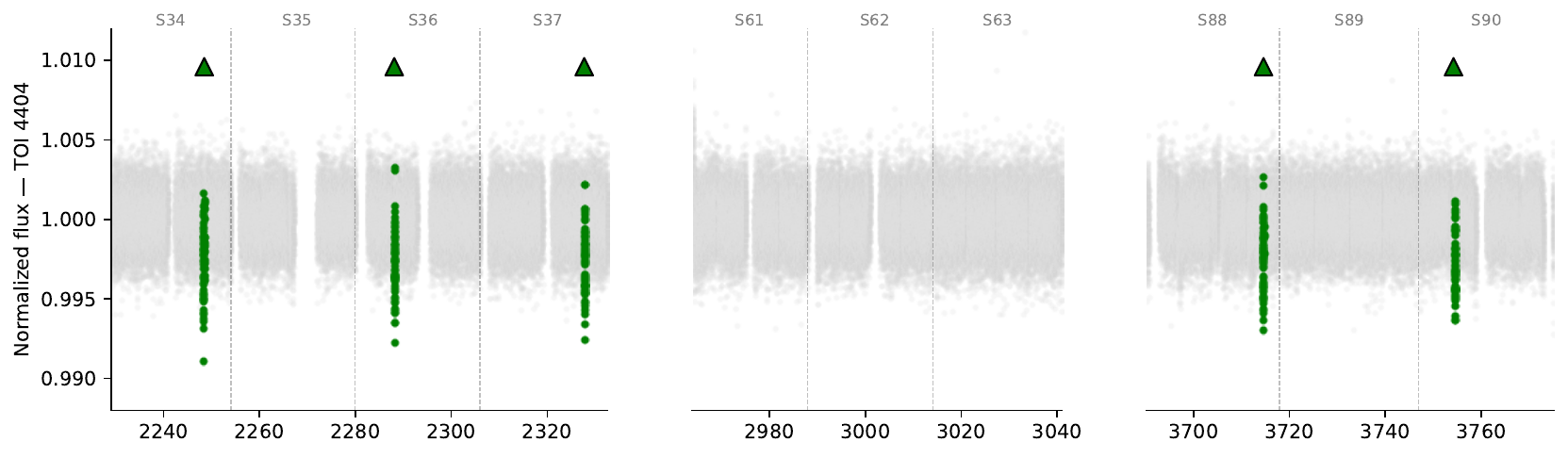}
\includegraphics[width=0.85\textwidth]{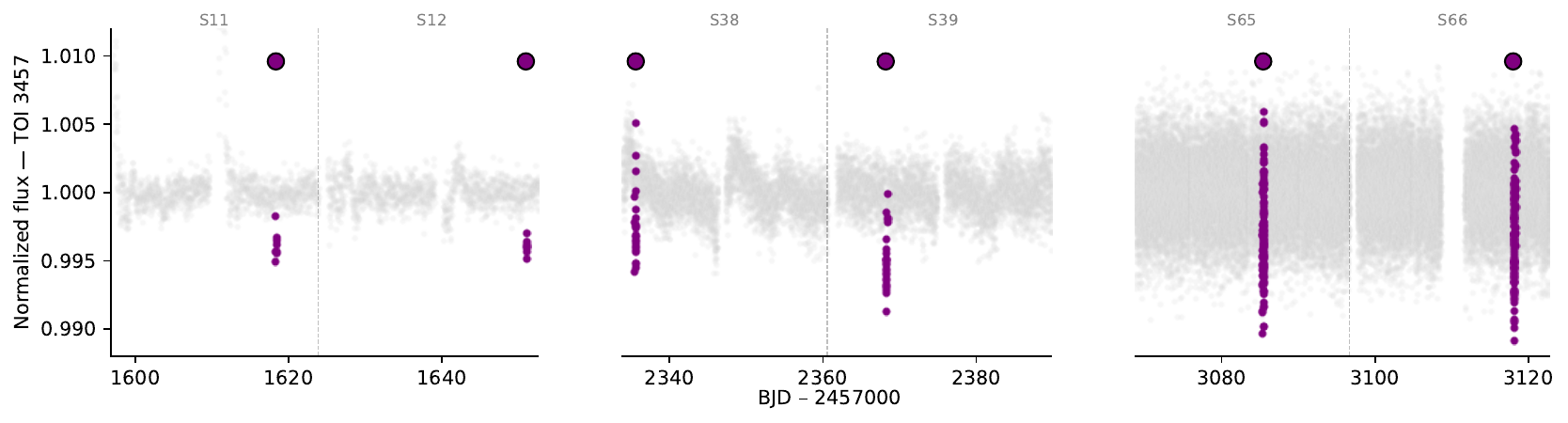}
\caption{TESS lightcurves of the five TOIs from 2018 to 2025. Each panel shows the normalised flux over time (grey) and highlights the transit events in colour. \edit{To improve readability for these long-period systems, only time intervals with available data are shown. As a result, predicted transits occurring during data gaps are not displayed.}
 Section~\ref{subsubsec:TESS_photometry} and Table~\ref{tab:tess_obs} provide further details.}
\label{fig:tess_lightcurves}
\end{figure*}

NASA's TESS mission launched in 2018 to detect exoplanets via stellar brightness variations. It scans the sky in ~27-day sectors, searching for periodic dips caused by transiting planets. The mission provides 2-minute cadence lightcurves and full-frame images (FFIs) at 30-minute intervals, later improved in extended missions. Stars with validated transit-like signals are classified as TESS Objects of Interest (TOIs; \citealt{2021ApJS..254...39G}), including the five targets analysed here. Observations are processed with the Science Processing Operations Center (SPOC) pipeline, which produces Presearch Data Conditioning-Simple Aperture Photometry (PDC-SAP) flux \citep{stumpe2012kepler,stumpe2014multiscale,2020RNAAS...4..201C}. 
For TOI-3457, we also used Quick-Look Pipeline (QLP) data when SPOC lightcurves are unavailable. Table~\ref{tab:tess_obs} summarises all TESS observations. We extracted and normalised PDC-SAP flux using \lightkurve{} \citep{2018ascl.soft12013L}, correcting instrumental effects and filtering flagged data. SPOC and QLP lightcurves were further detrended with \texttt{wotan} \citep{Hippke_2019} to remove long-term stellar variability and residual systematics. We then verified candidate parameters from ExoFOP and performed an automated search for additional transit-like signals using the Transit Least Squares (TLS) algorithm \citep{2019_TLS}. All lightcurves and signals are shown in Fig.~\ref{fig:tess_lightcurves}.
To assess contamination, we inspected TESS target pixel files with \texttt{tpfplotter} (Fig.~\ref{fig:tess_pix}). The absence of nearby sources supports the interpretation that the signals originate from the target stars and are not affected by blending or dilution.

\begin{table}
\scriptsize
\centering
\caption{Summary of TESS observations for the five TOIs.}
\label{tab:tess_obs}
\begin{tabular}{lcccc}
\toprule
\textbf{Target} & \textbf{Pipeline} & \textbf{Year} & \textbf{Observed Sectors} & \textbf{Exposure Time (s)} \\
\midrule

\multirow{7}{*}{TOI-4507} 
  & \multirow{7}{*}{SPOC} & 2018 & 02-03, 05-06 & 120 \\ 
  & & 2019 & 07-13 & 120 \\ 
  & & 2020 & 27-30, 32-33 & 120 \\ 
  & & 2021 & 34-39 & 120 \\ 
  & & 2023 & 61-69 & 120 \\ 
  & & 2024 & 87 & 120 \\ 
  & & 2025 & 88-90 & 120 \\
\midrule

\multirow{6}{*}{TOI-2404} 
  & \multirow{6}{*}{SPOC} & 2019 & 07 & 120 \\ 
  & & 2020 & 28-31, 33 & 120 \\ 
  & & 2021 & 34-39 & 120 \\ 
  & & 2023 & 61-69 & 120 \\ 
  & & 2024 & 87 & 120 \\ 
  & & 2025 & 88-90 & 120 \\
\midrule

\multirow{7}{*}{TOI-707} 
  & \multirow{7}{*}{SPOC} & 2018 & 01-06 & 120 \\ 
  & & 2019 & 07-09, 11-13 & 120 \\ 
  & & 2020 & 28-33 & 120 \\ 
  & & 2021 & 34-36, 38-39 & 120 \\ 
  & & 2023 & 61-62, 64-67, 69 & 120 \\ 
  & & 2024 & 87 & 120 \\ 
  & & 2025 & 88-89 & 120 \\ 
\midrule
\multirow{3}{*}{TOI-4404} 
  & \multirow{3}{*}{SPOC} & 2021 & 34-37 & 120 \\ 
  & & 2023 & 61-63 & 120 \\ 
  & & 2025 & 88-90 & 120 \\

\midrule

\multirow{3}{*}{TOI-3457} 
  & \multirow{2}{*}{QLP} & 2019 & 11-12 & 1800 \\ 
  & & 2021 & 38-39 & 600 \\ [0.1cm]
  
 & SPOC & 2023 & 65, 66 & 120 \\
\bottomrule
\end{tabular}
\end{table}

\subsubsection{ASTEP photometry}
\label{subsubsec:ASTEP_photometry}
To interpret the origin of the TOI signals, we conducted ground-based photometric follow-up observations using ASTEP \citep{Crouzet2008,Guillot2015,Crouzet2020}, located at Concordia Station in Antarctica. The facility operates a 0.4 m telescope optimised for high-precision time-series photometry under exceptionally stable atmospheric conditions \citep{mekarnia_2016}. 
Since its 2022 upgrade, ASTEP+ performs simultaneous observations in red (R) and blue (B) bands, corresponding to wavelengths of 800 nm (R) and 550 nm (B), respectively. The cameras offer pixel scales of 1.05 arcseconds (R) and 1.30 arcseconds (B), yielding wide fields of view of approximately 36 × 36 arcminutes (R) and 44 × 44 arcminutes (B) \citep{Schmider_Abe_2022}. We processed the data using a combination of IDL-based \citep{mekarnia_2016} and Python-based \citep{Dransfield_Triaud_2022} aperture photometry pipelines, which correct for systematics, extract robust lightcurves, and ensure compatibility with other datasets for joint modelling. 
Table~\ref{obs-ground} summarises all observations.

In addition to ASTEP, several ground-based facilities in the TESS Follow-up Observing Program (TFOP; \citealt{2019_tfop}) provided complementary photometric coverage, capturing additional transits and broadening the wavelength range. Combining ASTEP’s R and B bands with these multi-filter data helps enhance planet validation via achromatic behaviour and reveal eclipsing binaries through chromatic trends.

\subsubsection{LCO photometry}
\label{subsubsec:LCO_photometry}
We utilised the Las Cumbres Observatory (LCO; \citealt{2013_lco}) 1.0\,m telescope located at Cerro Tololo Inter-American Observatory (CTIO), equipped with the 4096$\times$4096  SINISTRO camera, to obtain photometric observations of three candidates in \textit{i'}, \textit{g'}, and \textit{z$_s$'} bandpasses (see Table~\ref{obs-ground}). 
All observations have typical cadences of 200-250\,s, airmass values ranging from 1.5 to 1.6, and seeing conditions around 0.389$\arcsec$. 
The data calibration was performed using the standard {\tt BANZAI} pipeline \citep{McCully_2018SPIE10707E} and photometric extraction was performed using {\tt AstroImageJ} software \citep{Collins_2017}.

\subsubsection{PEST photometry}
\label{subsubsec:PEST_photometry}
The Perth Exoplanet Survey Telescope (PEST; \citealt{PEST}) is located near Perth, Australia. The 0.3 m telescope is equipped with a $5544\times3694$ QHY183M camera.  Images are binned 2x2 in software giving an image scale of 0$\farcs$7 pixel$^{-1}$ resulting in a $32\arcmin\times21\arcmin$ field of view. A custom pipeline based on {\tt C-Munipack}\footnote{http://c-munipack.sourceforge.net} was used to calibrate the images and extract the differential photometry. 
In particular, TOI-4404.01 was observed in the \textit{r'}-band, providing full transit coverage (see Table~\ref{obs-ground}).

\subsubsection{MoanaES photometry}
\label{subsubsec:Moana_photometry}
MoanaES \citep[see][]{Trifonov2023, Brahm2023} is a station of the Observatoire Moana located at the El Sauce Observatory in the Río Hurtado Valley, Chile, at an altitude of 1570\,m. The Observatoire Moana operates a global network of small-aperture robotic telescopes for time-series photometry and transit follow-up. The El Sauce station hosts a 0.6\,m corrected Dall-Kirkham telescope and an Andor iKon-L 936 deep-depletion 2k × 2k CCD, delivering a pixel scale of 0\farcs67. 
We observed partial transits of TOI-2404's candidates in the \textit{r'} band (see Table~\ref{obs-ground}).

\begin{table}
\scriptsize
\centering
\caption{Summary of ground-based photometric follow-up from ASTEP, LCO-CTIO, PEST and MoanaES.}
\label{obs-ground}
\begin{tabular}{cccc}
\toprule
\textbf{Target} & \textbf{Telescope (filter)} & \textbf{Date} & \textbf{Coverage} \\
\midrule
TOI-4507.01 & ASTEP (R,B)  & 2023-05-23 & Egress \\
            & ASTEP (R,B)  & 2024-07-14 & - \\
            & ASTEP (R)    & 2025-05-24 & Full \\
\midrule
TOI-2404.01 & MoanaES (I)   & 2022-11-10 & - \\
            & MoanaES (R)   & 2023-02-20 & Ingress \\[0.1cm]
TOI-2404.02 & ASTEP (R)     & 2022-09-04 & Full \\
            & ASTEP (R,B)   & 2023-04-16 & Egress \\
            & ASTEP (R,B)   & 2023-06-29 & - \\
            & ASTEP (R,B)   & 2023-09-12 & Ingress \\
            & MoanaES (R)   & 2023-01-31 & Egress \\
            & LCO-CTIO (ip) & 2023-02-01 & Egress \\[0.1cm]
TOI-2404.03 & -            & -         & - \\
\midrule
TOI-707.01  & ASTEP (R,B)   & 2022-05-18 & Full \\[0.1cm]
TOI-707.02  & -            & -         & - \\
\midrule
TOI-4404.01 & ASTEP (R)     & 2021-09-28 & Ingress \\
            & ASTEP (R,B)   & 2023-06-24 & Full \\
            & ASTEP (R,B)   & 2024-05-06 & Full \\
            & LCO-CTIO (gp,zs) & 2022-03-06 & Full \\
            & PEST (rp)     & 2022-01-25 & Full \\
\midrule
TOI-3457.01 & ASTEP (R)     & 2021-09-08 & - \\
            & ASTEP (R,B)   & 2023-06-22 & Full \\
            & ASTEP (R,B)   & 2024-05-13 & Full \\
            & MoanaES (R)   & 2022-05-26 & Ingress \\
            & LCO-CTIO (ip) & 2022-05-26 & Ingress \\
\bottomrule
\end{tabular}

\end{table}

\subsection{Reconnaissance spectroscopy}
\label{subsec:Reconnaissance_spectroscopy}
To vet planetary candidates and refine stellar parameters, we leveraged reconnaissance spectroscopy using MINERVA and CHIRON, both contributing to the TFOP network. MINERVA (MINiature Exoplanet Radial Velocity Array) is a dedicated array of 0.7-meter telescopes located at the Fred Lawrence Whipple Observatory. It provides precise radial velocity measurements and moderate-resolution spectra, tailored for identifying spectroscopic binaries and estimating stellar properties. It provided three observations of TOI-4507 for initial vetting. CHIRON, mounted on the 1.5-meter telescope at Cerro Tololo Inter-American Observatory, is a fiber-fed Echelle spectrograph offering high-resolution ($\mathrm{R} \approx 80,000$) spectra. Its stability and wavelength coverage make it well-suited for stellar classification. CHIRON provided one observation of TOI-2404, two of TOI-707, one of TOI-4404, and 20 of TOI-3457 for initial vetting, and partly supported stellar characterisation (see Section~\ref{sec:stellar_characterisation}).

\subsection{Radial velocity observations}
\label{subsec:Radial_velocity_observations}

\subsubsection{FEROS spectroscopy}
\label{subsubsec:FEROS_spectroscopy}
The Fiber-fed Extended Range Optical Spectrograph (FEROS; \citealt{FEROS}) is an Echelle spectrograph on the 2.2 m MPG/ESO telescope at La Silla Observatory, Chile. It offers a resolving power of $\mathrm{R} \approx 50,000$, covering the visible spectrum from approximately 350 nm to 920 nm. Its radial velocity precision reaches about 10\,m\,s$^{-1}$, which makes it suitable for confirming massive exoplanets detected by TESS in the southern hemisphere. Since 2020, FEROS has been routinely used within the TFOP framework to measure TOI radial velocities. As part of the WINE survey (Warm gIaNts with tEss; \citealt{wine1,wine2,wine3,wine4}), FEROS obtained eleven out-of-transit spectra for TOI-4507, 19 for TOI-2404, and seven for TOI-3457. 
Tables~\ref{tab:rv_measurements_toi-4507}, \ref{tab:rv_measurements_toi-2404}, and \ref{tab:rv_measurements_toi-3457} list the observation dates, exposure times, and uncertainties.

\subsubsection{CORALIE spectroscopy}
\label{subsubsec:CORALIE_spectroscopy}
We also performed spectroscopic vetting of the targets with the CORALIE spectrograph installed at the Swiss 1.2m Euler telescope at La Silla observatory, Chile. CORALIE is a fiber-fed high resolution spectrograph with a resolution of $\mathrm{R} \approx 60,000$ \cite{Queloz2000}. All targets were observed with the science fiber together with the second fiber connected to the simultaneous Fabry-Pérot étalon. 
We obtained four observations of TOI-2404, and one observation of TOI-4404.
The spectra were processed using the standard calibration reduction pipeline and the radial velocities were derived by cross-correlation with the appropriate stellar mask for each target \citep{Pepe2002}. The radial velocities for TOI-2404 are listed in Table~\ref{tab:rv_measurements_toi-2404}. The observation of TOI-4404 immediately revealed it as a double-lined spectroscopic binary (see below).

\subsubsection{HARPS spectroscopy}
\label{subsubsec:HARPS_spectroscopy}
The High Accuracy Radial velocity Planet Searcher (HARPS) spectrograph \citep{2003_Harps} operates on the ESO 3.6\,m telescope at La Silla Observatory in Chile. This fiber-fed, cross-dispersed Echelle spectrograph delivers a resolving power of $\mathrm{R} \approx 120,000$ and maintains long-term stability at the $\sim$1\,m\,s$^{-1}$ level, enabling the characterisation of low-mass exoplanets. As part of the WINE survey, HARPS provided 26 out-of-transit spectra for TOI-4507, which serve to rule out a stellar-mass companion and validate the planetary nature of the signal (Table~\ref{tab:rv_measurements_toi-4507}). \edit{In-transit spectroscopy and Rossiter-McLaughlin modelling for this system have been published in \citet{Espinoza-Retamal_2026}}. Additional HARPS observations were obtained for TOI-2404, complementing the FEROS and CORALIE data (Table~\ref{tab:rv_measurements_toi-2404}).

\subsection{High-resolution imaging}
\label{subsec:High-resolution_imaging}
Blended stellar companions can mimic planetary transits or dilute transit depths \citep[e.g.][]{refId0,Ciardi_2015}. To rule out such contamination, we conducted high-resolution speckle imaging of our targets. The 4.1\,m SOAR telescope \citep{Tokovinin_2008} observed TOI-4507, TOI-2404, TOI-707, and TOI-4404 in the I band (\edit{879 nm}), with sensitivity curves and auto-correlation functions extracted following \citealt{Ziegler_2020}. For TOI-3457, we used Zorro on the GEMINI South 8\,m telescope \citep{10.3389/fspas.2021.716560}, which provides dual-channel imaging in 562\,nm and 832\,nm, reduced via the standard pipeline \citep{Howell_2011}. As shown in Fig.~\ref{fig:imaging}, no bright companions were detected within 0.5$\arcsec$-3.0$\arcsec$ (SOAR) or 0.2$\arcsec$-1.2$\arcsec$ (Zorro), supporting the interpretation that the transit signals originate from the target stars without significant blending.

\begin{table}
\scriptsize
\centering
\caption{Imaging observation summary.}
\label{tab:imaging}
\begin{tabular}{ccccc}
\toprule
\textbf{Target} & \textbf{Telescope} & \textbf{Instrument} & \textbf{Filter (nm)} & \textbf{Date} \\
\midrule
TOI-4507 & SOAR & HRCam & 879 & 2021-11-20 \\[0.2cm]
TOI-2404 & SOAR &HRCam &879 &2020-12-03\\[0.2cm]
TOI-707 & SOAR  & HRCam & 879 & 2019-11-09 \\[0.2cm]
TOI-4404 &SOAR& HRCam & 879 & 2022-03-20 \\[0.2cm]
TOI-3457 & Gemini  &Zorro&832 - 562   & 2023-04-08 \\

\bottomrule
\end{tabular}
\end{table}

\section{Stellar characterisation}
\label{sec:stellar_characterisation}
Characterising the host/primary star properties is pivotal for deriving accurate companion parameters. Our analysis started from the TESS Input Catalogue (TIC, \citealt{Stassun2018}), along with Gaia DR3 \citep{2023_gaiaDR3}, 2MASS \citep{2003_2MASS}, and other catalogues, and was refined by our observations.

\subsection{Blends and stellar multiplicity}
\label{subsec:blends and stellar multiplicity}
\begin{figure}
    \centering
    \includegraphics[width=\columnwidth]{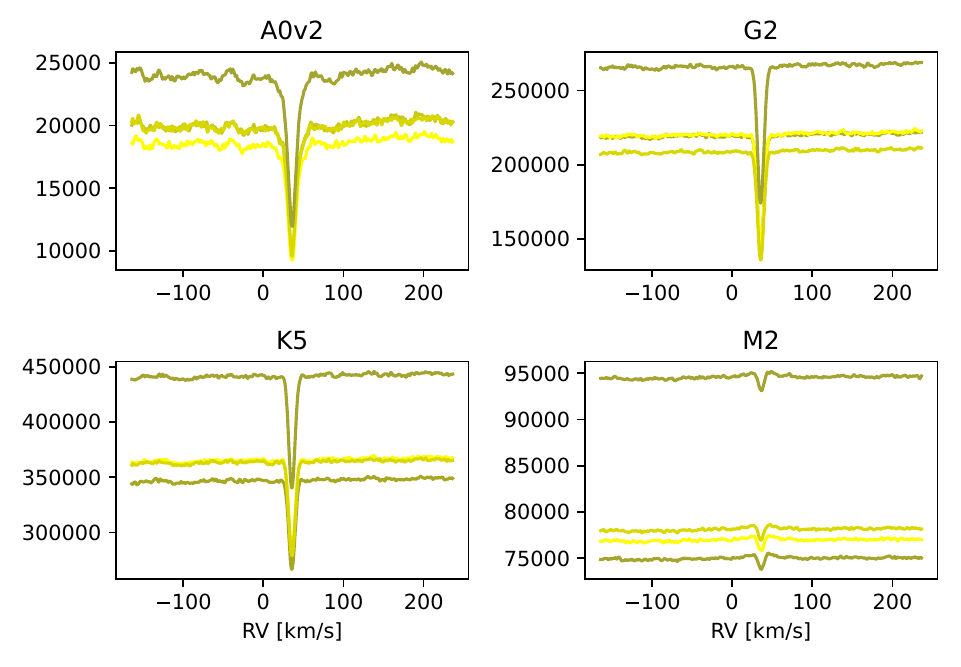}
    \caption{CORALIE CCFs of TOI-2404 show no evidence of multiple star\edit{s}, presenting a puzzling contrast with its photometric features showing a potential planetary transit alongside clear eclipsing binary signatures.}
    \label{fig:coralie_ccf_toi-2404}
\end{figure}

\begin{figure}
    \centering
    \includegraphics[width=\columnwidth]{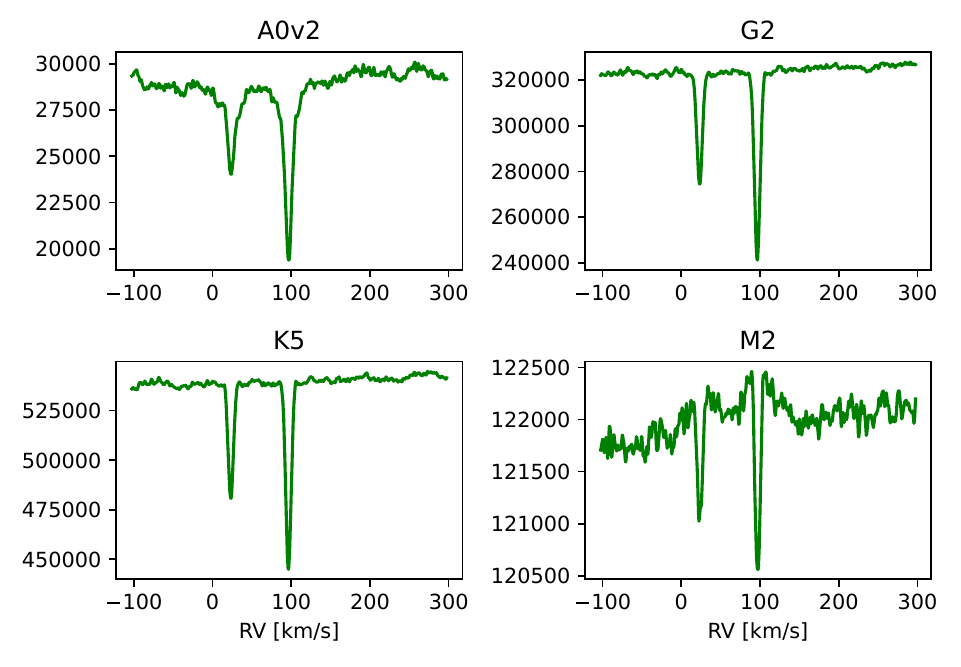}
    \caption{CORALIE CCF of TOI-4404 unveil\edit{s} it to be a double-lined spectroscopic binary.}
    \label{fig:coralie_ccf_toi-4404}
\end{figure}

Given the low field density and the absence of nearby companions in our high-resolution imaging, the risk of blended sources is low (see Section~\ref{subsec:High-resolution_imaging} and later Section~\ref{sec:validation}).
Nonetheless, TOI-2404 and TOI-4404 warrant caution, as both appear to host multiple stars.
For TOI-2404, the four CORALIE spectra and their cross-correlation functions reveal only a single, isolated peak (Fig.~\ref{fig:coralie_ccf_toi-2404}), indicating no direct spectroscopic evidence of multiplicity. 
However, photometric features suggest a potential planetary transit together with a heavily diluted eclipsing binary within a multi-star system (see Sections~\ref{sec:validation} and \ref{sec:results_and_discussion}). Accordingly, we interpret the derived stellar properties as those of the dominant star, not the binary components.
In contrast, TOI-4404 is clearly identified as a double-lined spectroscopic binary in our CORALIE data (Fig.~\ref{fig:coralie_ccf_toi-4404}). Throughout the analysis, we assume the primary star dominates the system’s light and the reported stellar parameters describe this primary component.

\subsection{Spectroscopic parameters}
\label{subsec:spectroscopic_parameters}
We first used high-resolution spectroscopy (see Sections~\ref{subsec:Reconnaissance_spectroscopy} and \ref{subsec:Radial_velocity_observations}) to refine the stellar effective temperature ($T_\mathrm{eff}$), surface gravity ($\log{g}$), metallicity ([Fe/H]), and projected rotational velocity ($v\sin{i}$).
All results are summarised in Table~\ref{tab:stellar-params}. 
Notably, the reported uncertainties do not account for potential systematic limitations and should be regarded as lower limits.

For TOI-4507 we leveraged 26 HARPS spectra, for TOI-2404 we used 19 FEROS spectra, and for TOI-3457 we used seven FEROS spectra. We co-added the spectra of each star and used the \texttt{zaspe} code \citep{zaspe} to derive precise stellar atmospheric parameters. This procedure compares the co-added spectrum to a grid of synthetic models in the regions of the spectrum that are most sensitive to changes in the atmospheric parameters. For TOI-707 we used two CHIRON spectra and for TOI-4404 we used a single CHIRON spectrum, which were extracted via the official pipeline \citep{2021AJ....162..176P}. The spectral analysis was performed as per \citet{2021AJ....161....2Z}; briefly, line profiles were obtained from each spectrum via a least-squares deconvolution against a synthetic spectral template, from which radial and rotational broadening velocities were derived. Stellar atmosphere parameters were determined by matching each observed spectrum against a library of observed spectra previously classified by the Spectroscopic Parameter Classification tool \citep{2010ApJ...720.1118B}, and interpolated via a gradient boosting regressor.

\subsection{SED modelling}
\label{subsec:sed_modelling}
Next, we performed spectral energy distribution (SED) modelling, in which the observed broadband fluxes, spanning from the ultraviolet to the mid-infrared, are fitted with synthetic stellar atmosphere models. We followed the procedures of \citet{2017_Stassun} and \citet{Stassun_2018}, modelling each star’s SED with Kurucz stellar atmospheres \citep{1979_Kurucz} constrained by the photometry and Gaia parallax. For TOI-4507, TOI-2404, and TOI-707 spectroscopic priors are used as input; for TOI-4404 and TOI-3457 the SEDs are freely sampled.
The SED fits provide the visual extinction ($A_V$), the bolometric flux ($F_{\rm bol}$), the bolometric luminosity ($L_{\rm bol}$), and stellar radius ($R_\star$). We combined these results with empirical mass-radius relations \citep{Torres2010} to obtain self-consistent estimates of the stellar mass ($M_\star$). Using spectroscopic priors, they also allow us to constrain the projected rotation period ($P_{\mathrm{rot}}/\sin i$) and stellar age.
Table~\ref{tab:stellar-params} presents the results and Fig.~\ref{fig:seds} shows the corresponding best-fit SEDs.
As above, the reported uncertainties do not capture possible systematic biases and should be interpreted as lower limits.
\subsection{Hertzsprung-Russell diagram}
\label{subsec:HR_diagram}
\begin{figure}
  \centering
  \includegraphics[width=\columnwidth]{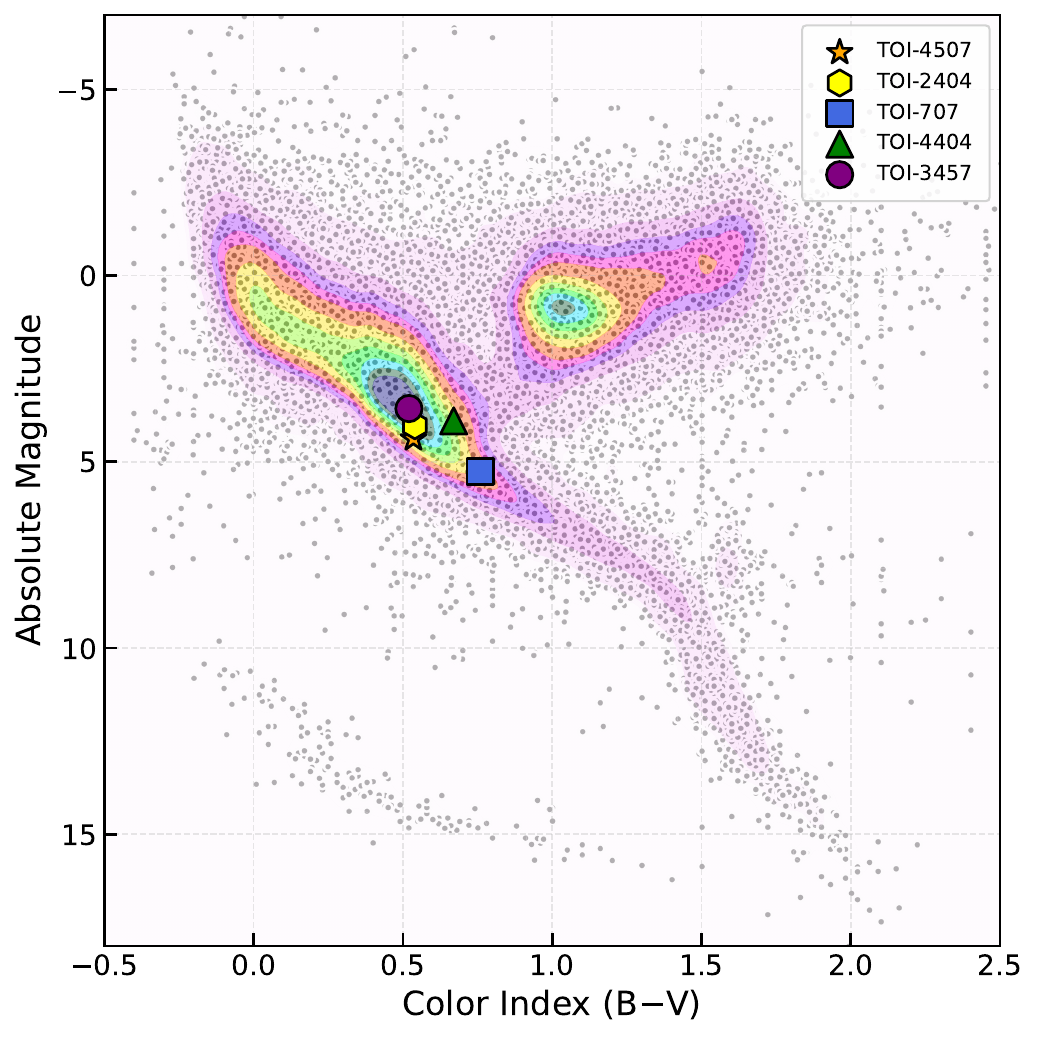}
    \caption{Hertzsprung-Russell diagram showing that all targets cluster in the same region, close to the main-sequence turn-off. The x-axis represents the Color Index (B-V), the y-axis the absolute magnitude, and colour maps the population density\protect\footnotemark.}
   \label{fig:hr_diagram}
\end{figure}

Fig.~\ref{fig:hr_diagram} shows the five host/primary stars placed on the Hertzsprung-Russell diagram, with absolute magnitude on the vertical axis and colour index on the horizontal axis. All five targets lie along the main sequence, consistent with their classification as mid-F to late-G dwarfs, and their positions align well with the radii, masses, and ages inferred from the SED analysis.

\begin{table*}
    \scriptsize
    \centering
        \renewcommand{\arraystretch}{1.6} 
\setlength{\tabcolsep}{10pt} 
    \caption{Summary of stellar parameters (of the exoplanet hosts or primary stars in multi-systems).
    }
    
    \label{tab:stellar-params}
    \renewcommand{\arraystretch}{1.2}
    \begin{tabular}{lccccc}
        \toprule
        
            & \textbf{TOI-4507} 
            & \textbf{TOI-2404} 
            & \textbf{TOI-707} 
            & \textbf{TOI-4404} 
            & \textbf{TOI-3457} \\
        
        \midrule
        
        TIC\textsuperscript{(1)} 
            & 179582003 
            & 142087638 
            & 167342439 
            & 342314656 
            & 357312511 \\
        
        Gaia DR3\textsuperscript{(2)} 
            & 4657949756881138176 
            & 5262245367587966208 
            & 5280307324275126016 
            & 5315690050833397248 
            & 5787442146160163200 \\
        
        RA\textsuperscript{(2)} 
            & 05h21m48.330s 
            & 06h33m46.114s 
            & 06h34m55.100s 
            & 08h14m53.998s 
            & 12h12m49.997s \\
        
        DEC\textsuperscript{(2)} 
            & -69d59m17.58s 
            & -74d11m24.35s 
            & -67d32m14.19s 
            & -57d25m59.10s 
            & -79d45m25.93s \\
        
        PM [RA] (mas/yr)\textsuperscript{(2)} 
            & $22.003 \pm 0.0158$ 
            & $-14.338 \pm 0.0123$ 
            & $-14.750 \pm 0.0142$ 
            & $26.172 \pm 0.0143$ 
            & $-39.544 \pm 0.0132$ \\
        
        PM [DEC] (mas/yr)\textsuperscript{(2)} 
            & $3.972 \pm 0.0169$ 
            & $6.035 \pm 0.0154$ 
            & $-11.073 \pm 0.0154$ 
            & $-46.221 \pm 0.0158$ 
            & $15.071 \pm 0.0130$ \\
        
        Parallax (mas)\textsuperscript{(2)} 
            & $5.643 \pm 0.0123$ 
            & $3.389 \pm 0.0109$ 
            & $7.676 \pm 0.0107$ 
            & $3.582 \pm 0.0117$ 
            & $2.653 \pm 0.0117$ \\
        
        Distance (pc)\textsuperscript{(1)} 
            & $178 \pm 1.0$ 
            & $296 \pm 2.2$ 
            & $131 \pm 0.4$ 
            & $286 \pm 1.7$ 
            & $371 \pm 3.1$ \\[0.2cm]
        
        Gaia RUWE\textsuperscript{(2)} 
            & 0.783 
            & 0.807 
            & 0.856 
            & 0.833 
            & 0.981 \\
        
        Gaia AEN (mas)\textsuperscript{(2)} 
            & 0.0327 
            & 0.0844 
            & 0.0705 
            & 0.1004 
            & 0.0635 \\[0.2cm]
        
        $m_{TESS}$\textsuperscript{(1)} 
            & $10.230 \pm 0.006$ 
            & $11.464 \pm 0.006$ 
            & $10.091 \pm 0.006$ 
            & $10.711 \pm 0.006$ 
            & $11.615 \pm 0.006$ \\
        
        $m_{Gaia G}$\textsuperscript{(2)} 
            & 10.567 
            & 11.561 
            & 10.589 
            & 11.196 
            & 12.153 \\
        
        $m_{Gaia BP}$\textsuperscript{(2)} 
            & 10.848 
            & 11.865 
            & 10.983 
            & 11.571 
            & 12.584 \\
        
        $m_{Gaia RP}$\textsuperscript{(2)} 
            & 10.133 
            & 11.084 
            & 10.037 
            & 10.652 
            & 11.547 \\
        
        $m_B$\textsuperscript{(3)} 
            & $11.52 \pm 0.09$ 
            & $12.11 \pm 0.14$ 
            & $11.42 \pm 0.07$ 
            & $12.42 \pm 0.23$ 
            & $13.195 \pm 0.03$ \\
        
        $m_V$\textsuperscript{(3)} 
            & $10.81 \pm 0.07$ 
            & $11.53 \pm 0.11$ 
            & $10.71 \pm 0.06$ 
            & $11.33 \pm 0.10$ 
            & $12.344 \pm 0.02$ \\
        
        $m_J$\textsuperscript{(4)} 
            & $9.689 \pm 0.023$ 
            & $10.533 \pm 0.026$ 
            & $9.444 \pm 0.023$ 
            & $10.033 \pm 0.023$ 
            & $10.839 \pm 0.022$ \\
        
        $m_H$\textsuperscript{(4)} 
            & $9.432 \pm 0.024$ 
            & $10.252 \pm 0.022$ 
            & $9.062 \pm 0.023$ 
            & $9.704 \pm 0.025$ 
            & $10.516 \pm 0.021$ \\
        
        $m_{Ks}$\textsuperscript{(4)} 
            & $9.381 \pm 0.025$ 
            & $10.212 \pm 0.025$ 
            & $8.972 \pm 0.025$ 
            & $9.605 \pm 0.021$ 
            & $10.400 \pm 0.019$ \\
        
        $m_{W1}$\textsuperscript{(5)} 
            & $9.344 \pm 0.023$ 
            & $10.170 \pm 0.023$ 
            & $8.944 \pm 0.022$ 
            & $9.559 \pm 0.022$ 
            & $10.423 \pm 0.019$ \\
        
        $m_{W2}$\textsuperscript{(5)} 
            & $9.272 \pm 0.020$ 
            & $10.193 \pm 0.020$ 
            & $9.006 \pm 0.020$ 
            & $9.610 \pm 0.020$ 
            & $10.376 \pm 0.057$ \\
        
        $m_{W3}$\textsuperscript{(5)} 
            & $9.227 \pm 0.058$ 
            & $10.175 \pm 0.044$ 
            & $8.937 \pm 0.021$ 
            & $9.588 \pm 0.031$ 
            & $9.186$ \\
        
        $m_{W4}$\textsuperscript{(5)} 
            & $9.961$ 
            & $9.277$ 
            & $9.096 \pm 0.229$ 
            & $9.210 \pm 0.466$ 
            & -- \\[0.2cm]

        $T_{\mathrm{eff}}$ (K) 
            & $6235 \pm 100$\textsuperscript{(6)} 
            & $6230 \pm 120$\textsuperscript{(7)} 
            & $5424 \pm 61$\textsuperscript{(8)} 
            & $5732 \pm 100 $\textsuperscript{(8)} 
            & $6300 \pm 120$\textsuperscript{(7)} \\
        
        $\log g$ (cgs) 
            & $4.44 \pm 0.15$\textsuperscript{(6)} 
            & $4.40 \pm 0.15$\textsuperscript{(7)} 
            & \edit{$4.39 \pm 0.07 $\textsuperscript{(1)}} 
            & $4.36 \pm 0.1 $\textsuperscript{(8)} 
            & $4.28 \pm 0.05$\textsuperscript{(7)} \\
        
        {[Fe/H] (dex)} 
            & $-0.08 \pm 0.05$\textsuperscript{(6)} 
            & $0.14 \pm 0.05$\textsuperscript{(7)} 
            & $0.16 \pm 0.05 $\textsuperscript{(8)} 
            & $-0.18 \pm 0.1 $\textsuperscript{(8)} 
            & $0.28 \pm 0.05$\textsuperscript{(7)} \\
        
        $v\sin{i}$ (km/s) 
            & $4.91 \pm 0.3$\textsuperscript{(6)} 
            & $4.2 \pm 0.5$\textsuperscript{(7)} 
            & $4.9 \pm 1.0$\textsuperscript{(8)} 
            & $5.3 \pm 1.0$\textsuperscript{(8)} 
            & $6.2 \pm 0.5$\textsuperscript{(7)} \\[0.2cm]
        
        $A_V$ (mag)\textsuperscript{(9)} 
            & $0.14 \pm 0.03$ 
            & $0.14 \pm 0.06$ 
            & $0.05 \pm 0.04$ 
            & $0.15 \pm 0.05$ 
            & $0.36 \pm 0.03$ \\
        
        $F_{\rm bol}$ ($10^{-9}$ cgs)\textsuperscript{(9)} 
            & $1.498 \pm 0.034$ 
            & $0.635 \pm 0.015$ 
            & $1.473 \pm 0.034$ 
            & $0.874 \pm 0.031$ 
            & $0.443 \pm 0.016$ \\
        
        $L_{\rm bol}$ ($L_\odot$)\textsuperscript{(9)} 
            & $1.466 \pm 0.034$ 
            & $1.724 \pm 0.041$ 
            & $0.779 \pm 0.018$ 
            & $2.124 \pm 0.075$ 
            & $1.962 \pm 0.069$ \\
        
        $R_\star$ ($R_\odot$)\textsuperscript{(9)} 
            & $1.039 \pm 0.036$ 
            & $1.203 \pm 0.025$ 
            & $0.978 \pm 0.021$ 
            & $1.581 \pm 0.040$ 
            & $1.474 \pm 0.037$ \\
        
        $M_\star$ ($M_\odot$)\textsuperscript{(9)} 
            & $1.15 \pm 0.07$ 
            & $1.16 \pm 0.07$ 
            & $0.99 \pm 0.06$ 
            & $1.03 \pm 0.06$ 
            & $1.04 \pm 0.06$ \\
        
        $\rho_\star$ ($\mathrm{g/cm^3}$)\textsuperscript{(9)} 
            & $1.44 \pm 0.19$ 
            & $0.939 \pm 0.086$ 
            & $1.49 \pm 0.14$ 
            & $0.368 \pm 0.035$
            & $0.458 \pm 0.040$ \\
        
        $P_{\rm rot}/\sin i_\star$ (d)\textsuperscript{(9)} 
            & $10.7 \pm 0.8$ 
            & $10.1 \pm 1.7$ 
            & $9.9 \pm 2.0$ 
            & --
            & -- \\
        
        Age ($P\sin i$) (Gyr)\textsuperscript{(9)} 
            & $1.8 \pm 0.2$ 
            & $1.1 \pm 0.3$ 
            & $0.6 \pm 0.2$ 
            & -- 
            & -- \\

        \bottomrule
    \end{tabular}
    
    \caption*{Sources: 
        \textsuperscript{(1)} TESS Input Catalog v8.2, \citep{Stassun2018, Stassun2019, Paegert2022}.
        \textsuperscript{(2)} Gaia DR3 \citep{2023_gaiaDR3}.
        \textsuperscript{(3)} TOI-4507, TOI-2404, and TOI-707 from Tycho-2 \citep{2000_Tycho2}; TOI-4404 and TOI-3457 from UCAC4 Catalogue \citep{2012_UCAC4_Catalogue}.
        \textsuperscript{(4)} 2MASS \citep{2003_2MASS}.
        \textsuperscript{(5)} WISE \citep{2010_WISE}.
        \textsuperscript{(6)} this work; spectroscopic parameters from HARPS \citep{2003_Harps}.
        \textsuperscript{(7)} this work; spectroscopic parameters from FEROS \citep{FEROS}.
        \textsuperscript{(8)} this work; spectroscopic parameters from CHIRON \citep{2013_CHIRON}.
        \textsuperscript{(9)} this work; SED analysis following \cite{2017_Stassun} and \cite{Stassun_2018}; mass-radius relations from \cite{Torres2010}; for TOI-4507, TOI-2404, and TOI-707 spectroscopic priors are used as input; for TOI-4404 and TOI-3457 the SEDs are freely sampled.
        The shown parameters are the TESS Input Catalog (TID) identifier, the Gaia Data Release 3 (DR3) identifier, right ascension (RA), declination (DEC), proper motions (PM), parallax, distance, Gaia Renormalized Unit Weight Error (RUWE) and Astrometric Excess Noise (AEN) which may indicate binarity, the magnitudes $m$ in various bandpasses, stellar effective temperature $T_\mathrm{eff}$, surface gravity $g$, metallicity [Fe/H], projected rotational velocity $v \sin{i}$, visual extinction $A_V$, bolometric flux $F_\mathrm{bol}$, bolometric luminosity $L_\mathrm{bol}$, radius $R_{\star}$, mass $M_{\star}$, density $\rho_{\star}$, rotation period $P_\mathrm{rot}/\sin i_{\star}$, and estimated age.
        }
    
\end{table*}

\footnotetext{This plot is created via \url{https://github.com/RobertoIA/Hertzsprung-Russell} using the HYG database (v3) at \url{http://www.astronexus.com/hyg} and \url{https://github.com/astronexus/HYG-Database}.}

\section{Validation of the planets}
\label{sec:validation}
We investigated the statistical validation of all five candidates using \triceratops{} \citep{Giacalone_2020, Giacalone_Dressing_2020}\footnote{Tool for Rating Interesting Candidate Exoplanets and Reliability Analysis of Transits Originating from Proximate Stars; \url{https://github.com/stevengiacalone/triceratops}}, a Bayesian framework tailored to TOIs that estimates false positive probability (FPP) and nearby false positive probability (NFPP). It evaluates scenarios such as background eclipsing binaries, hierarchical triples, and grazing eclipses by combining TESS photometry, Gaia astrometry, stellar parameters, and contrast curves from high-resolution imaging. The analysis uses only phase-folded lightcurves cropped around the transit window to emphasise transit shape and exclude unrelated variability. For each target, we included only TESS sectors with clearly detected signals. Contrast curves were incorporated to exclude contaminating nearby sources within the aperture, significantly reducing the likelihood of unresolved false positives. For each of the five TOIs in our sample, we ran \triceratops{} using the corresponding TESS SPOC lightcurves, Gaia DR2 information, and our contrast curves derived from high-resolution imaging (see Section \ref{subsec:High-resolution_imaging}), which rule out contaminating stellar companions within a few arcseconds of the targets. Following the standard thresholds of \citet{Giacalone_Dressing_2020}, we consider a candidate statistically validated when FPP $<$ 1.5\% and NFPP $<$ 0.1\%.
Table~\ref{tab:triceratops_results} summarises the results. Three of our long-period candidates (TOI-4507.01, TOI-707.01, and TOI-3457.01) fall below the validation thresholds and are statistically confirmed as planets. In contrast, TOI-4404.01 and TOI-2404.02/03 exceed the thresholds and are classified as false positives. This outcome aligns with independent evidence: CORALIE measurements confirm TOI-4404.01 as a double-lined spectroscopic binary, while the signals from TOI-2404.02/03 match the primary and secondary eclipses of an eccentric eclipsing binary. Although not part of our observing campaign, we also assessed the shorter-period signals TOI-2404.01 and TOI-707.02. TOI-2404.01 is confidently validated as a planetary signal, presenting an intriguing case for future study (see Section~\ref{subsec:TOI-2404_and_TOI-4404}). In contrast, TOI-707.02 remains unvalidated, and its true nature will require continued follow-up. \edit{A brief discussion of its possible nature is provided in Section \ref{subsec:TOI-707}}.

\begin{table}
\scriptsize
\centering
\caption{Results of the \triceratops{} statistical validation analysis, which validates candidates when FPP<1.5\% and NFPP<0.1\%.}
\label{tab:triceratops_results}
\begin{tabular}{lccc}
\toprule
\textbf{Target} & \textbf{FPP (\%)} & \textbf{NFPP (\%)} & \textbf{Validation Status} \\
\midrule
TOI-4507.01 & $0.14 \pm 0.01$     & $0.001$ & Validated \\[0.1cm]
TOI-2404.01 & $0.56 \pm 0.35$     & $0.000$ & Validated \\
TOI-2404.02 & $99.98 \pm 0.08$    & $0.000$ & Rejected \\
TOI-2404.03 & $56.85 \pm 14.95$   & $7.04$ & Rejected \\[0.1cm]
TOI-707.01  & $0.62 \pm 0.18$     & $0.000$ & Validated \\
TOI-707.02  & $64.16 \pm 0.27$      & $3.26$ & Rejected \\[0.1cm]
TOI-4404.01 & $99.90 \pm 0.11$    & $0.070$ & Rejected \\[0.1cm]
TOI-3457.01 & $0.56 \pm 0.98$     & $0.000$ & Validated \\
\bottomrule
\end{tabular}
\end{table}

\section{Photometric and radial velocity analysis}
\label{sec:Analysis}
We computed planetary and eclipsing binary parameters using \allesfitter{} \citep{allesfitter-code,allesfitter-paper}, an open-source framework that integrates tools such as \texttt{ellc} \citep{Maxted_2016} for lightcurve and RV modelling, and \texttt{celerite} \citep{2017AJ....154..220F} for Gaussian Process (GP) regression. \allesfitter{} handles planetary transits, eclipsing binaries, and stellar variability across photometric and spectroscopic datasets, with parameter estimation via MCMC (\texttt{emcee}; \citealt{Foreman_Mackey_2013}) and Nested Sampling (\texttt{dynesty}; \citealt{10.1093/mnras/staa278}). We jointly modelled TESS, ASTEP, and other ground-based lightcurves, along with RV data when available. Stellar activity and systematics were captured using Gaussian Processes (GPs) and hybrid splines, and models were optimised via MCMC following \citet{allesfitter-paper}. For each target, we followed a consistent multi-step process with minor adjustments. We began by recovering TESS transits using \texttt{transitleastsquares} \citep{Hippke_2019} and defining transit windows spanning roughly three transit durations. A GP with a Matern 3/2 kernel was trained on out-of-window data to set priors for its hyperparameters and white noise term. We then fitted the in-window data with a transit model, uniformly sampling quadratic limb-darkening parameters in q-space \citep{Kipping2013}, while modelling systematics using normal priors from the pre-trained GP and error scaling. We then incorporated ASTEP and other ground-based photometry into a joint fit. Due to their lower signal-to-noise, we applied fixed quadratic limb-darkening coefficients derived from stellar atmosphere models, model systematics using hybrid splines, and uniformly sample the error scaling. Specifically, we adopted coefficients from the PHOENIX/1D grid (\citealt{2013_Claret}; Table~\ref{tab:posteriors_extra}) assuming solar composition, a microturbulent velocity of 2\,km\,s$^{-1}$, a mixing-length parameter of 2.0, and spanning the relevant ranges in temperature, gravity, and metallicity. Where FEROS, CORALIE, or HARPS RVs were available, we performed a joint photometric-spectroscopic fit to further constrain system parameters, particularly eccentricity, uniformly sampling RV offsets and jitter (added in quadrature to reported errors). For each target, we first iteratively leveraged MCMC approaches to explore the parameter space and refine the initial guesses. For this, we typically use $\sim$200 walkers and a series of shorter, consecutive runs ($\sim$1000-10000 steps in each case).
We then conducted a series of final runs using Nested Sampling, in order to compare the Bayesian evidence $Z$ of various models. 
For each, the optimisation is counted as converged once the default stopping criterion of $\Delta \log{Z} = 0.01$ is reached \citep{allesfitter-paper}. 
\edit{For TOI-2404, the dilution of the TESS lightcurves was included as a free parameter in the modelling, adopting a uniform prior between 0 and 1, with a resulting value of $D_{\mathrm{TESS}} = 0.34^{+0.22}_{-0.19}$. No explicit dilution parameter was included for the other systems, where no significant contamination was identified.}
In all cases, we then also considered the statistical diagnostics described in \citet{allesfitter-paper}, and performed a series of visual inspections to confirm the stability and reliability of the results.

\section{Results and discussion}
\label{sec:results_and_discussion}
\begin{figure}
  \centering
  \includegraphics[width=\columnwidth]{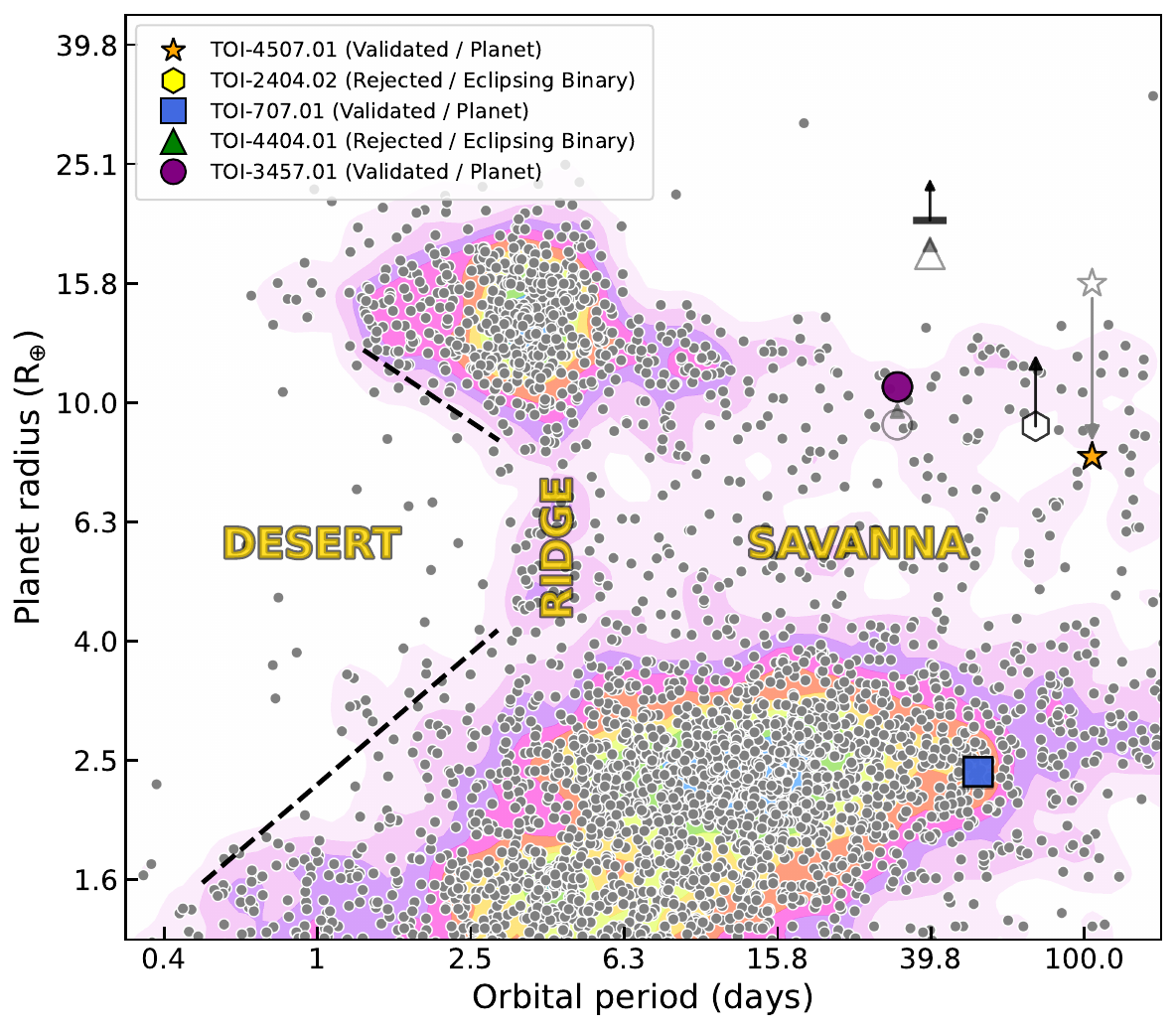}
  \caption{Three validated planets and two eclipsing binaries (coloured symbols) put into the context of all known planets (grey points) drawn from the NASA Exoplanet Archive. The grey markers represent the previously reported values prior to our analysis, while the arrows indicate lower limits on the stellar radii in the case of eclipsing binaries. Note the targets' intriguing positions in the radius-period parameter space relative to the most densely populated regions \citep{CastroGonzalez_2024}\protect\footnotemark.}
   \label{fig:savanna_plot}
\end{figure}

We present here the results for each target, including their derived physical and orbital properties interpreted in the context of planetary formation and evolution. Tables~\ref{tab:tois_alles} and \ref{tab:posteriors_extra} summarise the posterior parameters from model fits, constraining companion radius, orbital period, and other system characteristics. 
Fig.~\ref{fig:savanna_plot} places our results in the broader demographic context of the known exoplanet population.
\edit{The phase-folded lightcurves for each target are presented in this section, together with their best-fit models, with radial velocity data shown where available.}

\begin{table*}
   \scriptsize
   \centering
    \caption{Summary of the companions’ posterior and derived parameters estimated in this work.}
    \label{tab:tois_alles}    
    \renewcommand{\arraystretch}{1.2} 
    \begin{tabular}{lccccc}
        \toprule
         & \textbf{TOI-4507.01} & \textbf{TOI-2404.02\,/\,.03} & \textbf{TOI-707.01} & \textbf{TOI-4404.01} & \textbf{TOI-3457.01} \\
        \midrule

        \textbf{Modeled as} & \textbf{Exoplanet} & \textbf{Eclipsing Binary} & \textbf{Exoplanet} & \textbf{Eclipsing Binary} & \textbf{Exoplanet} \\ [0.25cm] 

        \multicolumn{6}{l}{\emph{Fitted parameters:}} \\ [0.1cm]

        $R_\square / R_\star$ 
        & {\brrA} 
        & unconstrained 
        & {\brrE} 
        & {\BrrB} 
        & {\brrD} \\ [0.1cm]

        $(R_\star + R_\square) / a_\square$ 
        & {\brsumaA} 
        & {\BrsumaC}
        & {\brsumaE} 
        & {\BrsumaB} 
        & {\brsumaD} \\ [0.1cm]

        $\cos{i_\square}$ 
        & {\bcosiA} 
        & {\BcosiC} 
        & {\bcosiE} 
        & {\BcosiB} 
        & {\bcosiD} \\ [0.1cm]

        $P_\square$ (days) 
        & {\bperiodA} 
        & {\BperiodC} 
        & {\bperiodE} 
        & {\BperiodB} 
        & {\bperiodD} \\ [0.1cm]

        $T_{0;\square}$ (BJD) 
        & {\bepochA} 
        & {\BepochC} 
        & {\bepochE} 
        & {\BepochB} 
        & {\bepochD} \\ [0.1cm]

        $\sqrt{e_\square} \cos{\omega_\square}$ 
        & {\bfcA} 
        & {\BfcC} 
        & --
        & --
        & {\bfcD} \\ [0.1cm]

        $\sqrt{e_\square} \sin{\omega_\square}$ 
        & {\bfsA} 
        & {\BfsC} 
        & --
        & -- 
        & {\bfsD} \\ [0.1cm]

        \edit{$K_\square$ (km/s) [99.9\%]} 
        & $<0.009$
        & $<0.008$ (non-detection)
        & -- 
        & -- 
        & $<0.200 $\\ [0.1cm]

        $J_\square$ 
        & -- 
        & unconstrained 
        & -- 
        & {\BsbratioTESSB} 
        & -- \\ [0.1cm]

        Dilution [99.9\%]
        & -- 
        & {\dilTESSC}
        & -- 
        & --
        & -- \\ [0.25cm]

        \multicolumn{6}{l}{\emph{Derived parameters:}} \\ [0.1cm]

        $R_\star/a_\square$ 
        & {\bRstaraA} 
        & unconstrained 
        & {\bRstaraE} 
        & {\BRstaraB} 
        & {\bRstaraD} \\ [0.1cm]

        $a_\square/R_\star$ 
        & {\baRstarA} 
        & unconstrained 
        & {\baRstarE} 
        & {\BaRstarB} 
        & {\baRstarD} \\ [0.1cm]

        $R_\square/a_\square$ 
        & {\bRcompanionaA} 
        & unconstrained 
        & {\bRcompanionaE} 
        & {\BRcompanionaB} 
        & {\bRcompanionaD} \\ [0.1cm]

        $R_\square$ ($\mathrm{R_{\oplus}}$) 
        & {\bRcompanionRearthA} 
        & unconstrained 
        & {\bRcompanionRearthE} 
        & {\BRcompanionRearthB} 
        & {\bRcompanionRearthD} \\ [0.1cm]

        $R_\square$ ($\mathrm{R_{jup}}$) 
        & {\bRcompanionRjupA} 
        & unconstrained 
        & {\bRcompanionRjupE} 
        & {\BRcompanionRjupB} 
        & {\bRcompanionRjupD} \\ [0.1cm]

        $a_\square$ ($\mathrm{R_{\odot}}$)  
        & {\baRsunA} 
        & unconstrained 
        & {\baRsunE} 
        & {\BaRsunB} 
        & {\baRsunD} \\ [0.1cm]

        $a_\square$ (AU)  
        & {\baAUA} 
        & unconstrained
        & {\baAUE} 
        & {\BaAUB} 
        & {\baAUD} \\ [0.1cm]

        $i_\square$ (deg) 
        & {\biA} 
        & {\BiC} 
        & {\biE} 
        & {\BiB} 
        & {\biD} \\ [0.1cm]

        $e_\square$ 
        & -- 
        & {\BeC}
        & --
        & --
    & \edit{ $> 0.59$ [$3\sigma$]} \\ [0.1cm]
        
        $\omega_\square$ 
        & -- 
        & {\BwC}
        & -- 
        & -- 
        & {\bwD} \\ [0.1cm] 

        $b_\mathrm{tra;\square}$ 
        & {\bbtraA} 
        & unconstrained 
        & {\bbtraE} 
        & {\BbtraB} 
        & {\bbtraD} \\ [0.1cm] 

        $T_\mathrm{tot;\square}$ (h) 
        & {\bTtratotA} 
        & {\BTtratotC} 
        & {\bTtratotE} 
        & {\BTtratotB} 
        & {\bTtratotD} \\ [0.1cm] 

        $T_\mathrm{full;\square}$ (h) 
        & {\bTtrafullA} 
        & {\BTtrafullC} 
        & {\bTtrafullE} 
        & {\BTtrafullB}
        & {\bTtrafullD} \\ [0.1cm] 

        $T_\mathrm{eq;\square | A_b=0}$ (K) 
        & {\TeqA} 
        & -- 
        & {\TeqE} 
        & -- 
        & {\TeqD} \\ [0.1cm]
        
        $T_\mathrm{eq;\square | A_b=0.3}$ (K) 
        & {\bTeqA} 
        & -- 
        & {\bTeqE} 
        & -- 
        & {\bTeqD} \\ [0.1cm] 

        $\rho_\mathrm{\star; orbital}$ (cgs) 
        & {\combinedhostdensityA} 
        & -- 
        & {\combinedhostdensityE} 
        & -- 
        & {\combinedhostdensityD} \\ [0.1cm]

       \edit{$M_{\mathrm{\Box}}$ ($M_{\mathrm{Jup}}$) [99.9\%] }
       & $< 0.24$ 
       & -- 
       & -- 
       & -- 
       & $< 2.51$ \\ [0.1cm]

        \bottomrule
    \end{tabular}

    \caption*{
        We use the suffix $\square$ as placeholder for all planet (b) and secondary star (B) identifiers. 
        Shown are the companion radius $R_\square$, host/primary star radius $R_\star$, orbital semi-major axis $a$, inclination $i$, orbital period $P$, epoch $T_0$, eccentricity $e$, argument of periastron $\omega$, RV semi-major amplitude $K$, surface brightness ratio $J$, dilution, impact parameter $b$, total transit time $T_\mathrm{tot}$, full transit time $T_\mathrm{full}$, equilibrium temperature $T_\mathrm{eq}$ evaluated at certain Bond albedos $A_b$, fitted stellar density $\rho_\star$, and companion mass $M_\square$.
        }
\end{table*}

\footnotetext{This plot was generated using \texttt{nep-des} (\url{https://github.com/castro-gzlz/nep-des}).}

\subsection{Demographic context - planetary diversity in the period-radius parameter space}
 \label{subsec:demographic_context}

Before examining individual targets in detail, we first zoom out and situate our validated three planets (TOI-4507\,b, TOI-3457\,b, and TOI-707\,b) within the broader demographic landscape of known exoplanets. This contextual view illuminates how these systems contribute to our understanding of planetary diversity. Fig.~\ref{fig:savanna_plot} places our targets within the period-radius diagram, showing their positions relative to the wider population of confirmed exoplanets \citep[e.g.][]{CastroGonzalez_2024}. The diagram reveals two dominant clusters. The first comprises short-period giant planets at radii above $\sim$10~$R_{\oplus}$ and periods shorter than $\sim$10 days, \edit{commonly referred to as the hot Jupiter pile-up \citep{UdrySantos2007}}. The second consists of short-period small planets, mainly super-Earths and sub-Neptunes between $\sim$1-4~$R_{\oplus}$ and periods under $\sim$30 days, many located on either side of the radius valley ($\sim$1.6-1.8~$R_{\oplus}$), likely sculpted by atmospheric escape. \edit{Between} these dense clusters, \edit{the distribution in the super-Neptune/sub-Jovian domain shows a distinct structure, reminiscent of that observed for Jupiter-\edit{sized} planets: an overdensity of dense Neptunes at orbital periods of 3-6 days, known as the Neptunian ridge \citep{CastroGonzalez_2024,CastroGonzalez_2024b}, separates the \edit{Neptunian desert} \citep{2011ApJ...727L..44S} from the sparsely populated savanna \citep{Bourrier_2023}.} \edit{The savanna} connects the hot, compact populations to the cold giants and long-period sub-Neptunes. Two of our validated planets, TOI-4507\,b and TOI-3457\,b, reside in this sparse region. Both occupy the warm Saturn regime, bridging the gap between hot giants close to their stars and distant cold gas planets. They also lie in a transitional zone between Neptune- and Jupiter-sized worlds, probing a poorly sampled domain of warm giant planets at long orbital periods. Our third validated planet, TOI-707\,b, differs slightly in nature. It sits at the outskirts of the sub-Neptune population, marking the gradual shift toward cooler planets that may still retain substantial volatile envelopes. Importantly, all three validated planets lie beyond the clusters of hot Jupiters or short-period radius valley, and with orbital periods longer than one month, they are part of a rare group more likely to preserve much of their primordial atmospheres.

\subsection{\texorpdfstring
  {TOI-4507\,b \& TOI-3457\,b - giants of the savanna}
  {\detokenize{TOI-4507 b & TOI-3457 b - giants of the savanna}}}

\label{subsec:TOI-4507_and_TOI-3457}

\begin{figure}
\centering
\includegraphics[width=0.98\columnwidth]{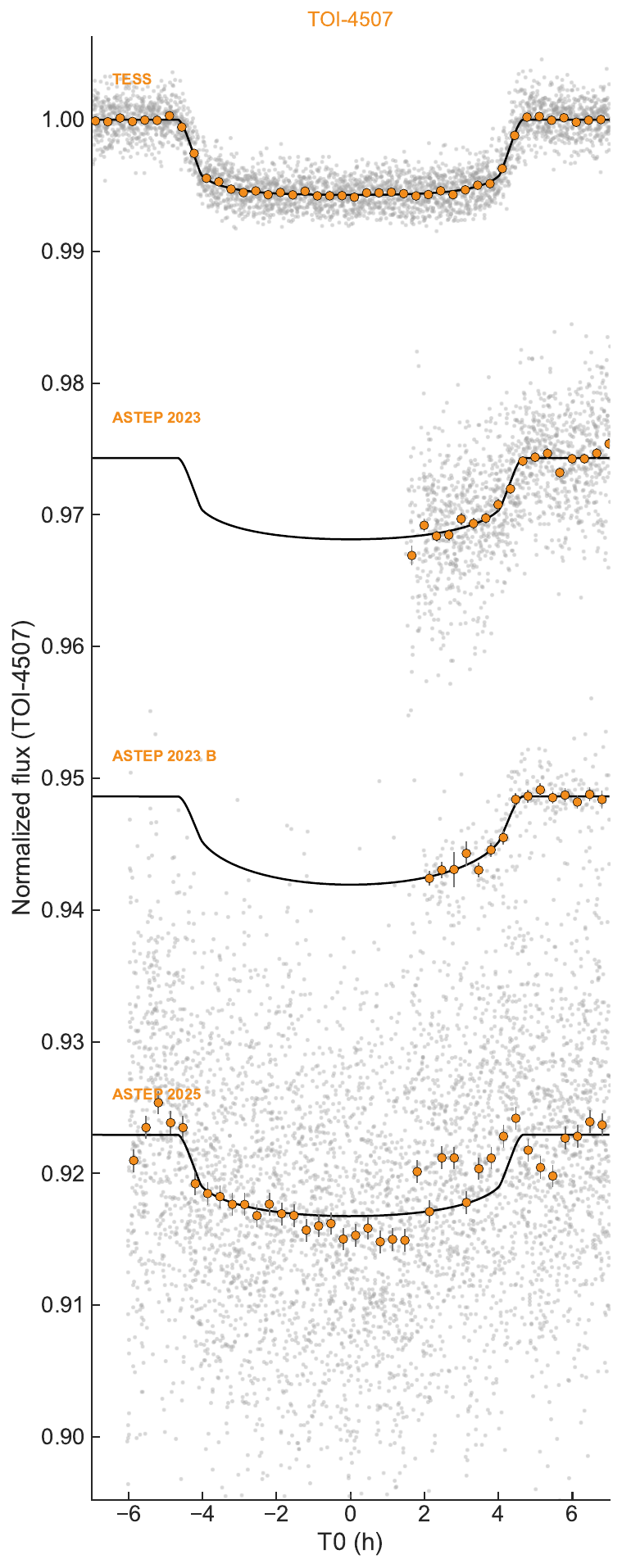}
\caption{Phase-folded lightcurve of TOI-4507 from TESS and ground-based photometry. The best-fit model is generated with \allesfitter{}.}
\label{fig:toi4507_phot}
\end{figure}

\begin{figure}
\centering
\includegraphics[width=\columnwidth]{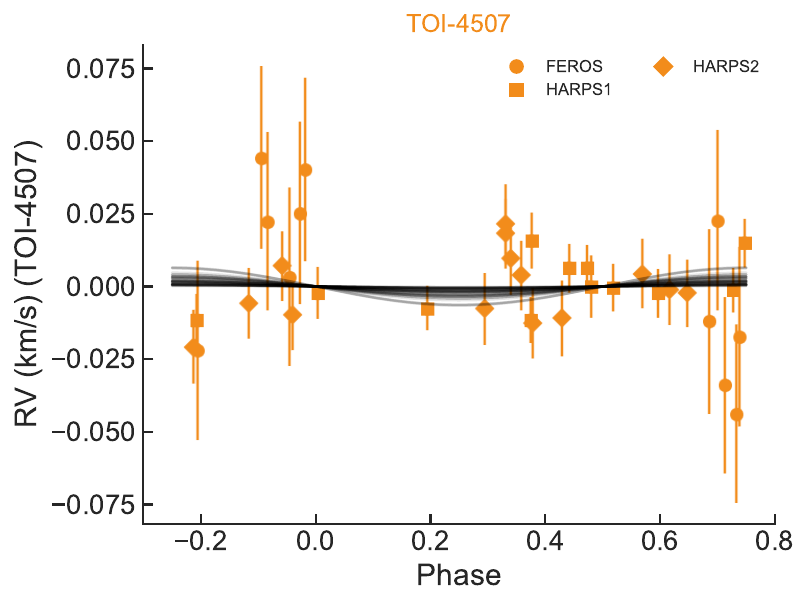}
\caption{Radial velocity measurements for TOI-4507 with the best-fit model.}
\label{fig:toi4507_rv}
\end{figure}

Among the studied systems, TOI-4507\,b and TOI-3457\,b stand out as giant-sized, long-period planets whose moderate irradiation and wide separations set them apart from the more common short-period giants. TESS photometry first revealed both signals, and our extensive ground-based follow-up validates them by combining high-precision, multi-colour ASTEP and TFOP-network photometry, FEROS and HARPS radial velocity measurements, and statistical validation. For TOI-4507\,b, the joint modelling of TESS and ASTEP photometry with out-of-transit FEROS and HARPS radial velocities (Fig.~\ref{fig:toi4507_phot} and \ref{fig:toi4507_rv}) yields a warm Saturn with radius {\bRcompanionRearthA}\,$R_\oplus$. Its orbital period of 104.61\,d (semi-major axis of 0.45\,AU), and equilibrium temperature of $<$615\,K place it well beyond the reach of strong tidal forces \citep{Jackson2008} and intense stellar heating \citep{Demory_2011}. 
Our final reported values adopt a circular orbital model with a linear trend in the RV data. 
When comparing circular and eccentric models, the Bayesian evidence ($\Delta Z < 0$) still supports the simpler circular solution, although it is worth noting the more complex model would have yielded a high eccentricity inconsistent with zero.
We also tested more complex models for the evident long-term RV trend, yet the Bayesian evidence ($\Delta Z < 3$) still favors a simple linear slope given the current data. \edit{A Generalized Lomb-Scargle (GLS) analysis of the RV data did not reveal significant periodicities at the orbital period of the planet, consistent with the sparse sampling and expected RV semi-amplitude.}
It remains to be seen whether stellar activity or a long-period companion may be contributing.
\edit{We refer the reader to \citet{Espinoza-Retamal_2026} for a detailed analysis of this target,} in which additional Rossiter-McLaughlin effect measurements can provide further insights on the system's eccentricity, obliquity, and thus evolutionary story.

\begin{figure}
\centering
\includegraphics[width=0.98\columnwidth]{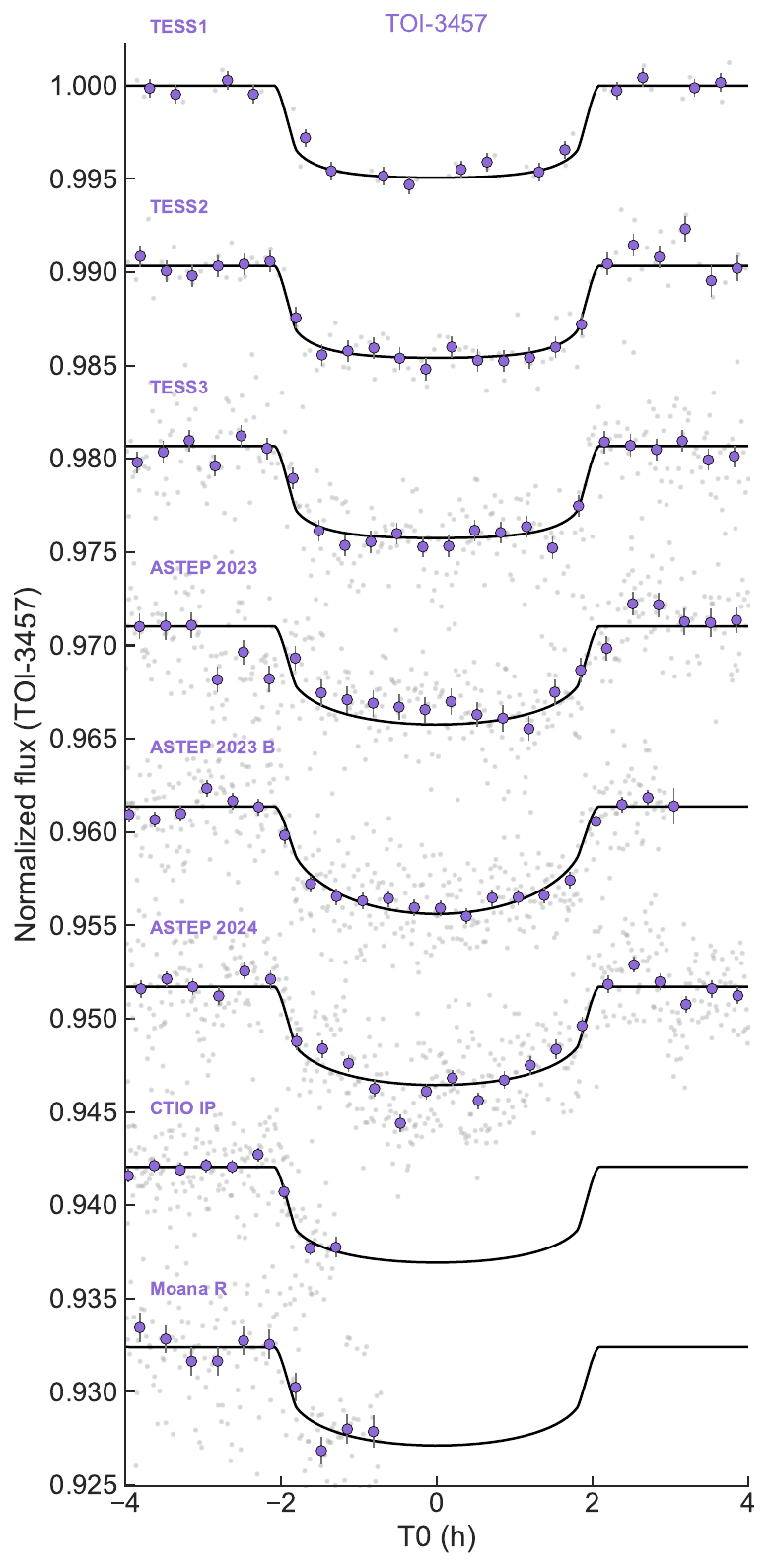}
\caption{Phase-folded lightcurve of TOI-3457 from TESS and ground-based photometry. The best-fit model is generated with \allesfitter{}.}
\label{fig:toi3457_phot}
\end{figure}

\begin{figure}
\centering
\includegraphics[width=\columnwidth]{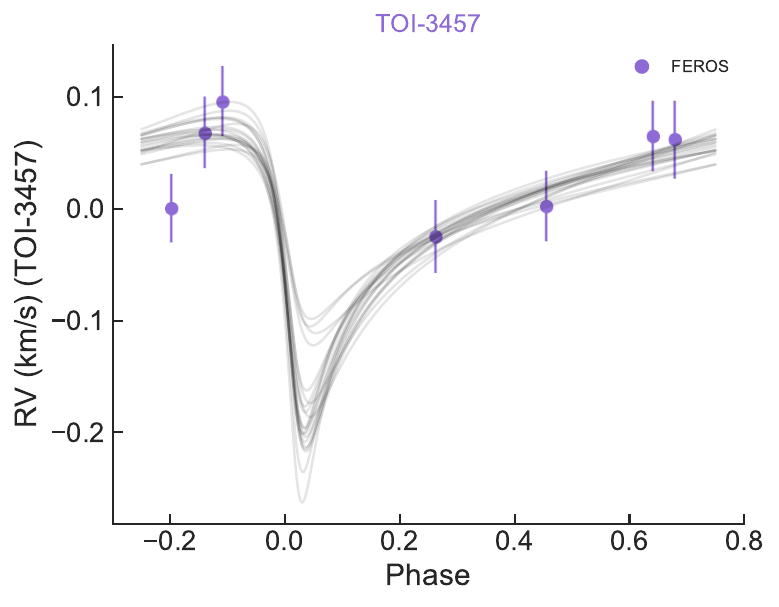}
\caption{Radial velocity measurements for TOI-3457 with the best-fit model.}
\label{fig:toi3457_rv}
\end{figure}

TOI-3457\,b, observed with ASTEP, MoanaES ($r$), and LCO-CTIO ($i'$) in addition to TESS, and monitored in radial velocity with FEROS (Fig.~\ref{fig:toi3457_phot} and \ref{fig:toi3457_rv}), fits into the same warm giant category. The \allesfitter{} run indicates a planet of radius {\bRcompanionRearthD}\,$R_\oplus$, orbiting every 33.60\,days at 0.20\,AU and an equilibrium temperature in the $<$700\,K range. This also makes it consistent with minimal atmospheric escape \citep{Owen_2019}.
\edit{We report the eccentric model, which is strongly supported by the Bayesian evidence ($\Delta \ln Z \approx 54$). This solution favours a non-circular orbit, for which we conservatively report a lower limit of $e > 0.59$ at 3$\sigma$. We note, however, that the current dataset remains limited, and that the eccentricity should not be interpreted as an independent constraint from the RV data alone. In particular, while the RV semi-amplitude seems reasonably well constrained, the parameters $\sqrt{e}\cos\omega$ and $\sqrt{e}\sin\omega$ do not provide a strong exclusion of the circular solution. The eccentricity exhibits a moderate correlation with transit-derived parameters such as $a/R_\star$ and impact parameter, which may indicate that the eccentricity constraint is driven by the transit duration in combination with the adopted stellar density prior.} Both planets reside in the sparsely populated savanna region of the period-radius diagram (Fig.~\ref{fig:savanna_plot}), bridging the gap between hot and cold giant populations. Their location beyond the reach of strong tidal forces and their probable early formation via core accretion \citep{Savvidou_2023} suggest migration to current orbits before disk dispersal \citep{Venturini_2016}. Theoretical models predict that eccentricity and inclination can help distinguish such migration mechanisms \citep{Nelson2018}. Smooth disk migration produces nearly circular orbits ($e \sim 0$). \edit{Similarly, planets undergoing high-eccentricity migration that are deposited close to the star, within the pile-up and ridge, are expected to have mostly circularized orbits due to tidal dissipation \citep{Dawson2018,CastroGonzalez_2026}.} \edit{In contrast,} high-eccentricity migration \edit{that leads to longer-period final orbits} \edit{is less affected by tidal circularization and can} leave planets with $e > 0.2$ \citep[e.g.][]{juric2008dynamical}. 
\edit{For TOI-4507\,b and TOI-3457\,b, the available RV data do not yet allow a fully independent characterisation of their orbital properties and may be affected by long-term trends.} Continued monitoring will be essential to constrain their eccentricities, which may eventually hint at dynamically excited configurations. This could clarify their migration histories, potentially consistent with the view that warm Saturns form beyond the snow line \citep{Pollack1996} and migrate inward early \citep{Winn_2015}. Warm giants like TOI-4507\,b and TOI-3457\,b likely retain much of their primordial atmospheres, as transmission spectroscopy often reveals extended hydrogen-helium envelopes \citep{Muller_2023}. Similar planets have recently been identified by the WINE survey, including TOI-6695\,b ($\approx$80\,d period) \citep{wine4} and several warm Jupiters and sub-Saturns with periods of 10-20\,d \citep{wine3}, as well as by the Next-Generation Transit Survey (NGTS; \citealt{Wheatley_2017}), which has uncovered a number of longer-period warm giants in a comparable regime \citep{gill2024,battley2024,ulmermoll2025}. Future atmospheric characterisation of these targets may shed further light on their primordial fingerprints. In particular, TOI-4507\,b appears promising given its high transmission spectroscopy metric (see \edit{Appendix}~\ref{app:atmospheric_characterisation_prospects}).

\subsection{\texorpdfstring{TOI-707\,b - a sub-Neptune at the population edge} {TOI-707 b - a sub-Neptune at the population edge}}
\label{subsec:TOI-707}

\begin{figure}
\centering
\includegraphics[width=\columnwidth]{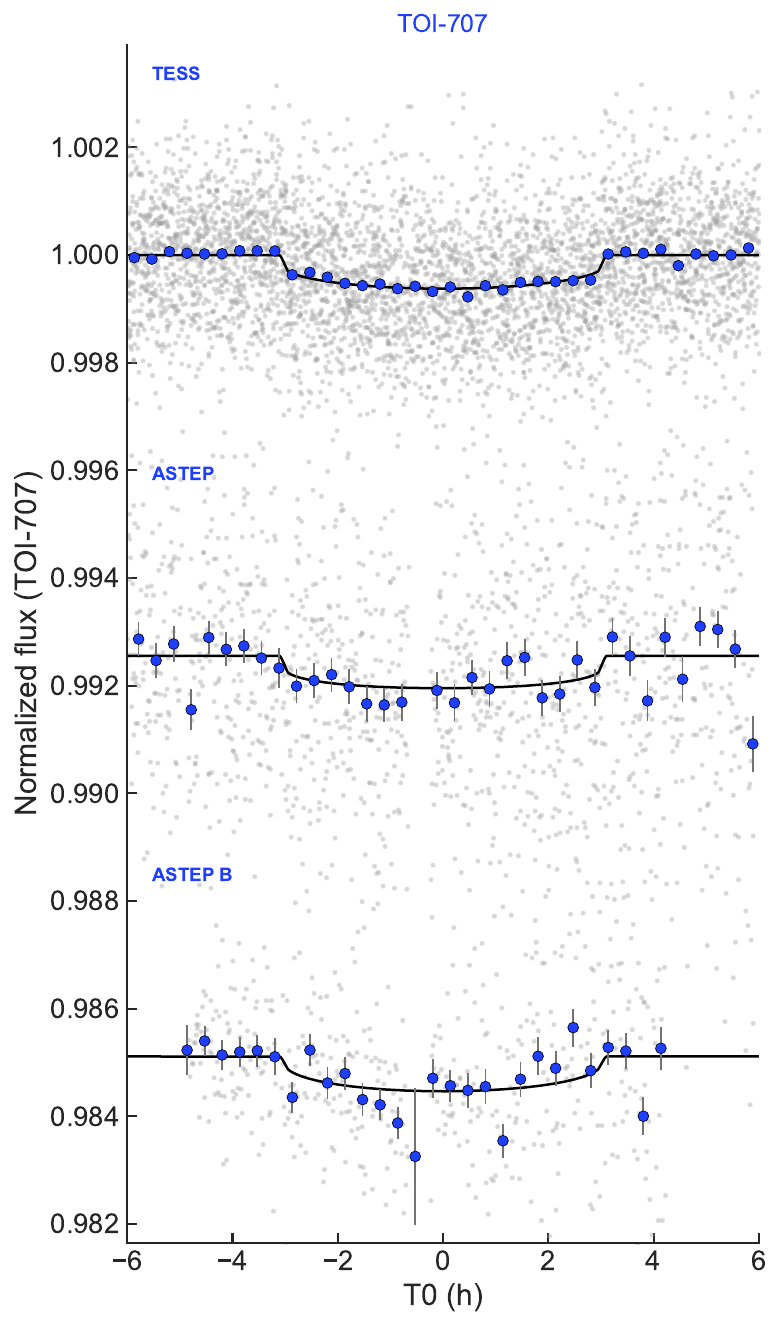}
\caption{Phase-folded lightcurve of TOI-707 from TESS and ASTEP photometry. The best-fit model is generated with \allesfitter{}.}
\label{fig:toi707_phot}
\end{figure}

We validate TOI-707\,b as a sub-Neptune with a radius of 2.4\,$R_\oplus$, an orbital period of 52.80\,d and a semi-major axis of 0.28\,AU. It orbits the coolest and smallest star in our sample and sits at the long-period outskirts of the densely populated short‑period sub‑Neptune region.
Only photometric data from TESS and ASTEP are available in this study for this target (Fig.~\ref{fig:toi707_phot}), leaving the eccentricity and mass unconstrained. TOI-707\,b's placement in the radius-period distribution suggests that it may represent a transitional case between the more commonly detected short-period sub-Neptunes and those on larger orbits, where only few planets have been confirmed. 
Recent studies have highlighted that sub-Neptunes potentially exhibit a wide range of compositions, from volatile-rich mini-Neptunes to high-density super-Earths \citep{tang2024reassessingsubneptunestructureradii}.
Another key factor in sub-Neptune evolution is atmospheric escape. Low-mass sub-Neptunes with small H/He envelopes, $\le$ $0.1\%$, can undergo complete atmospheric loss within 10 Gyr, particularly if they receive strong irradiation \citep{tang2024reassessingsubneptunestructureradii}. Given that TOI-707\,b's host star is a relatively cool G star \edit{($T_\mathrm{eff}$= 5409 K)}, atmospheric escape may have been less efficient compared to planets orbiting hotter stars. If TOI-707\,b still preserves an atmosphere, it would provide evidence that sub-Neptunes can maintain their envelopes even at relatively long orbital periods. The target's equilibrium temperature of $<$500\,K places it in the range where atmospheric retention is still possible but subject to gradual mass loss. Also, models indicate that sub-Neptunes cool differently from terrestrial planets, with core cooling rates being regulated by the overlying envelope. This means that planets like TOI-707\,b could still be 'evolving' with their internal structures shifting over time \citep{tang2024reassessingsubneptunestructureradii}. TOI-707\,b shares key characteristics with other sub-Neptunes identified in recent years. For instance, HD~21520\,b is a slightly warmer sub-Neptune with a shorter orbital period of 25.1\,d, also transiting a bright G-type star \citep{nies2024hd21520bwarm}. With an estimated equilibrium temperature of 640\,K, HD~21520\,b is expected to retain a substantial atmospheric envelope. TOI-707\,b, by comparison, orbits at roughly twice the period and has a lower equilibrium temperature, placing it in a more temperate regime. These conditions suggest that TOI-707\,b is well-positioned to also preserve a significant atmosphere. Another interesting comparison is TOI-1437\,b, a transiting sub-Neptune discovered by TESS and characterised using both the HIRES instrument at Keck Observatory and the Levy Spectrograph on the Automated Planet Finder (APF) telescope \citep{pidhorodetska2024tesskecksurveyxxiisubneptune}. It also orbits a solar-mass star and has a well-constrained radius of $2.24 \pm 0.23\,R_{\oplus}$ and a mass of $9.6 \pm 3.9\,M_{\oplus}$, making it one of the most precisely characterised sub-Neptunes to date. With a similar radius and environment, TOI-707\,b might share a comparable bulk composition. However, in the absence of radial velocity measurements, its mass - and by extension, its density and internal structure - remains to be determined by future RV surveys. A final noteworthy comparison is Kepler-10\,c, a super-Earth/sub-Neptune hybrid with a radius of $2.2\,R_{\oplus}$ and an orbital period of 45\,d \citep{Fressin_2011}. Initially classified as a rocky super-Earth, subsequent observations suggested Kepler-10\,c may possess a higher density and potentially a substantial atmosphere. 
However, Kepler-10\,c orbits a significantly hotter host star, implying that its atmospheric evolution and retention history differ markedly from that of TOI-707\,b. If future radial velocity measurements of TOI-707\,b reveal a similarly elevated mass-to-radius ratio despite its more temperate environment, it could serve as a valuable case study for probing the transition between rocky super-Earths and volatile-rich sub-Neptunes. Future atmospheric observations will be challenging yet essential to determine whether TOI-707\,b retained a volatile-rich envelope or instead is the stripped core of a once-larger planet that underwent substantial atmospheric erosion (see also Section~\ref{app:atmospheric_characterisation_prospects}). \edit{Beyond the validated planet TOI-707\,b, we also note the presence of a further transit-like feature associated with TOI-707.02, which is characterised by a high false positive probability and is therefore not discussed as a validated planet candidate. Such a feature could arise from several astrophysical configurations, including a background eclipsing binary or a planet transiting a secondary star gravitationally bound to the system. Furthermore, the presence of the validated planet TOI-707\,b may influence the false positive probability estimation by introducing correlations in the lightcurve modelling. Given the current data, we cannot discriminate between these scenarios, and additional high-resolution imaging or spectroscopic follow-up would be required to further constrain the nature of TOI-707.02.}

\subsection{TOI-2404 \& TOI-4404 - eclipsing binaries masking as warm planets}
\label{subsec:TOI-2404_and_TOI-4404}

\begin{figure}
\centering
\includegraphics[width=0.98\columnwidth]{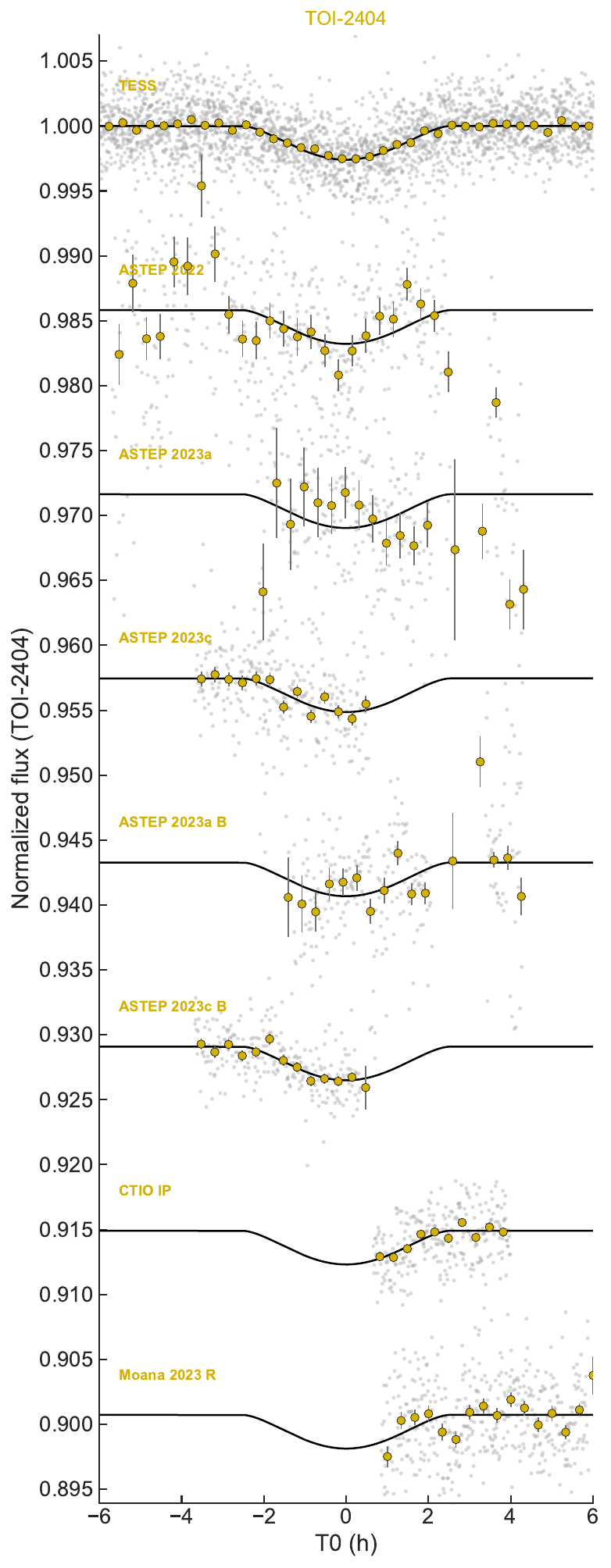}
\caption{Phase-folded lightcurve of TOI-2404 showing the eclipsing binary signal. The best-fit model is generated with \allesfitter{}.}
\label{fig:toi2404_phot}
\end{figure}

\begin{figure}
\centering
\includegraphics[width=\columnwidth]{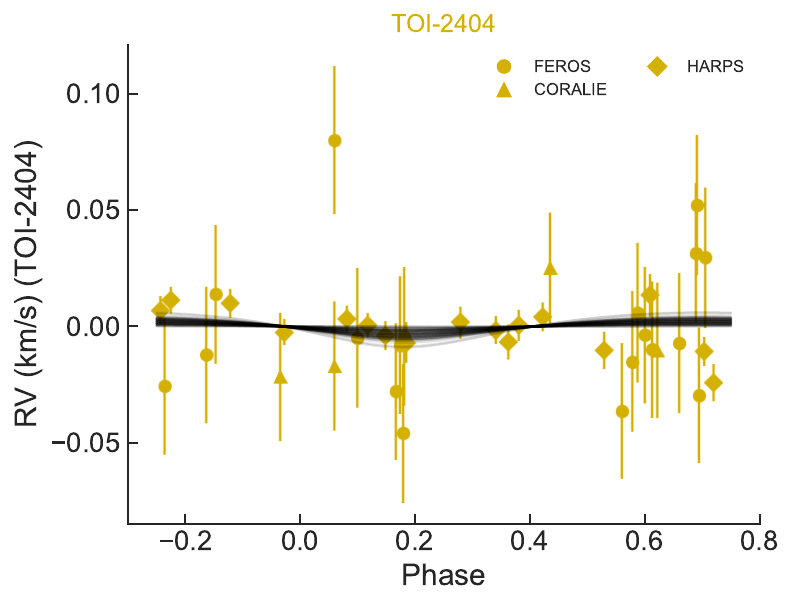}
\caption{Radial velocity measurements for TOI-2404.}
\label{fig:toi2404_rv}
\end{figure}

\begin{figure}
\centering
\includegraphics[width=0.90\columnwidth]{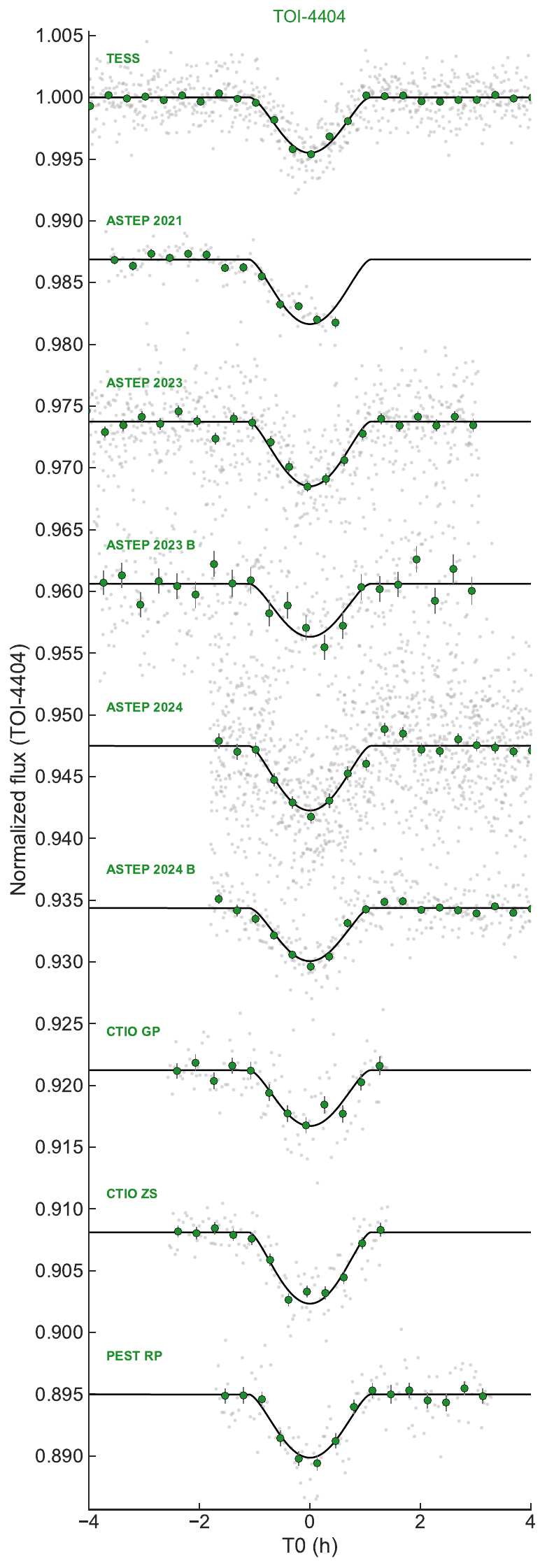}
\caption{Phase-folded lightcurve of TOI-4404 showing the eclipsing binary signal. The best-fit model is generated with \allesfitter{}.}
\label{fig:toi4404_phot}
\end{figure}

The candidate signals TOI-2404.02/03 (74.6\,d) and TOI-4404.01 (39.6\,d) show clear signs of eclipsing binaries. Both phase-folded light curves exhibit pronounced V-shaped profiles (Fig.~\ref{fig:toi2404_phot} and \ref{fig:toi4404_phot}), indicative of grazing stellar eclipses rather than flat-bottomed planetary transits. High-resolution imaging rules out unresolved blends bright enough to mimic the signals. Further, statistical validation with \triceratops{} yields high false positive probabilities for both candidates (Table~\ref{tab:triceratops_results}) and suggests the events occur on target, supporting their classification as multi-star systems.
For TOI-2404.02/03, the presence of two distinct V-shaped events offset from half the orbital phase suggests an eccentric binary orbit, with both primary and secondary eclipses. 
Yet, the spectroscopic data appear to be single-peaked, showing no sign of any stellar multiplicity.
Adding to the puzzle is the likely planetary nature of TOI-2404.01 (20.3\,d), whose orbit would probably not be stable within the binary system itself.
The multiple lines of evidence - including low field crowdedness, likely lack of blending, absence of spectroscopic binarity, and distinct photometric signals - suggest a potential triple-star configuration. 
In such a scenario, the dominant star might host the planet candidate TOI-2404.01 (20.3\,d), while the eclipsing binary TOI-2404.02/03 (74.6\,d) is a faint and thus highly diluted pair, possibly located at a greater distance.
For TOI-2404.02/03, we tried various models to probe these assumptions using \allesfitter{}, deriving constraints on their orbital and physical parameters. 
We cannot constrain the radius ratio nor surface brightness ratio due to these degeneracies, but the apparent transit depth lets us place an upper limit on the dilution of $<0.8$ (defined as dilution = 1 - flux target / flux aperture). 
Thanks to the primary and secondary eclipses, we can precisely identify the orbital parameters, \edit{yielding an eccentric orbit} of $e=0.2$.
No significant RV signal is detected (Fig.~\ref{fig:toi2404_rv}), with a non-detection threshold of $<8$\,m/s. \edit{A GLS analysis of the available RV datasets also does not recover significant peaks at the expected orbital periods of the system.} \edit{We also phase-folded the RV measurements at the 20.3-day period of the inner transiting candidate, but no coherent phase-dependent signal is detected within the current precision and phase coverage.}
In contrast, TOI-4404.01 exhibits clear spectroscopic evidence of binarity via its double-lined cross-correlation function (Section~\ref{sec:stellar_characterisation}). However, the grazing eclipse and absence of a secondary signature leave the radius ratio and orbital configuration highly degenerate. In our {\allesfitter} model, we therefore adopt a strong and simplifying assumption: that the signal arises from the primary eclipse and that the orbit is circular. This is more of a toy model than a physically constrained solution, intended to explore one plausible scenario. Under these conditions and uniformly sampled from photometry alone, the companion radius falls in the range 1.4-4.6\,$R_\mathrm{Jup}$ and the surface brightness ratio to around 1-2\,\%. \edit{However, the similar contrasts of the two components in the CCF are not consistent with a strongly unequal-brightness system. Statistical validation with \triceratops{} instead favours eclipsing binary scenarios, with single and double eclipsing binary configurations accounting for the majority of the probability (Table~\ref{tab:triceratops_results}), including solutions with either a lower-mass primary and faint companion or nearly equal-brightness stellar components. We also explored an eccentric grazing eclipsing binary configuration, which can naturally explain the absence of a secondary eclipse. This scenario is moderately favoured by the data ($\Delta \ln Z \approx +4$), although not at a level sufficient to uniquely distinguish it from other possible configurations.
Overall, the available photometric and spectroscopic constraints do not allow us to unambiguously determine the system architecture. Selected radial velocity measurements or a more detailed spectroscopic analysis could help further constrain the system.}
Together, these observational and statistical results support the classification of TOI-2404 and TOI-4404 as multi-star systems, whose eclipsing binaries initially mimicked the signals of warm giant planets.

\section{Conclusions}
\label{sec:conclusions}

This study highlights the crucial role of near-polar, ground-based facilities like ASTEP \edit{\citep{Guillot2015,Crouzet2020}} in the follow-up of long-period transiting TOIs, demonstrating its ability to refine planetary parameters in ways complementary to space-based missions \edit{such as TESS \citep{Ricker2015}}. The five candidates analysed in this work are TOI-4507.01, TOI-4404.01, TOI-2404.02\,/\,.03, TOI-3457.01 and TOI-707.01, all with orbital periods longer than one month, a regime where confirming transit events becomes increasingly challenging.\footnote{Two of these systems also host potential inner planets, TOI-2404.01 and TOI-707.02. Although our campaign focused on periods longer than one month, we assess their preliminary statistical validation using TESS photometry and direct imaging in Section~\ref{sec:validation}.} \edit{ASTEP’s nearly uninterrupted monitoring during the Antarctic dark season offers a key advantage for long-period exoplanets. Darkness sets in from March and becomes continuous from May to mid-August, enabling extended, nearly uninterrupted coverage of transits.} Its continuous observations provide crucial data to track these rare events with high precision. Such coverage is unmatched by other ground-based observatories at lower latitudes, highlighting the unique scientific value of Antarctic facilities for exoplanet research. While three candidates were statistically validated as planets in an underexplored regime (warm giants TOI-4507\,b and TOI-3457\,b, warm sub-Neptune TOI-707\,b), the other two \edit{were found to be most consistent with} eclipsing binaries (\edit{TOI-2404 and TOI-4404}) based on our data and analyses. By probing this underexplored region of parameter space, our study adds new empirical constraints on the occurrence and properties of warm giants and sub-Neptunes, which are critical for testing theories of planet formation and migration \edit{\citep[]{Dawson2018,Fortney2021}}. In particular, TOI-4507\,b and TOI-3457\,b enlarge the scarce sample of warm giants with multi-month orbits, likely formed beyond the snow line and migrated inward at early times \edit{\citep[]{Pollack1996,Lin1996}}. Their cooler environments and weaker tidal forces make them valuable probes of whether migration was governed by smooth disk processes or by high-eccentricity dynamical pathways. TOI-707\,b, instead, belongs to the poorly explored population of warm sub-Neptunes, whose internal structure and volatile content remain debated \edit{\citep[]{Raymond2022}}, and whose long period and moderate irradiation make it a rare opportunity to test whether such planets retain extended H/He envelopes or instead evolve into water-rich super-Earths \edit{\citep[]{Zeng2019}}. By extending the frontier of long-period transiting planets and offering benchmarks for theories of formation and evolution, these results underscore ASTEP’s contribution in refining planetary parameters while validating these precious needles in the haystack of false positives that are otherwise challenging to follow-up.

\section*{Data availability}
The photometric and radial velocity data used in this work,
together with the stellar parameters, companion parameters,
additional model parameters and priors adopted in the analysis,
are available in electronic form at the CDS via anonymous ftp to
cdsarc.cds.unistra.fr or via
\url{https://cdsarc.cds.unistra.fr/}.

\begin{acknowledgements}
We thank the anonymous referee for the thorough review and helpful comments, which further improved the quality of this study.
This paper includes public data collected by the \tess{} mission, which are publicly available from the Mikulski Archive for Space Telescopes (MAST). Funding for the \tess\ mission is provided by NASA's Science Mission Directorate. We acknowledge the use of public \tess\ data from pipelines at the \tess\ Science Office and at the \tess\ Science Processing Operations Center. Resources supporting this work were provided by the NASA High-End Computing (HEC) Program through the NASA Advanced Supercomputing (NAS) Division at Ames Research Center for the production of the SPOC data products. 
This work makes use of observations from the ASTEP telescope. ASTEP benefited from the support of the French and Italian polar agencies IPEV and PNRA in the framework of the Concordia station program, and from OCA, INSU, and Idex UCAJEDI (ANR-15-IDEX-01). We further acknowledge support from ESA through the Science Faculty of the European Space Research and Technology Centre (ESTEC). This research also received funding from the European Research Council (ERC) under the European Union's Horizon 2020 research and innovation programme (grant agreement No. 803193/BEBOP), from the Science and Technology Facilities Council (STFC; grant Nos. ST/S00193X/1, ST/W002582/1, and ST/Y001710/1), and from the ERC/UKRI Frontier Research Guarantee programme (CandY/EP/Z000327/1).
This work makes use of observations from the LCOGT network. Part of the LCOGT telescope time was granted by NOIRLab through the Mid-Scale Innovations Program (MSIP). MSIP is funded by the National Science Foundation (NSF).
This research has used data from the CTIO/SMARTS 1.5m telescope, which is operated
as part of the SMARTS Consortium by RECONS (www.recons.org) members Todd Henry, Wei-Chun Jao, Sebastian Carrazco-Gaxiola, Hodari-Sadiki Hubbard-James, Tim Johns, and Leonardo Paredes.
This work makes use of observations obtained with the Perth Exoplanet Survey Telescope (PEST), operated by T.G. Tan in Perth, Australia.
We acknowledge the use of FEROS at the MPG/ESO 2.2\,m telescope at La Silla Observatory, Chile. Based on observations collected under ESO programmes 0103.A-9008(A), 0104.A-9007(A), 0110.A-9011(A), 0108.A-9003(A), 0109.A-9003(A), 0111.A-9011(A), 112.265K.001. 
We acknowledge the use of HARPS at the ESO 3.6\,m telescope at La Silla Observatory, Chile. Based on observations collected under ESO programmes 0104.C-0413(A), 106.21ER.001, 1102.C-0923(A), 112.25W1.001, 114.27CS.001, 115.286G.001, 
112.261U.001, 112.261U.003, 108.22A8.001, 109.239V.001, 110.23YQ.001.
Based on observations obtained at the Southern Astrophysical Research (SOAR) telescope, which is a joint project of the Ministério da Ciência, Tecnologia e Inovações do Brasil (MCTI/LNA), the US National Science Foundation’s NOIRLab, the University of North Carolina at Chapel Hill (UNC), and Michigan State University (MSU).
Based on observations obtained at the international Gemini Observatory, a program of NSF NOIRLab, which is managed by the Association of Universities for Research in Astronomy (AURA) under a cooperative agreement with the U.S. National Science Foundation on behalf of the Gemini Observatory partnership: the U.S. National Science Foundation (United States), National Research Council (Canada), Agencia Nacional de Investigaci\'{o}n y Desarrollo (Chile), Ministerio de Ciencia, Tecnolog\'{i}a e Innovaci\'{o}n (Argentina), Minist\'{e}rio da Ci\^{e}ncia, Tecnologia, Inova\c{c}\~{o}es e Comunica\c{c}\~{o}es (Brazil), and Korea Astronomy and Space Science Institute (Republic of Korea).
This research has made use of the NASA Exoplanet Archive \citep{Akeson_2013,christiansen2025nasaexoplanetarchiveexoplanet} and the Exoplanet Follow-up Observation Program (ExoFOP; DOI: 10.26134/ExoFOP5) website, which are operated by the California Institute of Technology, under contract with the National Aeronautics and Space Administration under the Exoplanet Exploration Program.
This study includes data from targets proposed for observation by the TESS Asteroseismic Science Operations Center (TASOC) team through TESS Guest Investigator (GI) programs in Cycle 1: G011160 (PI Daniel Huber), G011155 (PI Daniel Huber), and G011188 (PI Marc Pinsonneault).
A.J. acknowledges support from Fondecyt project 1251439.
Foundation under grant number PCEFP2\_194576. The contribution of ML has been carried out within the framework of the NCCR PlanetS supported by the Swiss National Science Foundation under grants 51NF40\_182901 and 51NF40\_205606.
T.T. acknowledges support by the BNSF program "VIHREN-2021" project No.KP-06-DV/5.
Funding for KB was provided by the European Union (ERC AdG SUBSTELLAR, GA 101054354). 
KAC acknowledges support from the TESS mission via subaward s3449 from MIT. 
We acknowledge financial support from the Agencia Estatal de Investigaci\'on of the Ministerio de Ciencia e Innovaci\'on MCIN/AEI/10.13039/501100011033 and the ERDF “A way of making Europe” through project PID2021-125627OB-C32, and from the Centre of Excellence “Severo Ochoa” award to the Instituto de Astrofisica de Canarias.
This work has made use of the following software packages: 
\allesfitter{} \citep{allesfitter-code,allesfitter-paper},
\texttt{AstroImageJ} \citep{Collins:2017}, 
\texttt{nep-des} (\url{https://github.com/castro-gzlz/nep-des}),
\texttt{TAPIR} \citep{Jensen:2013}, and
\triceratops{} \citep{Giacalone_Dressing_2020}.
\end{acknowledgements}

\bibliographystyle{aa}
\bibliography{ref.bib}

\clearpage
\newpage

\begin{appendix}

\counterwithout{figure}{section}
\counterwithout{table}{section}

\renewcommand{\thefigure}{A.\arabic{figure}}
\setcounter{figure}{0}

\renewcommand{\thetable}{A.\arabic{table}}
\setcounter{table}{0}

\section{TESS target pixel file figures}

Given TESS's relatively large plate scale (21$\arcsec$ per pixel), the photometric aperture used to extract lightcurves often includes multiple Gaia sources, especially in dense stellar fields. To further assess potential contamination and verify the location of the transit source, we analysed the TESS target pixel files using the \texttt{tpfplotter} tool \citep{aller2020planetary}. 

\begin{figure*}
\centering
\begin{subfigure}[b]{0.30\textwidth}
  \centering
  \includegraphics[width=\linewidth]{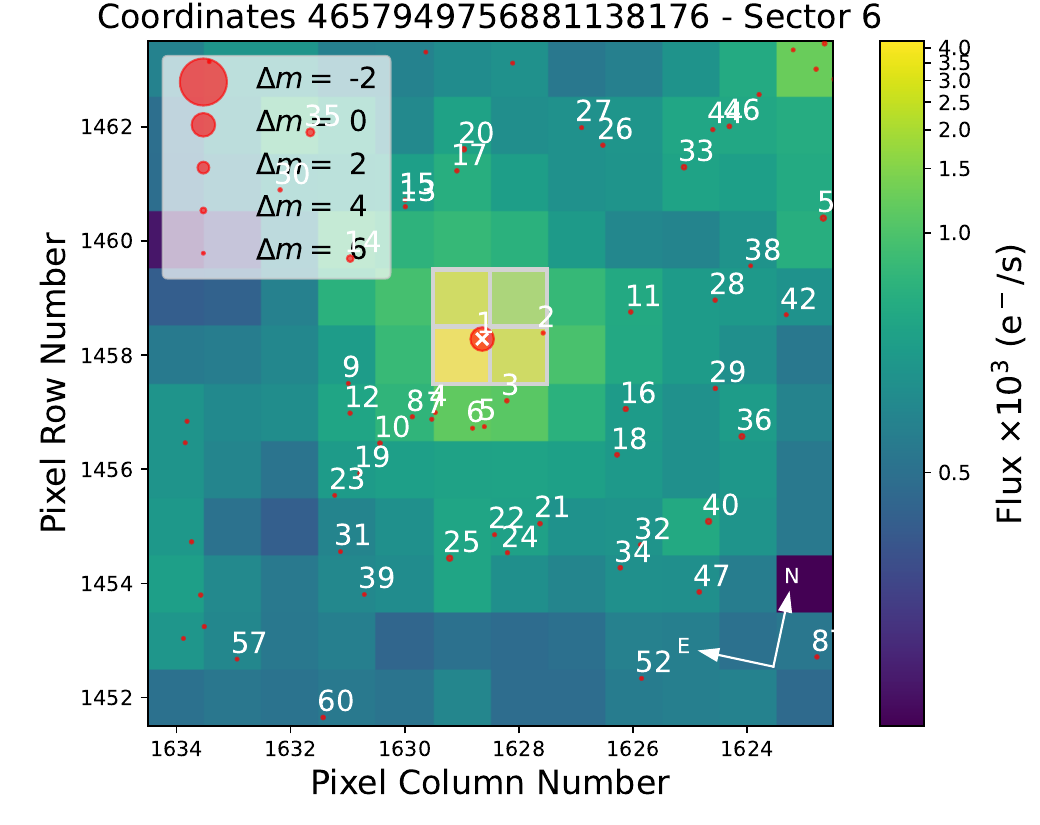}
  \caption{TOI-4507}
\end{subfigure}
\hspace{0.01\textwidth}
\begin{subfigure}[b]{0.30\textwidth}
  \centering
  \includegraphics[width=\linewidth]{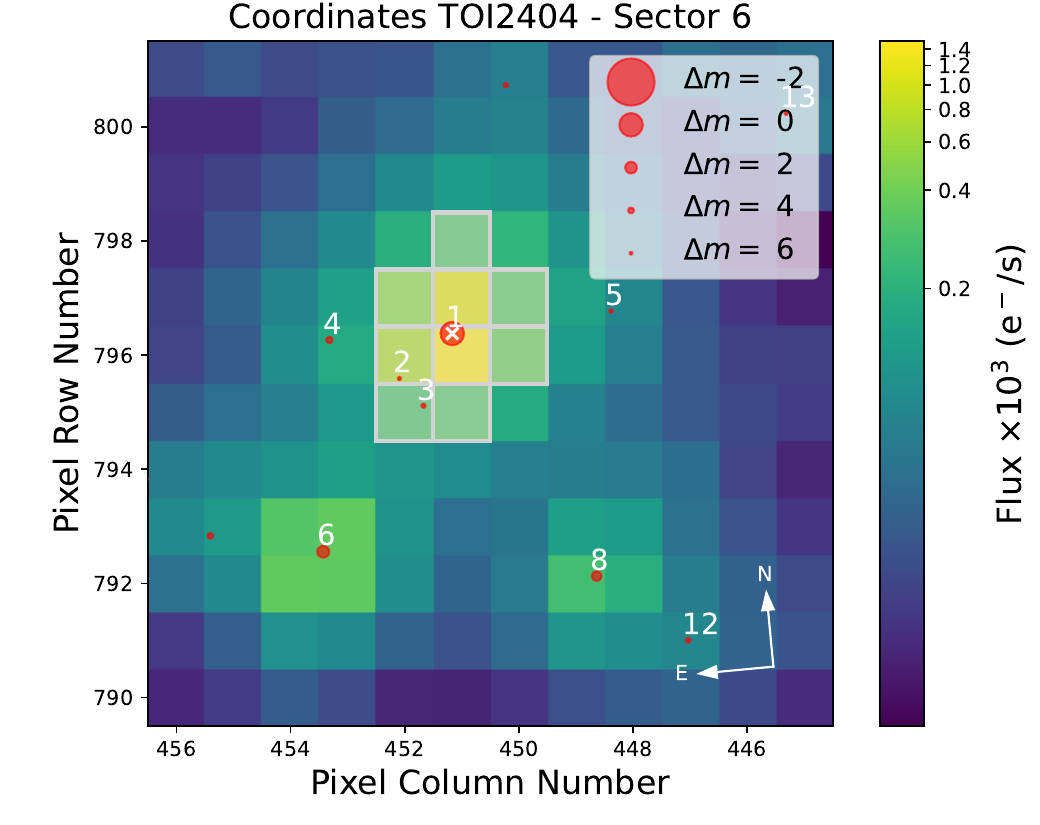}
  \caption{TOI-2404}
\end{subfigure}
\hspace{0.01\textwidth}
\begin{subfigure}[b]{0.30\textwidth}
  \centering
  \includegraphics[width=\linewidth]{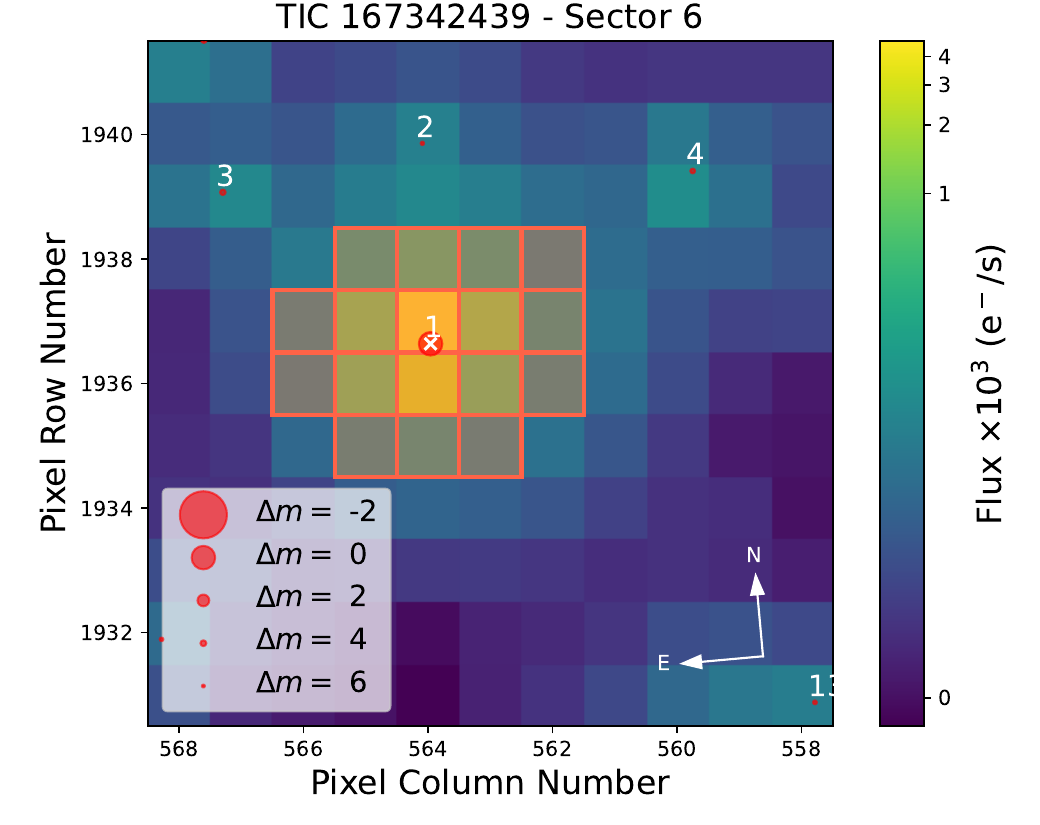}
  \caption{TOI-707}
\end{subfigure}
\hspace{0.01\textwidth}
\begin{subfigure}[b]{0.30\textwidth}
  \centering
  \includegraphics[width=\linewidth]{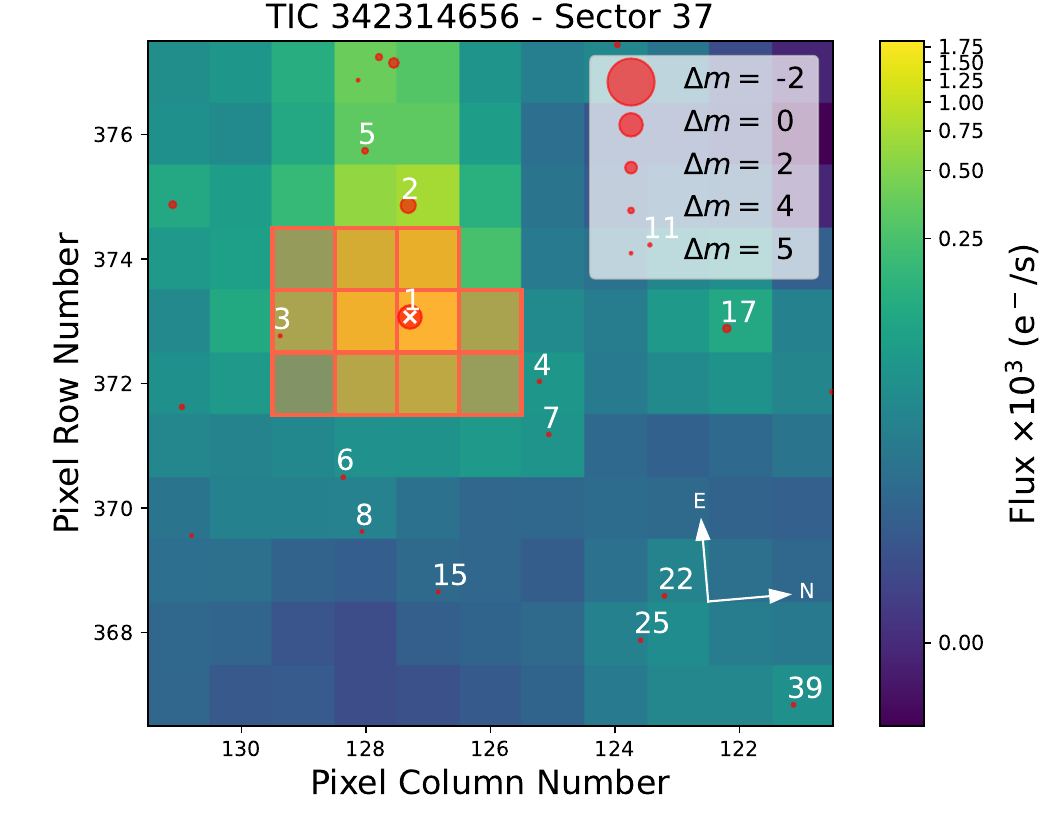}
  \caption{TOI-4404}
\end{subfigure}
\hspace{0.01\textwidth}
\begin{subfigure}[b]{0.30\textwidth}
  \centering
  \includegraphics[width=\linewidth]{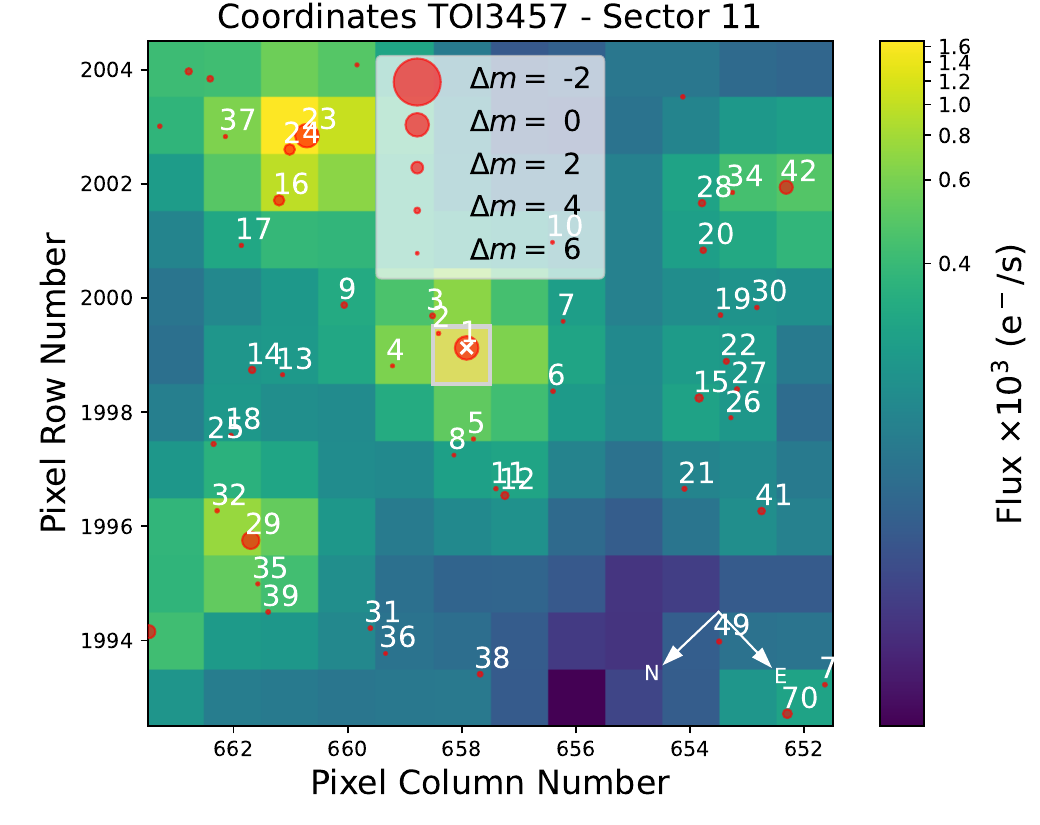}
  \caption{TOI-3457}
\end{subfigure}

\vspace{0.5em}
\caption{TESS target pixel images for the five targets using \texttt{tpfplotter} \citep{aller2020planetary}. Gaia DR2 sources within and around the TESS photometric aperture are overplotted as red dots, with nearby stars annotated by their magnitude differences. The photometric aperture is minimally contaminated by nearby sources, supporting the interpretation that the transit signals originate from the intended host stars. Panels: (a) TOI-4507, (b) TOI-4404, (c) TOI-2404, (d) TOI-3457, and (e) TOI-707.}
\label{fig:tess_pix}
\end{figure*}

\texttt{tpfplotter} overlays Gaia DR2 sources onto the calibrated TESS pixel images, providing a visual representation of the stellar environment. This allows for direct identification of any nearby sources falling within or near the TESS aperture. In addition, the tool marks the optimal photometric aperture used by the TESS pipeline and annotates the magnitude difference ($\Delta m$) of nearby stars relative to the target.
For all five targets, the analyses shown in Fig.~\ref{fig:tess_pix} confirm that the TESS photometric aperture is minimally affected by nearby sources. This analysis is essential to ensure that the observed transits are indeed associated with the intended host stars and not with nearby eclipsing binaries or background objects.

\section{Radial velocity measurements tables}
As explained in Section~\ref{subsec:Radial_velocity_observations}, ample RV data obtained from a combination of spectrographs and observing campaigns supports our analysis. Tables~\ref{tab:rv_measurements_toi-4507}, \ref{tab:rv_measurements_toi-2404}, and \ref{tab:rv_measurements_toi-3457} list our RV measurements for the respective targets, including the Barycentric Julian Date (BJD), measured RVs with uncertainties, bisector inverse spans (BIS) with uncertainties, and instrument identifiers.

\begin{table*}
\scriptsize
\centering
\setlength{\tabcolsep}{22pt} 
\caption{\edit{Radial velocity measurements of TOI-4507.}}

\begin{tabular}{lcccccc}

\toprule
\textbf{BJD} & \textbf{RV (m/s)} & \textbf{RV err (m/s)} & \textbf{BIS (m/s)} & \textbf{BIS err (m/s)} & \textbf{Instrument} \\
\midrule

2458717.9321 & 26094.5 & 11.0 & 33.0 & 10.0 & FEROS \\
2458718.8983 & 26072.7 & 9.0 & 33.0 & 9.0 & FEROS \\
2458722.8798 & 26053.6 & 9.3 & 39.0 & 9.0 & FEROS \\
2458724.9013 & 26075.4 & 11.3 & 54.0 & 11.0 & FEROS \\
2458725.8661 & 26090.3 & 11.4 & 21.0 & 11.0 & FEROS \\
2458800.7123 & 26071.4 & 10.3 & 37.0 & 10.0 & FEROS \\
2458804.6729 & 26031.6 & 9.5 & 25.0 & 9.0 & FEROS \\
2458810.8081 & 26026.6 & 9.5 & 25.0 & 9.0 & FEROS \\
2460263.7975 & 26009.3 & 12.4 & 17.0 & 11.0 & FEROS \\
2460266.7953 & 25987.4 & 8.1 & 20.0 & 8.0 & FEROS \\
2460268.8010 & 25977.5 & 9.0 & 20.0 & 9.0 & FEROS \\[0.1cm]

2458766.8297 & 26045.8 & 2.6 & 45.0 & 2.0 & HARPS \\
2458773.8446 & 26063.5 & 3.9 & 25.0 & 3.0 & HARPS \\
2458777.8408 & 26057.2 & 7.6 & 40.0 & 7.0 & HARPS \\
2458781.8231 & 26056.9 & 3.4 & 26.0 & 3.0 & HARPS \\
2458803.6903 & 26055.6 & 2.0 & 30.0 & 2.0 & HARPS \\
2458805.7769 & 26071.7 & 3.6 & 17.0 & 3.0 & HARPS \\
2458810.7656 & 26045.2 & 4.8 & 30.0 & 4.0 & HARPS \\
2458832.7768 & 26054.4 & 4.8 & 32.0 & 4.0 & HARPS \\
2458852.6945 & 26048.8 & 2.0 & 43.0 & 2.0 & HARPS \\
2458871.6149 & 26071.8 & 6.3 & 30.0 & 6.0 & HARPS \\
2458881.5876 & 26062.2 & 2.5 & 25.0 & 2.0 & HARPS \\
2458894.6413 & 26053.4 & 3.7 & 40.0 & 3.0 & HARPS \\
2459180.6397 & 26069.7 & 3.3 & 18.0 & 3.0 & HARPS \\
2459183.6347 & 26055.3 & 3.3 & 27.0 & 3.0 & HARPS \\
2459185.6363 & 26038.9 & 4.0 & 37.0 & 4.0 & HARPS \\
2459205.5873 & 26055.4 & 3.4 & 18.0 & 3.0 & HARPS \\
2459213.6455 & 26048.6 & 2.0 & 15.0 & 2.0 & HARPS \\
2459228.6015 & 26030.0 & 5.3 & 17.0 & 5.0 & HARPS \\
2459238.5614 & 26044.8 & 3.8 & 27.0 & 3.0 & HARPS \\
2459244.6436 & 26057.4 & 3.2 & 33.0 & 3.0 & HARPS \\
2459246.5703 & 26040.8 & 4.5 & 51.0 & 4.0 & HARPS \\
2459281.5894 & 26042.0 & 4.5 & 25.0 & 4.0 & HARPS \\
2459295.6066 & 26038.5 & 6.4 & 20.0 & 6.0 & HARPS \\
2460226.8824 & 26053.9 & 7.5 & 32.0 & 7.0 & HARPS \\
2460227.8712 & 26042.1 & 5.3 & 40.0 & 5.0 & HARPS \\
2460256.7031 & 26030.8 & 4.2 & 40.0 & 4.0 & HARPS \\

\bottomrule
\end{tabular}
\tablefoot{FEROS program IDs: 0103.A-9008(A), 0104.A-9007(A), 0110.A-9011; HARPS program IDs: 0104.C-0413(A), 106.21ER.001, 1102.C-0923(A), 112.25W1.001, 114.27CS.001, 115.286G.001.}

\label{tab:rv_measurements_toi-4507}
\end{table*}

\begin{table*}
\scriptsize
\centering
\setlength{\tabcolsep}{22pt} 
\caption{\edit{Radial velocity measurements of TOI-2404.}}

\begin{tabular}{lcccccc}

\toprule
\textbf{BJD} & \textbf{RV (m/s)} & \textbf{RV err (m/s)} & \textbf{BIS (m/s)} & \textbf{BIS err (m/s)} & \textbf{Instrument} \\
\midrule

2459505.8140038 & 35748.9 & 10.6   & 37.0   & 12.0   & FEROS \\
2459506.8494581 & 35726.4 & 11.6   & -40.0  & 12.0   & FEROS \\
2459517.8287537 & 35710.6 & 10.2   & -2.0   & 11.0   & FEROS \\
2459541.7908497 & 35688.8 & 9.9    & -14.0  & 11.0   & FEROS \\
2459646.6023768 & 35681.8 & 11.7   & 123.0  & 13.0   & FEROS \\
2459652.6523509 & 35689.7 & 10.4   & 11.0   & 11.0   & FEROS \\
2459682.4686043 & 35776.6 & 14.9   & 9.0    & 14.0   & FEROS \\
2459685.4918301 & 35692.0 & 10.4   & 30.0   & 11.0   & FEROS \\
2459691.5317308 & 35692.6 & 10.3   & 15.0   & 11.0   & FEROS \\
2459943.7332785 & 35660.5 & 8.3    & 13.0   & 10.0   & FEROS \\
2459945.7487735 & 35702.7 & 11.3   & -47.0  & 12.0   & FEROS \\
2459946.7221484 & 35693.2 & 8.5    & 8.0    & 10.0   & FEROS \\
2459947.6404759 & 35686.9 & 8.3    & 30.0   & 10.0   & FEROS \\
2459953.7195185 & 35667.3 & 8.5    & 15.0   & 10.0   & FEROS \\
2460033.5411028 & 35671.2 & 8.5    & 100.0  & 10.0   & FEROS \\
2460063.5319138 & 35669.0 & 9.2    & 31.0   & 11.0   & FEROS \\
2460064.5210173 & 35651.0 & 10.7   & 38.0   & 12.0   & FEROS \\
2460102.5327807 & 35728.5 & 10.0   & -45.0  & 11.0   & FEROS \\
2460262.7823104 & 35684.6 & 8.3    & 46.0   & 10.0   & FEROS \\[0.1cm]

2459675.52926672017 & 35672.829 & 27.543   & 1.576   & 38.951   & Coralie \\
2459682.48607122991 & 35677.319 & 27.726   & -24.059 & 39.211   & Coralie \\
2459710.50937693985 & 35719.551 & 23.726   & 98.147  & 33.554   & Coralie \\
2459724.50117135979 & 35684.185 & 29.203   & -37.653 & 41.299   & Coralie \\[0.1cm]

2460231.82314332016 & 35597.967 & 5.433   & -16.177  & 10.865   & HARPS \\
2460239.77621026989 & 35583.600 & 7.433   & -14.778  & 14.866   & HARPS \\
2460252.78869066993 & 35583.214 & 5.209   & -8.587   & 10.418   & HARPS \\
2460256.79602485988 & 35600.782 & 5.084   & -11.882  & 10.169   & HARPS \\
2460288.69165703980 & 35586.916 & 8.017   & -12.549  & 16.035   & HARPS \\
2460295.73969470989 & 35595.538 & 6.244   & -8.301   & 12.488   & HARPS \\
2460355.62906228984 & 35597.176 & 4.940   & 10.529   & 9.880    & HARPS \\
2460376.62453335989 & 35587.318 & 6.993   & 28.842   & 13.986   & HARPS \\
2460598.74740936980 & 35592.299 & 5.223   & -33.011  & 10.446   & HARPS \\
2460601.73171277996 & 35594.251 & 5.998   & -27.033  & 11.995   & HARPS \\
2460618.76715038018 & 35607.170 & 8.466   & -12.656  & 16.932   & HARPS \\
2460638.81724887993 & 35603.705 & 5.523   & -43.668  & 11.047   & HARPS \\
2460656.70732444013 & 35594.380 & 4.427   & -13.732  & 8.853    & HARPS \\
2460701.61303166021 & 35569.753 & 7.450   & -9.406   & 14.901   & HARPS \\
2460705.69758271985 & 35605.165 & 5.066   & -19.982  & 10.132   & HARPS \\
2460720.56952964980 & 35591.327 & 4.596   & -10.045  & 9.193    & HARPS \\
2460733.60477608023 & 35589.840 & 5.366   & -0.868   & 10.732   & HARPS \\

\bottomrule
\end{tabular}

\tablefoot{FEROS program IDs: 0108.A-9003(A), 0109.A-9003(A), 0110.A-9011(A), 0111.A-9011(A), 112.265K.001; HARPS program IDs: 112.25W1.001, 112.261U.001, 112.261U.003.}
\label{tab:rv_measurements_toi-2404}
\end{table*}

\begin{table*}
\scriptsize
\centering
\setlength{\tabcolsep}{22pt} 
\caption{\edit{Radial velocity measurements of TOI-3457.}}
\begin{tabular}{lcccccc}

\toprule
\textbf{BJD} & \textbf{RV (m/s)} & \textbf{RV err (m/s)} & \textbf{BIS (m/s)} & \textbf{BIS err (m/s)} & \textbf{Instrument} \\
\midrule

2459592.8313 & 10142.1 & 14.9 & -27.0 & 13.0 & FEROS \\
2459649.8014 & 10111.2 & 16.5 & -21.0 & 15.0 & FEROS \\
2459643.7859 & 10048.9 & 15.9 & 124.0 & 14.0 & FEROS \\
2459702.7092 & 10021.4 & 17.8 & 1.0 & 15.0 & FEROS \\
2459689.6489 & 10114.5 & 16.8 & -22.0 & 15.0 & FEROS \\
2459687.7445 & 10046.9 & 13.9 & 80.0 & 13.0 & FEROS \\
2459683.6790 & 10108.2 & 22.1 & -52.0 & 18.0 & FEROS \\

\bottomrule
\end{tabular}
\tablefoot{FEROS program IDs: 0108.A-9003(A), 0109.A-9003(A).}
\label{tab:rv_measurements_toi-3457}
\end{table*}

\section{High-resolution imaging figures}
\begin{figure*}
\centering
\begin{subfigure}[b]{0.3\textwidth}
  \centering
  \includegraphics[width=\linewidth]{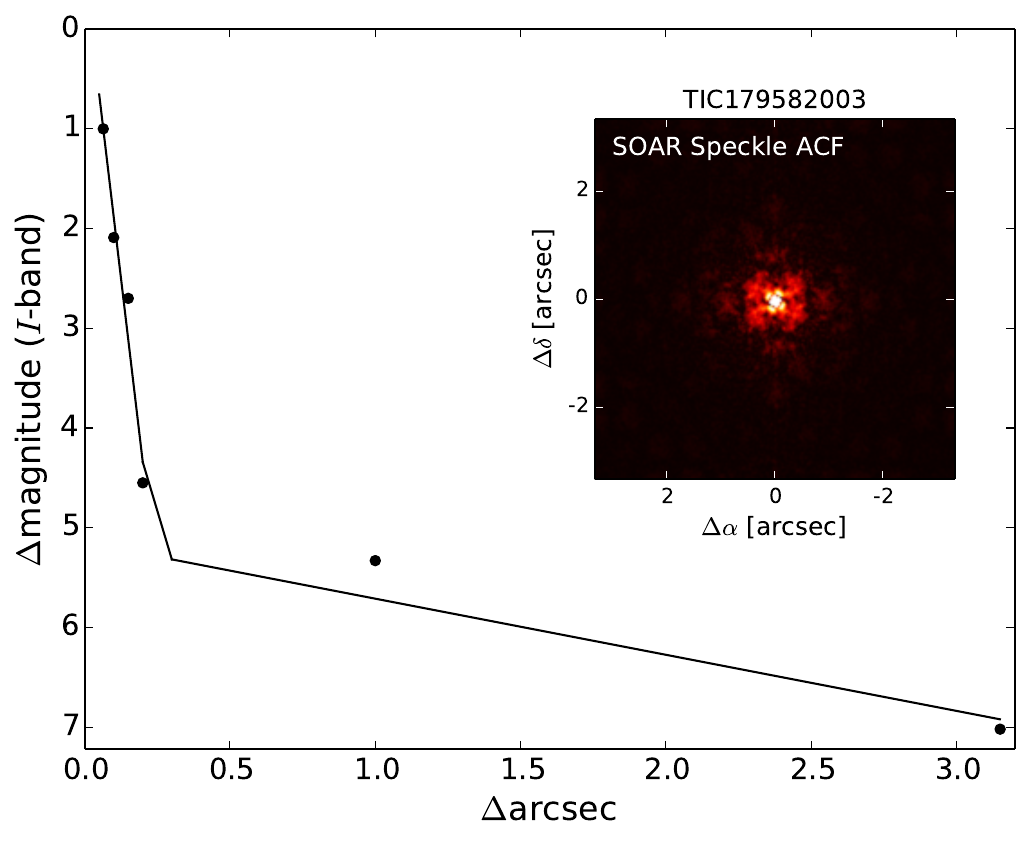}
  \caption{TOI-4507}
\end{subfigure}
\hspace{0.01\textwidth}
\begin{subfigure}[b]{0.3\textwidth}
  \centering
  \includegraphics[width=\linewidth]{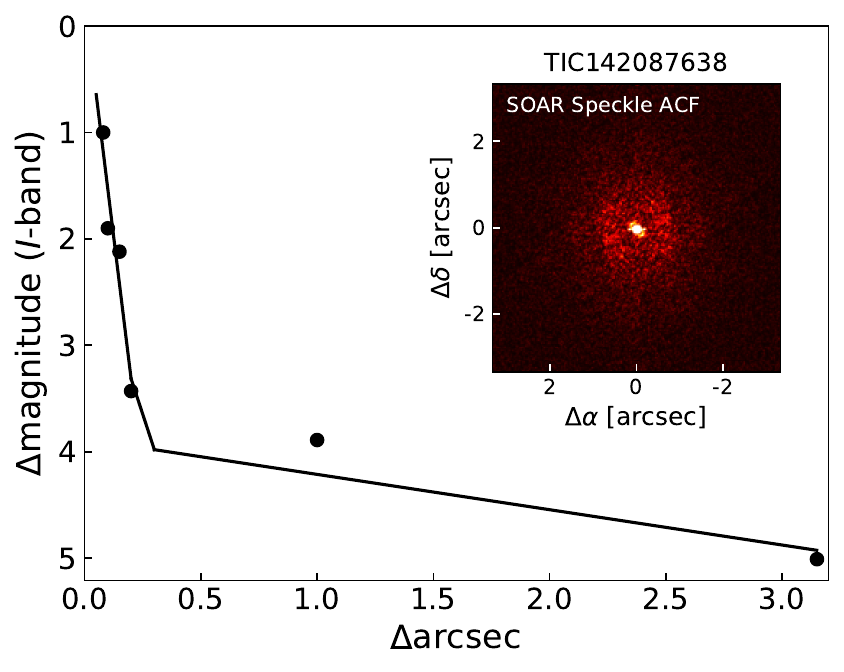}
  \caption{TOI-2404}
\end{subfigure}
\hspace{0.01\textwidth}
\begin{subfigure}[b]{0.3\textwidth}
  \centering
  \includegraphics[width=\linewidth]{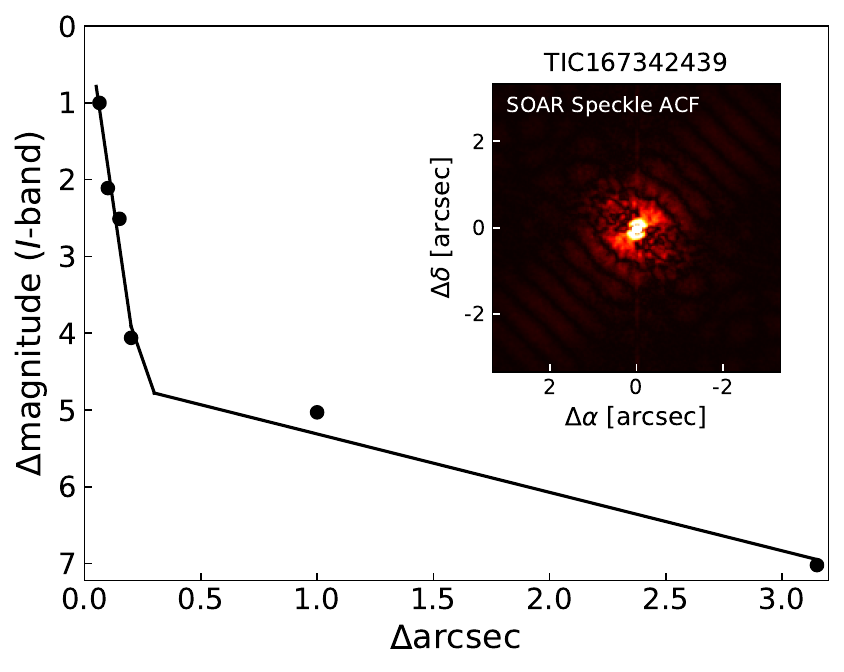}
  \caption{TOI-707}
\end{subfigure}
\hspace{0.01\textwidth}
\begin{subfigure}[b]{0.3\textwidth}
  \centering
  \includegraphics[width=\linewidth]{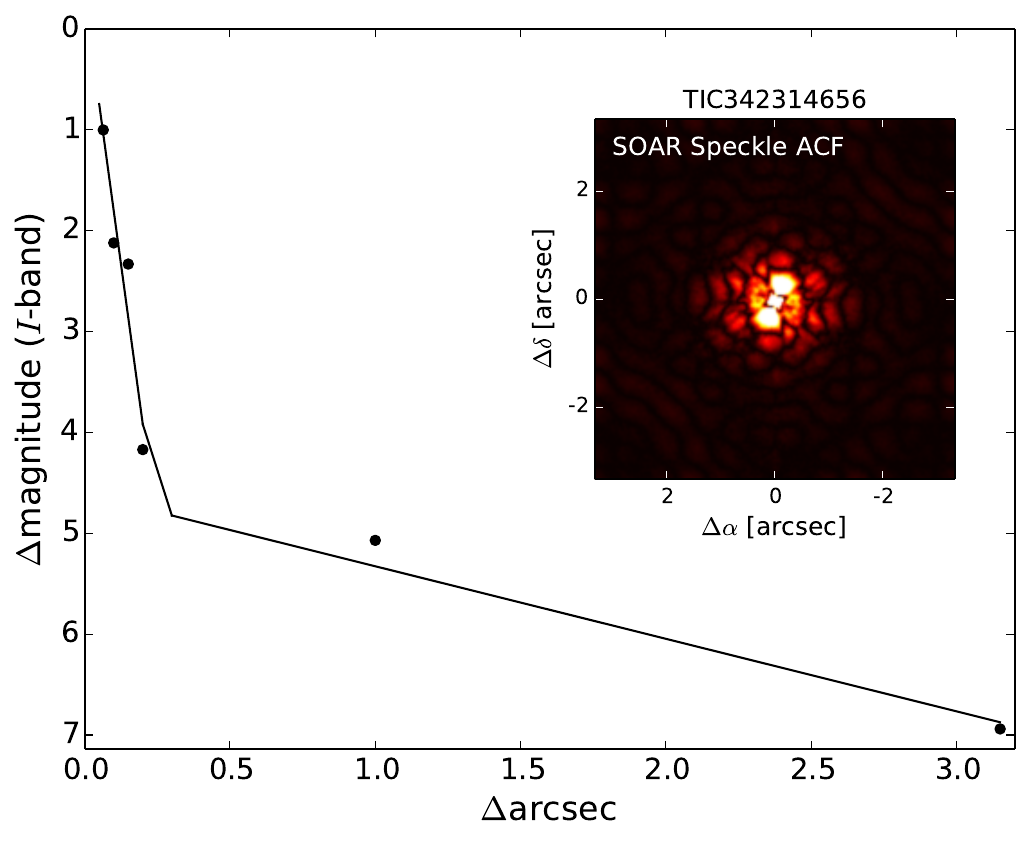}
  \caption{TOI-4404}
\end{subfigure}
\hspace{0.01\textwidth}
\begin{subfigure}[b]{0.3\textwidth}
  \centering
  \includegraphics[width=\linewidth]{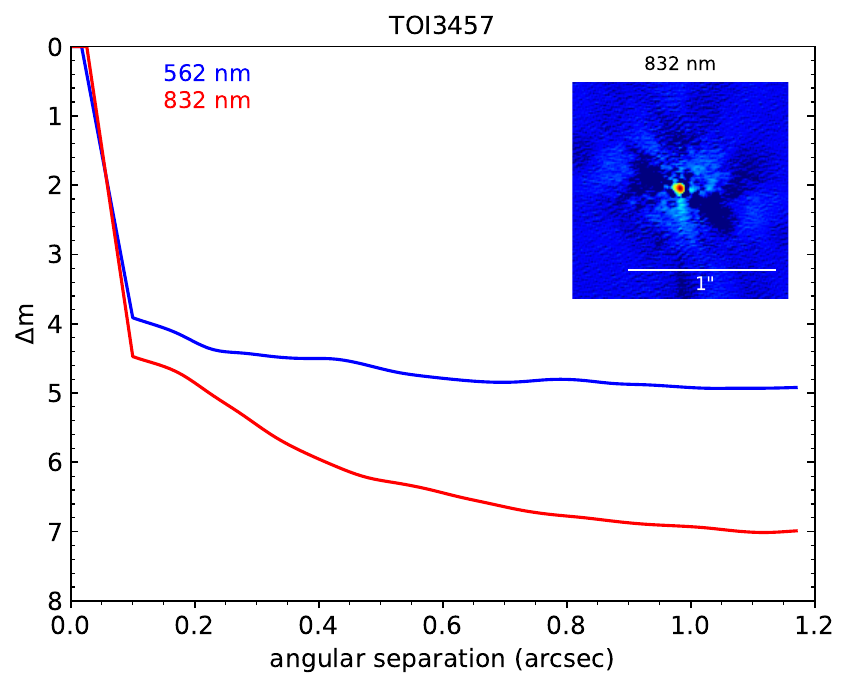}
  \caption{TOI-3457}
\end{subfigure}

\vspace{0.5em}
\caption{High-resolution speckle imaging of the five targets obtained with SOAR/HRCam in the I-band (879 nm) for TOI-4507, TOI-2404, TOI-707, and TOI-4404, and with Gemini/Zorro in the 562 nm (blue) and 832 nm (red) bands for TOI-3457. The resulting contrast curves and reconstructed images confirm the absence of nearby stellar companions within separations of 0.2-3.0\arcsec{} and down to magnitude differences of $\approx5$\,mag, supporting the planetary origin of the observed transit signals. 
}
\label{fig:imaging}
\end{figure*}

Section \ref{subsec:High-resolution_imaging} describes the details of the high-resolution imaging campaign, summarised in Table~\ref{tab:imaging} and Fig.~\ref{fig:imaging}. Here, we provide the full contrast curves and reconstructed images for each target. Observations were performed with SOAR/HRCam in the I-band (879 nm) and Gemini/Zorro in the 562 and 832 nm bands for TOI-3457. These data confirm the absence of nearby stellar companions within separations of 0.2-3.0\arcsec{} and magnitude differences up to $\approx5-7$\,mag.

\section{SED analysis figures}
\begin{figure*}
\centering
\begin{subfigure}[b]{0.3\textwidth}
  \centering
  \includegraphics[width=\linewidth]{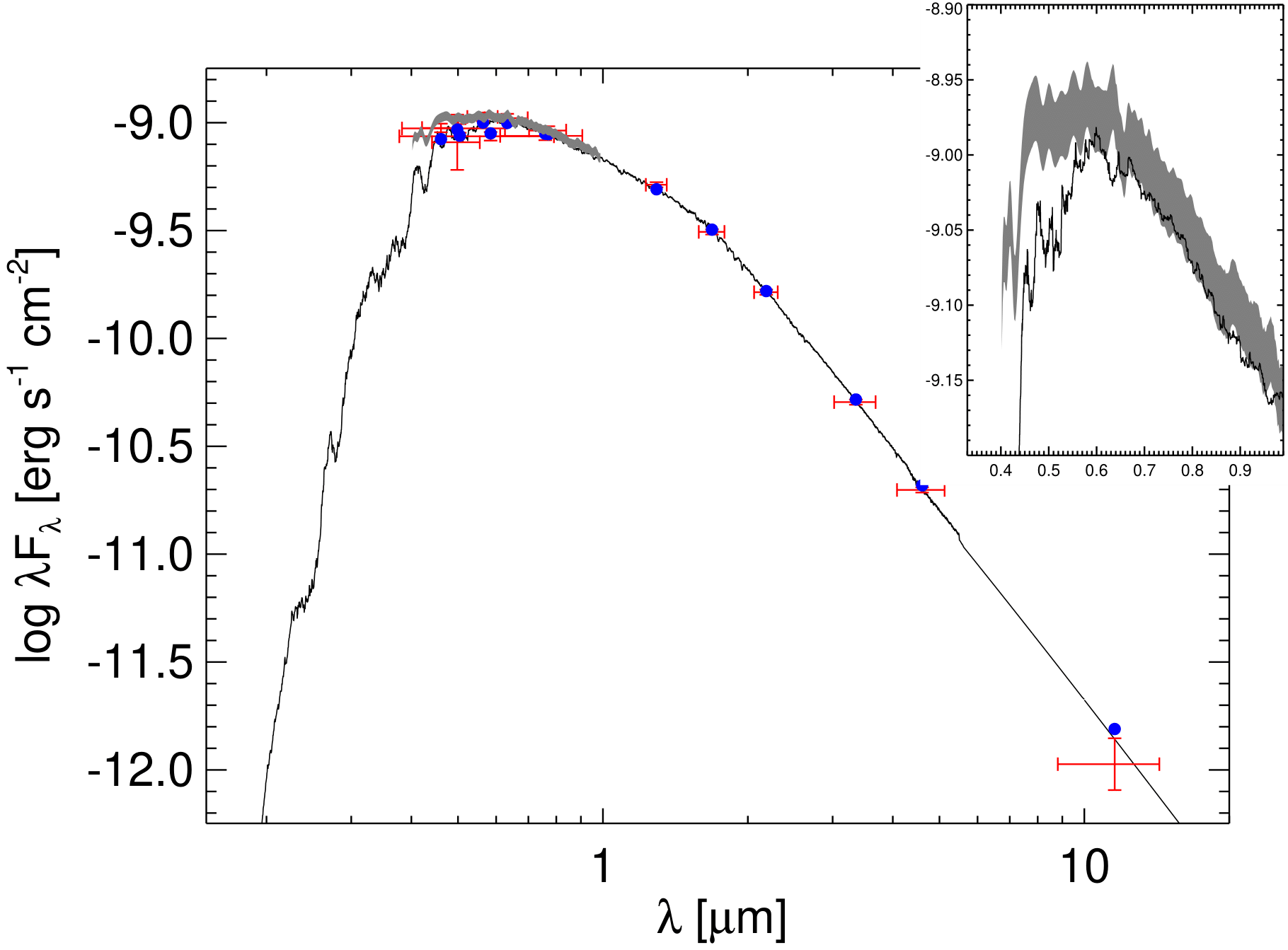}
  \caption{TOI-4507}
\end{subfigure}
\hspace{0.01\textwidth}
\begin{subfigure}[b]{0.3\textwidth}
  \centering
  \includegraphics[width=\linewidth]{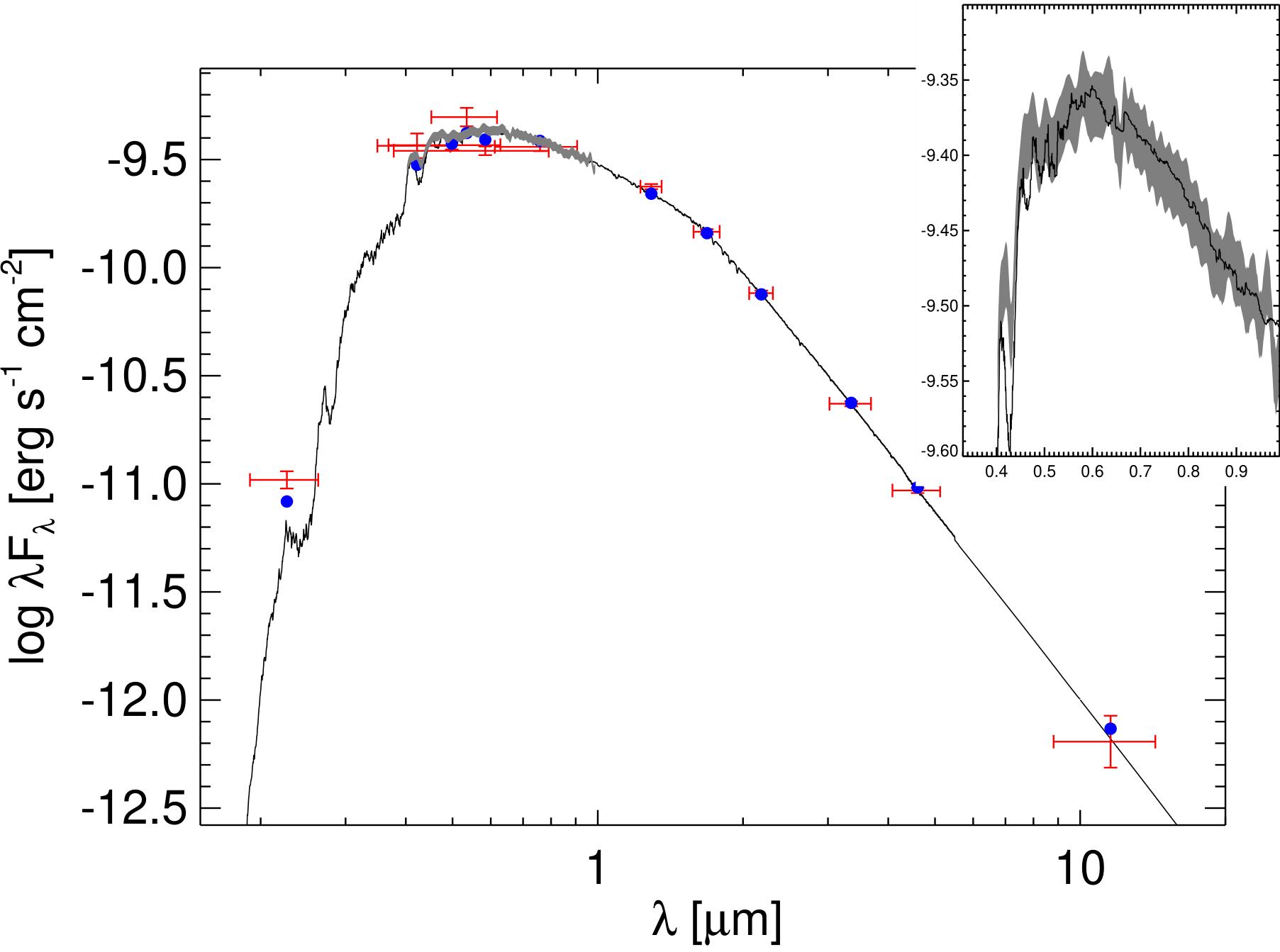}
  \caption{TOI-2404}
\end{subfigure}
\hspace{0.01\textwidth}
\begin{subfigure}[b]{0.3\textwidth}
  \centering
  \includegraphics[width=\linewidth]{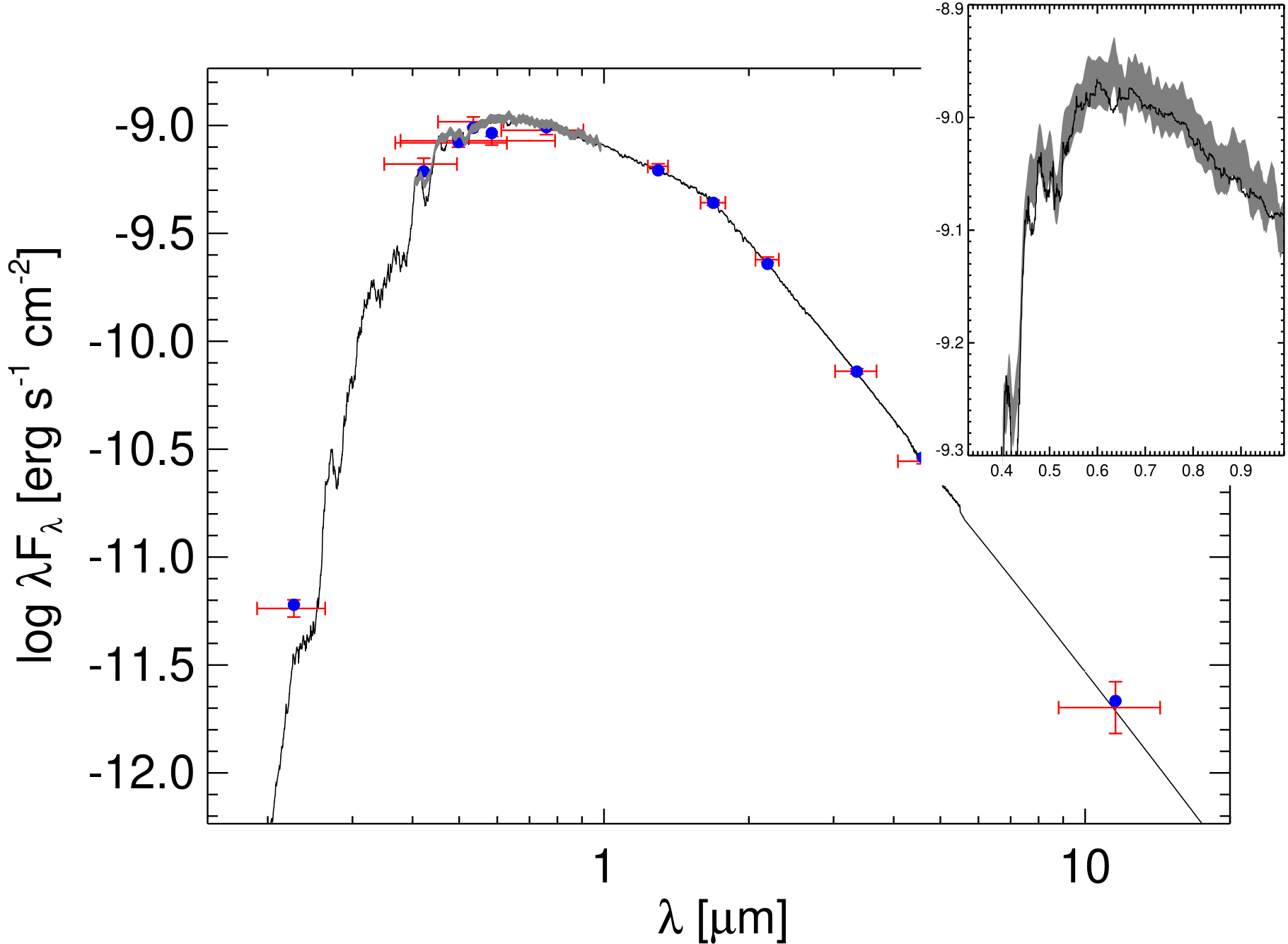}
  \caption{TOI-707}
\end{subfigure}
\hspace{0.01\textwidth}
\begin{subfigure}[b]{0.3\textwidth}
  \centering
 \includegraphics[width=\linewidth]{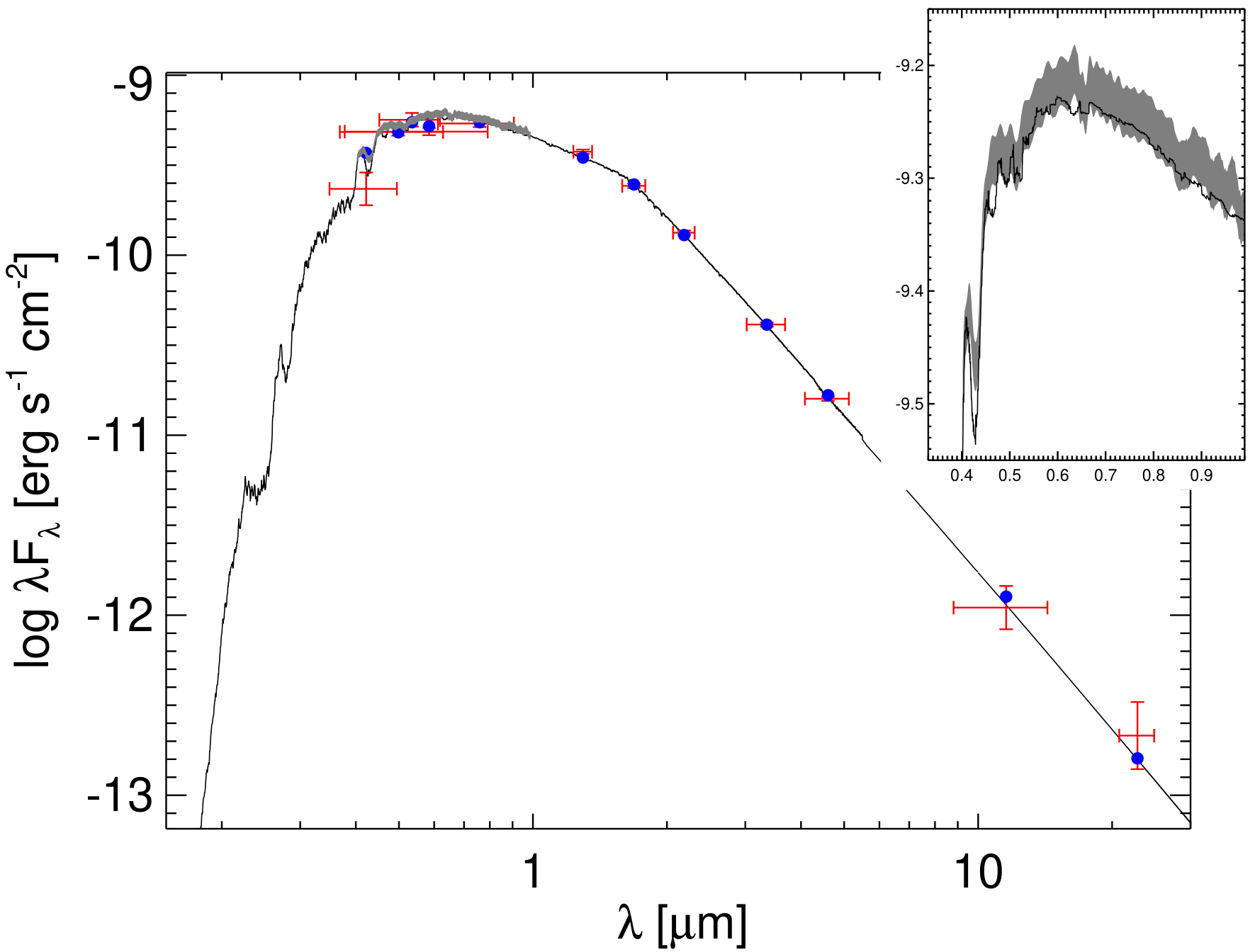}
  \caption{TOI-4404}
\end{subfigure}
\hspace{0.01\textwidth}
\begin{subfigure}[b]{0.3\textwidth}
  \centering
 \includegraphics[width=\linewidth]{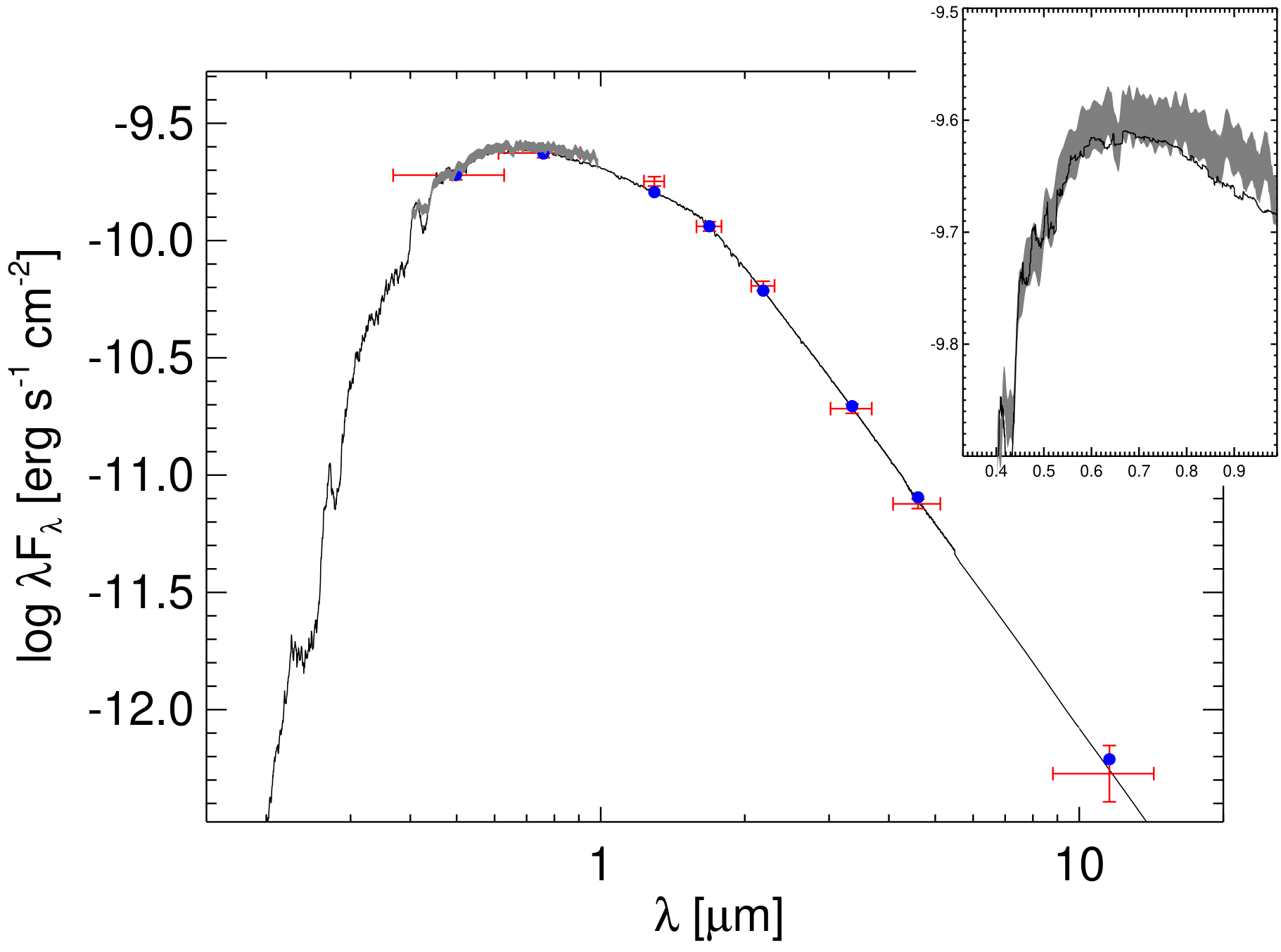}
  \caption{TOI-3457}
\end{subfigure}

\vspace{0.5em}
\caption{Spectral energy distribution (SED) fits for the five host stars. Each panel shows the observed broadband photometry and the best-fit stellar atmosphere model with corresponding band-integrated model fluxes. The SED-derived parameters are listed in Table~\ref{tab:stellar-params}. 
}
\label{fig:seds}
\end{figure*}

As explained in Section~\ref{sec:stellar_characterisation}, we performed SED analysis to further refine the stellar parameters. The resulting fits can be found in Fig.~\ref{fig:seds}.

\section{Priors adopted in the analysis}

\edit{Table~\ref{tab:priors_targets} summarises the priors adopted in the MCMC and nested sampling analyses for the main fitted parameters of each target.}

\begin{table*}
\scriptsize
\centering
\setlength{\tabcolsep}{6pt}
\caption{\edit{Priors adopted for the main fitted parameters of each target.}}
\begin{tabular}{lccccc}
\toprule
\textbf{Parameter} & \textbf{TOI-4507} & \textbf{TOI-2404} & \textbf{TOI-707} & \textbf{TOI-4404} & \textbf{TOI-3457} \\
\midrule

$P$ (d)
& 104.616 $\pm$ 0.005
& 74.607 $\pm$ 0.005
& 52.799 $\pm$ 0.005
& 39.624 $\pm$ 0.005
& 32.600 $\pm$ 0.005 \\

$T_0$ (BJD)
& 2459669.2445 $\pm$ 0.005
& 2459678.079 $\pm$ 0.01
& 2458345.5245 $\pm$ 0.005
& 2459248.475 $\pm$ 0.005
& 2459531.1692 $\pm$ 0.005 \\

$R_{\mathrm{comp}}/R_\star$
& 0.071 [0,0.1]
& 0.229 [0,10]
& 0.029 [0,0.1]
& 0.55 [0,1]
& 0.068 [0,0.1] \\

$(R_\star+R_{\mathrm{comp}})/a$
& 0.0117 [0,0.1]
& 0.0192 [0,0.1]
& 0.0287 [0,0.1]
& 0.024 [0,0.1]
& 0.0336 [0,0.1] \\

$\cos i$
& 0.001 [0,0.1]
& 0.0183 [0,0.1]
& 0.0249 [0,0.1]
& 0.023 [0,0.1]
& 0.0168 [0,0.1] \\

$K$ (km\,s$^{-1}$)
& 0.0012 [0,0.01]
& 0 [0,0.01]
& --
& --
& 0.074 [0,0.2] \\

$\sqrt{e}\cos\omega$
& --
& -0.449 [-0.451,-0.443]
& --
& --
& 0 [-1,1] \\

$\sqrt{e}\sin\omega$
& --
& 0.002 [-0.1,0.1]
& --
& --
& 0 [-1,1] \\

Dilution
& --
& 0 [0,1]
& --
& --
& -- \\

\bottomrule
\end{tabular}
\tablefoot{Values correspond to the initial guesses, while the numbers shown in brackets or after the $\pm$ symbol define the adopted uniform prior bounds.}
\label{tab:priors_targets}
\end{table*}

\section{Additional posterior parameters table}
Section~\ref{sec:Analysis} outlines the model fitting and analysis approach used in this study. Table~\ref{tab:posteriors_extra} presents the posterior distributions for the limb darkening coefficients, systematic noise parameters, and white noise levels that were included in the fit. These values provide insight into the photometric precision and model assumptions adopted for each target.

\begin{table*} 
   \scriptsize
   \centering
    \renewcommand{\arraystretch}{1.6} 
    \setlength{\tabcolsep}{10pt} 
    \caption{\edit{Summary of additional model parameters.}}
    \begin{tabular}{lccccc}
        \toprule
         & \textbf{4507.01}
         & \textbf{2404.02/03}  
         & \textbf{707.01} 
         & \textbf{4404.01} 
         & \textbf{3457.01} \\
        \midrule

        \multicolumn{6}{l}{\textit{Limb-darkening coefficients}} \\
        $a_{\mathrm{R}}$ 
            & 0.4422 
            & 0.4363 
            & 0.4883 
            & 0.5031 
            & 0.4923 \\
        $b_{\mathrm{R}}$ 
            & 0.1932 
            & 0.1956 
            & 0.1737 
            & 0.1607 
            & 0.1646 \\
        
        $a_{\mathrm{B}}$ 
            & 0.6754 
            & 0.6652 
            & 0.7697 
            & 0.8086 
            & 0.7871 \\
        $b_{\mathrm{B}}$ 
            & 0.1411 
            & 0.1554 
            & 0.0619 
            & 0.0302 
            & 0.0485 \\
        
        $a_{\mathrm{i'}}$ 
            & -- 
            & 0.3652 
            & -- 
            & -- 
            & 0.4086 \\
        $b_{\mathrm{i'}}$ 
            & -- 
            & 0.1976 
            & -- 
            & -- 
            & 0.1809 \\
        
        $a_{\mathrm{g'}}$ 
            & -- 
            & -- 
            & -- 
            & 0.7411 
            & -- \\
        $b_{\mathrm{g'}}$ 
            & -- 
            & -- 
            & -- 
            & 0.0591 
            & -- \\
        
        $a_{\mathrm{z'}}$ 
            & -- 
            & -- 
            & -- 
            & 0.3458 
            & -- \\
        $b_{\mathrm{z'}}$ 
            & -- 
            & -- 
            & -- 
            & 0.1875 
            & -- \\
        
        $a_{\mathrm{r'}}$ 
        & -- 
            & -- 
            & -- 
            & 0.5281 
            & -- \\
        $b_{\mathrm{r'}}$ 
            & -- 
            & -- 
            & -- 
            & 0.1603 
            & -- \\
            
        $q_{1 -\mathrm{TESS}}$ 
            & \hostldcqoneTESSA 
            & \hostldcqoneTESSC 
            & \hostldcqoneTESSE 
            & \hostldcqoneTESSB 
            & \hostldcqoneTESSthreeD \\
        
        $q_{2 -\mathrm{TESS}}$ 
            & \hostldcqtwoTESSA 
            & \hostldcqtwoTESSC 
            & \hostldcqtwoTESSE 
            & \hostldcqtwoTESSB 
            & \hostldcqtwoTESSthreeD \\[0.1cm]

         \multicolumn{6}{l}{\textit{Gaussian-process hyperparameters (Matérn-3/2)}} \\
        $\ln{\mathrm{GP}_{\sigma; \mathrm{TESS}}}$
            & \baselinegpmaternthreetwolnsigmafluxTESSA 
            & \baselinegpmaternthreetwolnsigmafluxTESSC 
            & \baselinegpmaternthreetwolnsigmafluxTESSE 
            & \baselinegpmaternthreetwolnsigmafluxTESSB 
            & \baselinegpmaternthreetwolnsigmafluxTESSthreeD \\
        
        $\ln{\mathrm{GP}_{\rho; \mathrm{TESS}}}$
            & \baselinegpmaternthreetwolnrhofluxTESSA 
            & \baselinegpmaternthreetwolnrhofluxTESSC 
            & \baselinegpmaternthreetwolnrhofluxTESSE 
            & \baselinegpmaternthreetwolnrhofluxTESSB 
            & \baselinegpmaternthreetwolnrhofluxTESSthreeD \\[0.1cm]
        \multicolumn{6}{l}{\textit{Photometric noise parameters}} \\

        $\log{\sigma_\mathrm{TESS}}$
            & \lnerrfluxTESSA 
            & \lnerrfluxTESSC 
            & \lnerrfluxTESSE 
            & \lnerrfluxTESSB 
            & \lnerrfluxTESSthreeD \\
        $\log{\sigma_\mathrm{ASTEP; 2021; R}}$ 
            & -- 
            & --
            & -- 
            & {\lnerrfluxASTEPtwozerotwooneB} 
            & -- \\

        $\log{\sigma_\mathrm{ASTEP; 2022; R}}$ 
            & -- 
            & {\lnerrfluxASTEPtwozerotwotwoC} 
            & {\lnerrfluxASTEPE} 
            & -- 
            & -- \\
        
        $\log{\sigma_\mathrm{ASTEP; 2022; B}}$ 
            & -- 
            & --
            & {\lnerrfluxASTEPBE} 
            & -- 
            & -- \\
            
        $\log{\sigma_\mathrm{ASTEP; 2023; R}}$ 
            & {\lnerrfluxASTEPtwozerotwothreeA} 
            & {\lnerrfluxASTEPtwozerotwothreeaC}
            & -- 
            & {\lnerrfluxASTEPtwozerotwothreeB} 
            & {\lnerrfluxASTEPtwozerotwothreeD} \\

        $\log{\sigma_\mathrm{ASTEP; 2023; B}}$ 
            & {\lnerrfluxASTEPtwozerotwothreeBA} 
            & {\lnerrfluxASTEPtwozerotwothreeaBC}
            & -- 
            & --
            & {\lnerrfluxASTEPtwozerotwothreeBD} \\

        $\log{\sigma_\mathrm{ASTEP; 2023 (2); R}}$ 
            & -- 
            & {\lnerrfluxASTEPtwozerotwothreecC}
            & -- 
            & -- 
            & -- \\

        $\log{\sigma_\mathrm{ASTEP; 2023 (2); B}}$ 
            & --
            & {\lnerrfluxASTEPtwozerotwothreecBC} 
            & -- 
            & -- 
            & -- \\
            
        $\log{\sigma_\mathrm{ASTEP; 2024; R}}$ 
            & --
            & --
            & -- 
            & {\lnerrfluxASTEPtwozerotwofourB}
            & {\lnerrfluxASTEPtwozerotwofourD} \\
            
        $\log{\sigma_\mathrm{ASTEP; 2024; B}}$ 
            & -- 
            & --
            & -- 
            & $-6.486\pm0.032$ 
            & -- \\
            
        $\log{\sigma_\mathrm{ASTEP; 2025; R}}$ 
            & {\lnerrfluxASTEPtwozerotwofiveA}
            & --
            & -- 
            & --
            & -- \\
            
        $\log{\sigma_\mathrm{MoanaES; r}}$ 
            & --
            & {\lnerrfluxMoanatwozerotwothreeaRC}
            & -- 
            & --
            & {\lnerrfluxMoanaRD} \\
            
        $\log{\sigma_\mathrm{LCO-CTIO; g}}$ 
            & --
            & --
            & -- 
            & {\lnerrfluxCTIOGPB}
            & -- \\
            
        $\log{\sigma_\mathrm{LCO-CTIO; ip}}$ 
            & --
            & {\lnerrfluxCTIOIPC}
            & -- 
            & --
            & {\lnerrfluxCTIOIPD} \\
            
        $\log{\sigma_\mathrm{PEST}}$ 
            & --
            & --
            & -- 
            & --
            & -- \\[0.1cm]
        \multicolumn{6}{l}{\textit{Logarithmic representation of the RV jitter $\sigma_{\mathrm{RV}}$ (km\,s$^{-1}$)}} \\
        $\log{\sigma_\mathrm{FEROS}}$ 
            & \lnjitterrvFEROSA 
            & --
            & -- 
            & --
            & \lnjitterrvFEROSD \\

        $\log{\sigma_\mathrm{HARPS-1}}$ 
            & \lnjitterrvHARPSoneA 
            & --
            & -- 
            & --
            & -- \\

        $\log{\sigma_\mathrm{HARPS-2}}$ 
            & \lnjitterrvHARPStwoA 
            & --
            & -- 
            & --
            & -- \\[0.1cm]
          \multicolumn{6}{l}{\textit{Instrument zero-point offset (km\,s$^{-1}$)}} \\
        $\Delta \mathrm{RV_{FEROS}}$
            & \baselineoffsetrvFEROSA 
            & --
            & -- 
            & --
            & \baselineoffsetrvFEROSD \\

        $\Delta \mathrm{RV_{HARPS-1}}$
            & \baselineoffsetrvHARPSoneA 
            & --
            & -- 
            & --
            & -- \\

        $\Delta \mathrm{RV_{HARPS-2}}$
            & \baselineoffsetrvHARPStwoA 
            & --
            & -- 
            & --
            & -- \\[0.1cm]
            \multicolumn{6}{l}{\textit{Linear RV slope (m\,s$^{-1}$\,day$^{-1}$)}} \\
        RV slope
            & \stellarvarslopervA 
            & --
            & -- 
            & --
            & -- \\

        \bottomrule
    \end{tabular}
    \tablefoot{Shown are the quadratic limb-darkening coefficients $a$ and $b$ for the ground-based instruments' various bandpasses from \citet{2013_Claret}; the white noise parameters $\sigma$ and GP Matérn 3/2 hyperparameters $\mathrm{GP}_{\sigma}$ and $\mathrm{GP}_{\rho}$ for the respective instruments as derived in this work. \edit{All logarithms denoted as \textit{log} are base 10, while \textit{ln} indicates the natural logarithm.}}
    \label{tab:posteriors_extra}
\end{table*}

\section{Empirical masses and predicted RV semi-amplitudes}\label{app:empirical_masses}

Since TOI-707\,b lacks radial-velocity data, its mass was inferred from empirical mass-radius relations using the results from \citet{ChenKipping2017}. For consistency, we also computed empirical-mass values for TOI-4507\,b and TOI-3457\,b. The adopted empirical masses are
$M_p=50.6\,M_\oplus$ (TOI-4507\,b), $6.43\,M_\oplus$ (TOI-707\,b), and $80.0\,M_\oplus$ (TOI-3457\,b). Using these empirical masses together with the orbital solutions of Table \ref{tab:tois_alles} (circular for TOI-4507\,b and TOI-707\,b; $e=0.684$ for TOI-3457\,b), we compute the expected RV semi-amplitude with the standard Keplerian expression \citep{1999Cumming}:
\begin{equation*}
    K_{\rm pred}=\left(\frac{2\pi G}{P}\right)^{1/3}\frac{M_p\sin i}{(M_\star+M_p)^{2/3}}\frac{1}{\sqrt{1-e^2}}\, .
\end{equation*}
The resulting semi-amplitudes are 
$K_{\rm pred}=6.26\ {\rm m\,s^{-1}}$ (TOI-4507\,b), 
$1.10\ {\rm m\,s^{-1}}$ (TOI-707\,b), and 
$21.4\ {\rm m\,s^{-1}}$ (TOI-3457\,b), which can be directly compared with the $K_\square$ upper limits reported in Table \ref{tab:tois_alles}.

\section{Atmospheric characterisation prospects}\label{app:atmospheric_characterisation_prospects}

For the three validated planets, TOI-4507\,b, TOI-707\,b, and TOI-3457\,b, the observational prospects for atmospheric characterisation with JWST and \textit{Ariel} are estimated at first order using the Transmission (TSM) and Emission Spectroscopy Metrics (ESM) \citep{Kempton_2018}. When empirical masses are required (see Appendix~\ref{app:empirical_masses}), they are used consistently in these estimates. Adopting empirical masses, the TSM values for TOI-707\,b and TOI-3457\,b are modest and lie below commonly used follow-up thresholds ($\geq 90$). For TOI-4507\,b the empirical-mass TSM is also below threshold (of order $\sim\!60$, depending on the albedo assumption). In contrast, when the RV-derived mass for TOI-4507\,b is used, the TSM becomes high ($\sim\!200$, or $\sim\!180$ for $A_{\rm B}=0.3$), exceeding standard giant-planet cutoffs. 
For TOI-3457\,b, adopting its RV-derived mass does not change the qualitative conclusion: its TSM remains below commonly used thresholds. 
All three planets yield ESM values below the typical threshold of 7.5, indicating that thermal-emission spectroscopy would be more challenging.
\end{appendix}

\end{document}